# Towards Refactoring of DMARF and GIPSY Case Studies

Afshin Somani (6765793 ) Concordia University Montreal, Canada

Ahmad Al-Sheikh Hassan (6735029) Concordia University Montreal, Canada

Anurag Reddy Pedditi (6862322) Concordia University Montreal, Canada

Challa Sai Sukesh Reddy (6847250) Concordia University Montreal, Canada

Vijay Nag Ranga (6745814) Concordia University Montreal, Canada

Saravanan Iyyaswamy Srinivasan (7090838) Concordia University Montreal, Canada

Hongyo Lao (6871240) Concordia University Montreal, Canada

Zhu Zhili (6954618) Concordia University Montreal, Canada

ABSTRACT

***Software Quality is a major concern in software engineering development in order to be competitive. Such a quality can be achieved by a possible technique called Refactoring where the systems external behaviour of the system is not changed. Initially we present our work by analyzing the case studies of ongoing researches of DMARF and GIPSY by understanding their needs and requirements involving the major components in their respective systems. Later sections illustrate the conceptual architecture of these case studies, for this we have referenced the original architecture to draw the important candidate concepts presented in the system, and analyzing their associations with other concepts in the system and then compared this conceptual architecture with the original architectures. Later the document throws light on identifying the code smells exist in the architectures to find them and resolve to minimize the deeper problems. Jdeodorant, SonarQube are the tools which we go across for identification and analyzing the source code quality, both these tools are available as an IDE plugin or as an open source platforms. Next is to identify the design patterns exist in the architectures along with their importance and need for existence in respective systems. Finally, the implication is towards introducing refactoring methods onto the smells which have been identified and possibly refactor them accordingly by applying appropriate refactoring methods and showcasing the respective tests to ensure that changes in the architecture does not change the behavior much.***

INTRODUCTION

This report primarily focuses on two case studies DMARF and GIPSY starting with understanding the needs and requirements, architecture design reconstruction, and actual architecture, architecture fusion with respect to the two case studies. Later throws light on design patterns recognition, code smells identifications, and the interrelated refactoring methods. JDeodorant, SonarQube are used to analyze the quality of the case studies with reference to its source code. ObjectAid UML Explorer has been used as a reverse engineering tool to derive the actual architecture of the two software. Finally, implemented four refactoring for each case study with supporting test cases and corresponding results are interpreted.

## I. BACKGROUND

*OSS CASE STUDIES*

### *A. DMARF*

Modular Audio Recognition Framework (MARF) is an open-source research platform written in Java with a collection of pattern recognition, signal processing, and natural language processing (NLP) algorithms. The main goal of MARF is to compare the algorithms and allow them for dynamic module selection based on the configurations given by the application.

Distributed MARF (DMARF) is based on classical MARF whose pipeline stages were made into distributed nodes and as a front-end. DMARF supports high volume processing of recorded audio, imagery or textual data for pattern recognition and bio metric applications as its domains. It emphasizes on audio processing, such as conference recordings to identify the speakers for forensic analysis to perform subject identification and classification.

DMARF is built on classic MARF where the difference between can be noticed in the pipeline structure.
Classic MARF, The pipeline stages [1]:

- sample loading
- preprocessing,
- feature extraction
- training/classification

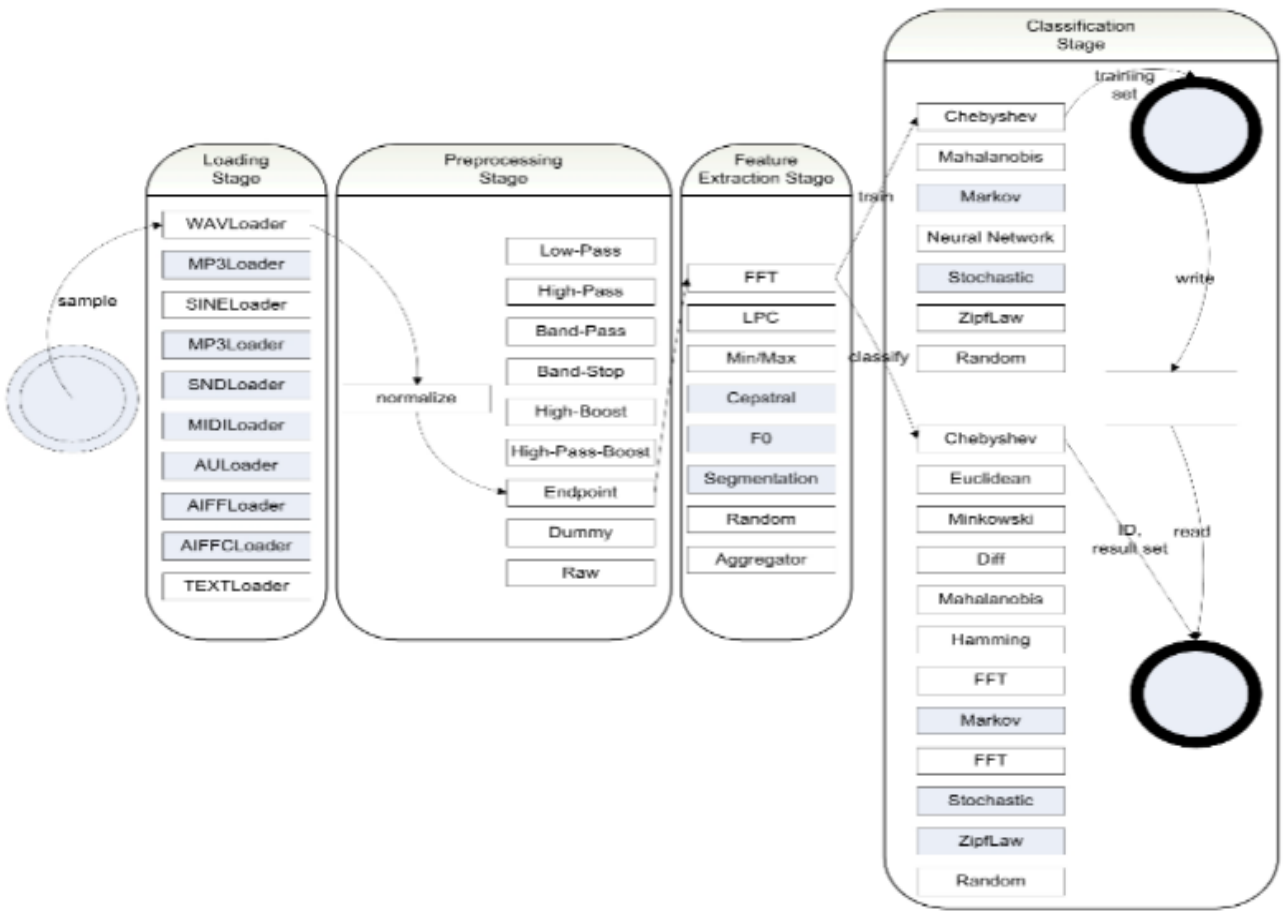

**Figure. 1. MARF's Pattern Recognition Pipeline [1].**

MARF has the ability of adding any module/algorithm implementation at any stage of the pipeline of pattern recognition Figure 1, [1]. In this regard, DMARF has been introduced as the distributed version of MARF as the stages run as distributed nodes as well as a front-end, Figure 2. The four stages and the front-end have been employed without backup recovery or hot swappable capabilities: the communication is to be done over Java RMI, CORBA], and XML-RPC web services.

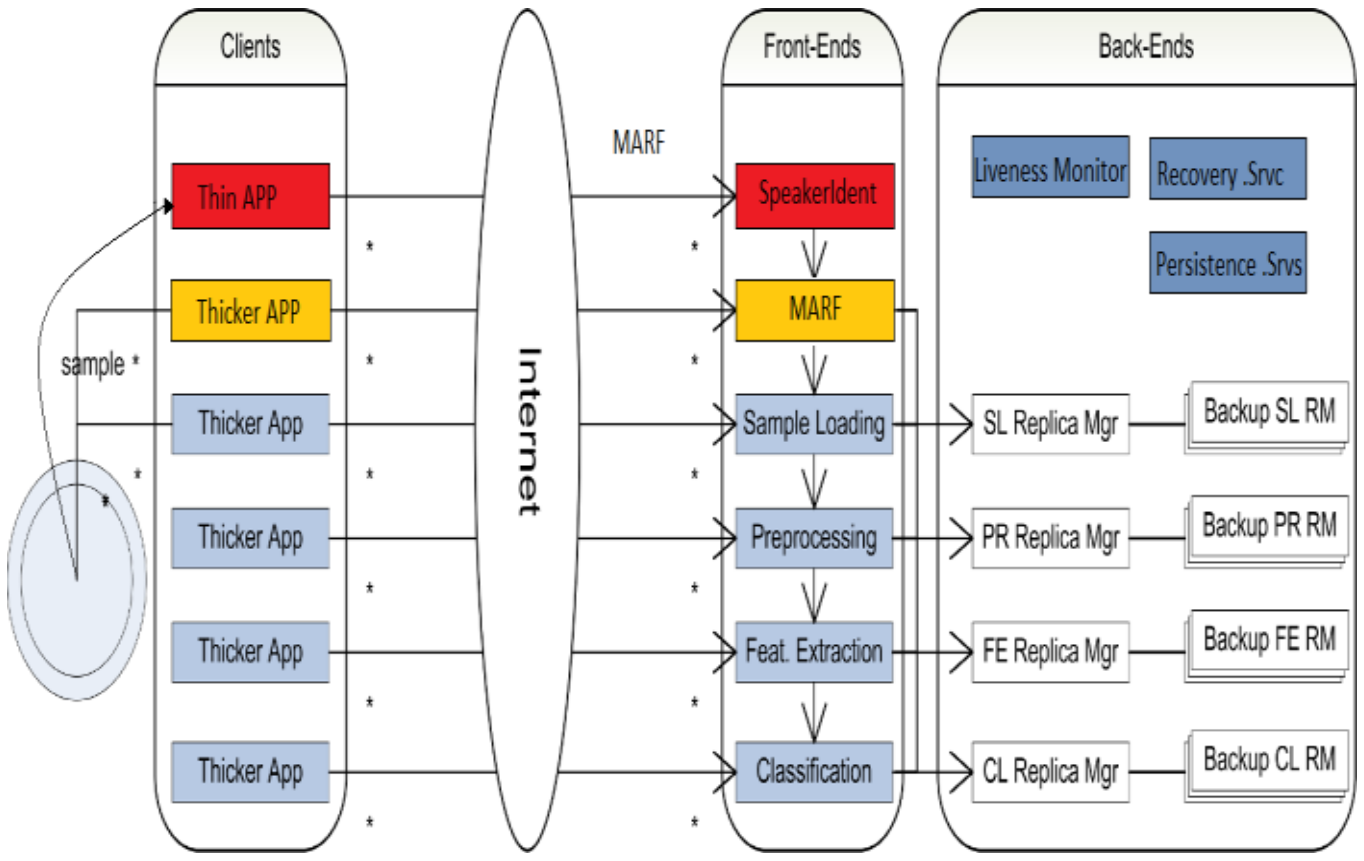

**Figure 2. DMARF's Pattern Recognition Pipeline [1].**

MARF is a desktop application which processes recorded audio (speaker identification service), textual data, or imagery data. In order to have this service more popular and more handy, DMARF has been introduced as a web application allowing this service to be available on line. The main focus is in audio, such as conference recordings to address what has been said to identities of speakers. This service can be used for security purposes as well [1].

**Requirements:**

- Different levels of **front ends**, from higher to lower; a client application may invoke. Some services may invoke other services via their front-ends at the same time while executing in a pipelined mode [1].
- The **back-ends** provide the actual servant applications and some other features such as primary backup replication, monitoring, and disaster recovery modules [1].

There are many distributed services that intercommunicate with each other in DMARF; some are of them are general services that expose the pipeline to the client and communicate with other service to perform the task, and some others are more specific front-end services based on existing non-distributed applications such as The Speaker Identification service that communicates with MARF service to perform the application tasks [1].

**System Architecture:**

**Module View:** The system is composed of **layers**

The **top layer** consists of a front-end and a back-end. The front-end exists on the client side (e.g. a web form/servlet collection of client classes that connect and query the servers), and it exists on the server side (MARF Pipeline), where all pipelines stages are concerned with data base and other storage sub functions figure (3). The service which connect the client are the back-end [1].

**Execution view**:

*Runtime entities:* Java Virtual Machine (JVM), and on the server side there must be a DNS running and a web servlet container (Apache Tomcat. The WS client require (JRE), a servlet container environment, and a browser to view and submit a web form [1].

*Communication Path:* The modules communicate through message passing between methods; a Java XML Remote Procedure Call (JAX-RPC 1.1)-based implementation over the Simple Object Access Protocol (SOAP) is used for Web Services (WS). Each terminal business logic module in MARF (StorageManager class) is responsible for communication to the Data Base. Java's reflection is used to reveal instantiation communication paths at run-time for pluggable modules [1].

*Execution Configuration*: The execution configuration is concerned with where its data/ and policies/ directories are. In the case of WS, it has to be where Tomcat's current directory [1].

**Proof-of-Concept Prototype Assumption [1]:**

- No garbage collection on the server (completely limiting the WAL size or outdated data in the training sets or any other DB.
- DMARF services do not apply nested transaction while pipelining.

- WAL functionality has been implemented only for the Classification Service.
- Services intercommunicate ONLY through the pipeline mode of operation.
- Replication is present in case of primary-backup absence.

Three designs namely Transactions, Recoverability and WAL (Write Ahead Log) are present. WAL gives you the ability to keep the record of every request, so when a process crashes, it can resume from the last working point. Transaction is a data structure which maintains transactions, id, and file name of the object, serializable value and time stamps [2].

The classical MARF's pipeline in Figure 1 is to distribute stages that are not directly present in the figure-sample loading and front-end application service (e.g. speaker identification service, etc.). It also implements some disaster recovery with replication techniques in the distributed system [3].

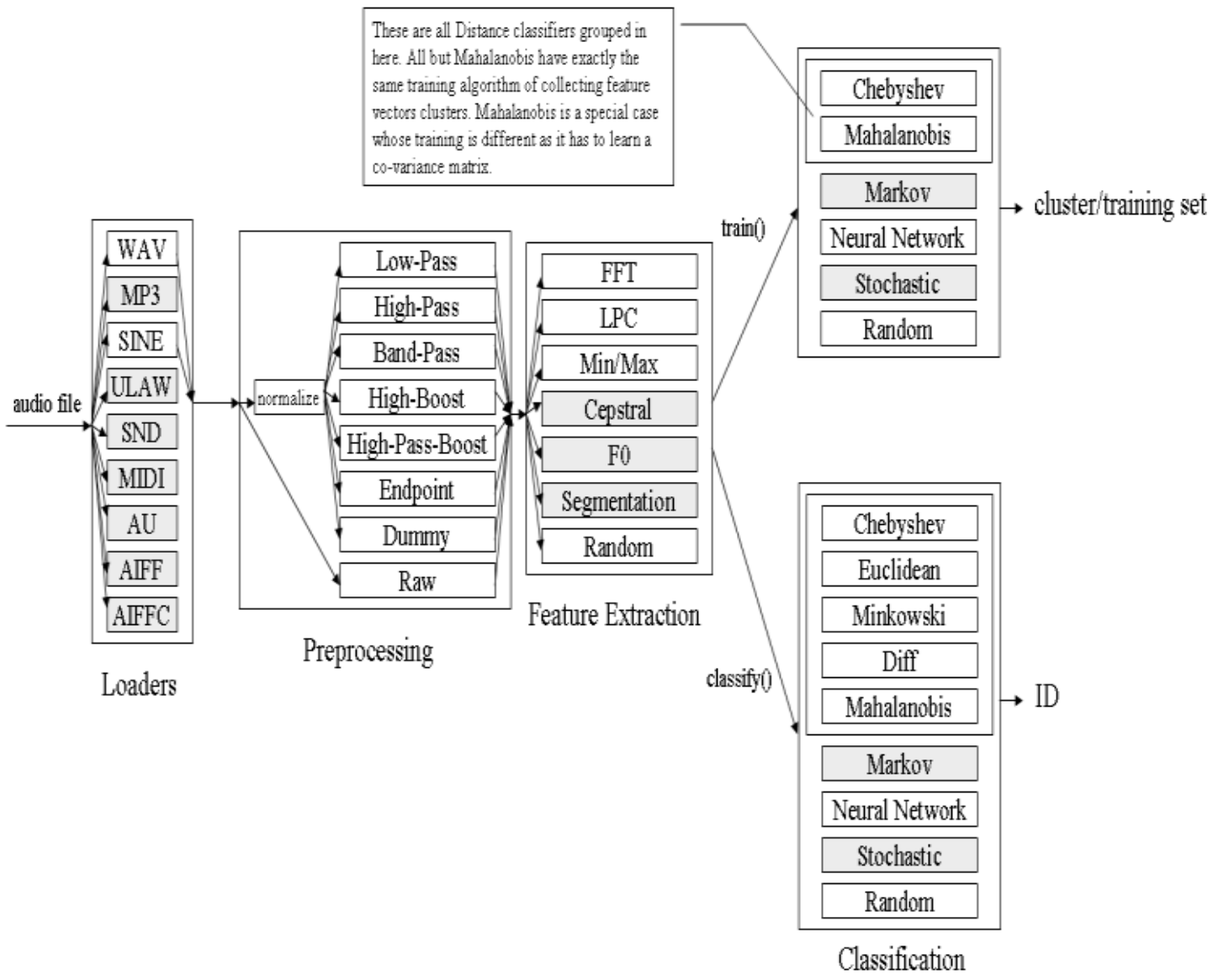

**Figure 3: The Core MARF Pipeline Data Flow [3]**

Some service types of MARF are,

- Application Services
- General MARF Pipeline Services
- Sample Loading Services
- Preprocessing Services
- Feature Extraction Services
- Training and Classification Services

These services are backed up by their corresponding server implementations in CORBA, Java RMI, and Web Services XML-RPC. Services can potentially be embedded into other application or hardware systems for speaker and language identification [3].

Three designs namely Transactions, Recoverability and WAL (Write Ahead Log) are present. WAL gives you the ability to keep the record of every request, so when a process crashes, it can resume from the last working point. Transaction is a data structure which maintains transactions, id, and file name of the object, serializable value and time stamps [3].

MARF has many applications, like SpeakerIdentApp and LangIdentApp for speaker and language identification.

Some research and implementation details to amend Distributed MARF [3]

- Finish proxy agents and instrumentation.
- Implement our own managers and the functions to compile new MIBs into the manager.
- Complete prototyped GUI for ease-of-use of our management applications (as-is MARF is mostly console-based).
- Complete full statistics MIB and implement RMON along with some performance management functions such as collecting statistics and plotting the results.
- Propose a possible RFC.
- Make a public release and a publication.
- Implement some fault management functions such as alarms reporting.
- Look into XML in Network Management (possibly for XML-RPC).
- Look more in detail at Java and network management, JMX (right now through AdventNet).
- Distributed Management of different DMARF nodes from various locations.
- Management of Grid-based Computing in DMARF.
- Analysis of CORBA and where it fits in Network Management in DMARF.
- Multimedia Management using SNMP.

On inclusion of autonomous property to DMARF, it can be extended to robotic systems that require less-to none human intervention for pattern analysis. Autonomic Computing (AC) is inspired from the human nervous system-the main idea is that the software system should be able to manage itself with the dynamic requirements and threats just as the human body does. This principle of autonomous computing can be used to solve various problems of distributed pattern recognition like security, availability etc. [4].

The Autonomic System Specification Language (ASSL) consists of three main tiers and some sub tiers, these tiers are helpful in providing specifications of the system varying in different levels of abstraction and also assist in reducing the complexity, overall improving the perception of the system. Autonomic System (AS) is the first tier, it represents a general and global outlook, and general autonomic rules are applied in terms of service level objectives, self-management features, topology and global actions. AS interaction protocol is the second tier, it is responsible for communication between AE (Autonomic Elements) and also consists of channels, communication functions and messages. Autonomic Elements is the third tier, it consists of AE rules (self-management policies), AE interaction protocol (AEIP), AE actions, AE events, AE metrics, AE friends (List of AE with a level of trust), in this tier individual AE with their own behavior in their interactive sets are defined. ASSL is not only applicable for distributed systems but also can be applied to pipelined distributed systems [4].

ASSL framework is used to develop and integrate self-management features to DMARF. This features enhances DMARF by introducing an autonomous middleware that is responsible for managing four stages of frameworks recognition pipeline in addition to pattern analysis, natural language processing and signal processing.
ASSL framework follows a multi tire architecture which includes autonomic properties like [4]

- Self-Configuration
- Self-Healing
- Self-Optimization
- Self-Protection

**ASSL Multi-Tier Model**

I. Autonomic System (AS)
* AS Service-level Objectives
* AS Self-managing Policies
* AS Architecture
* AS Actions
* AS Events
* AS Metrics
II. AS Interaction Protocol (ASIP)
* AS Messages
* AS Communication Channels
* AS Communication Functions
III. Autonomic Element (AE)
* AE Service-level Objectives
* AE Self-managing Policies
* AE Friends
* AE Interaction Protocol (AEIP)
- AE Messages
- AE Communication Channels
- AE Communication Functions
- AE Managed Elements
* AE Recovery Protocol
* AE Behavior Models
* AE Outcomes
* AE Actions
* AE Events
* AE Metrics

Self-healing property towards autonomic specification of DMARF with ASSL is discussed.

DMARF self-healing requirements:

In DMARF pipeline, there are four main pipeline stages, in that if one pipeline stage goes offline then pipeline halts. So to recover this situation there is need to replace the failed node, recovery of failed node or change route of failed node with different route but with the same functionality. In that situation the pipeline should have self-healing technique [5].

Self-healing in DMARF:
DMARF must be able to recover by providing at least one available pipeline. It has two replications, replication of service and replication within the node itself.
Adding the autonomic computing behavior to the DMARF behavior results in ADMARF. Autonomic DMARF is capable of self-management, self-healing is the property of self-management. Here ASSL is to specify the node replacement and node recovery of ADMARF [5].

Self-healing algorithm specifies:
- ADMARF monitors runtime performance
- Every stage in ADMARF analyze the problem
- If node is down then node-replacement algorithm is performed by the AM stage
- If node is not performing then node-recovery algorithm is performed

ASSL self-healing algorithm is spread on both systems AS tier and AE tier (sub tiers in ASSL specification model).
AS tier:
In AS tier, global ADMARF self-healing behavior is specified. The process used for self-healing is same as self-management policy structure.
AE tier:
In AE tier, self-healing for each ADMARF is specified. Autonomic managers are used for each DMARF stages Figure (4).

```
ASSELF_MANAGEMENT {
  SELF_HEALING {
    // a performance problem has been detected
    FLUENT inLowPerformance {
      INITIATED_BY { EVENTS.lowPerformanceDetected }
      TERMINATED_BY {EVENTS.performanceNormalized,
                     EVENTS.performanceNormFailed }
    }
    MAPPING {
      CONDITIONS { inLowPerformance }
      DO_ACTIONS { ACTIONS.startSelfHealing }
    }
  }
} // ASSELF_MANAGEMENT
```

**Figure 4. Self- healing [5]**

**ASSL Self-Protection Model for DMARF**

For securing data's integrity and confidentiality in DMARF, ASSL introduces an autonomic property as self-protection.

| Tier | Actions | Security Check | Initiating Event | Terminating Event | Metric |
|---|---|---|---|---|---|
| AS | checkPublicMessage | inSecurity Check | publicMessageIsComing | PublicMessageSecure or publicMessageInsecure | hereIsInsecurePublic Message |
| AE | checkPrivateMessage | | PrivateMessageIsComing | PrivateMessageSecure or PrivateMessageInsecure | hereIsInsecurePrivate Message |

**Table 1: AS and AE Specification [6]**

For this property DMARF system adheres to a specification of node to node identification using proxy certificates and sender's digital signature. ASSL self-protection property involves changes in specification of the systems AS, AE, ASIP and AEIP tiers where all events, actions and metrics are performed Table (1).

**Specifications**

ASIP and ASEP tiers involve protocol specifications for public and private message communications respectively. Having a single bi-directional communication channel and two functions (send and receive message) the ASSL specification in these tiers search for proxy certificate of each message with their sender's digital signature. Incoming message failing to carry the depicted information is considered to be insecure and is discarded by the system with the help of a metric [6].
As like IP tier, AS and AE follow a self-protecting policy specification for public and private messages respectively. By initiating a security check fluent each message is tested for security figure (5). Checking whether the incoming message is an instance of ASSL and finding the sender's information is kept as the criteria in this policy. Message being instance of ASSL and having a valid digital signature of the sender clears the check else all the IO operations over the message is blocked [6].
These ASSL specifications in various tiers emphasizes DMARF to a self-protecting ADMARF (Autonomic DMARF) system. The proxy certificate validation is done using Java Data Security Framework (JDSF).

| Tier | Communication Channel | Message | Communication Function | Metric |
|---|---|---|---|---|
| ASIP | Public-Link | publicMessage | receivePublicMessages | thereIsInsecurePublicMessage |
| ASEP | Private-Link | privateMessage | receivePrivateMessages | thereIsInsecurePrivateMessage |

**Table 2: ASIP and ASEP Specification [6]**

```
SELF_PROTECTING {
  // a new incoming message has been detected
  FLUENT inSecurityCheck {
    INITIATED_BY { EVENTS.publicMessageIsComing }
    TERMINATED_BY { EVENTS.publicMessageSecure,
                    EVENTS.publicMessageInsecure }
  }
  MAPPING {
    CONDITIONS { inSecurityCheck }
    DO_ACTIONS { ACTIONS.checkPublicMessage }
  }
}
```

**Figure 5. Self-protecting [6]**

Self-optimization property in autonomic specification of distributed MARF with ASSL [7] is discussed.

MARF Self-Optimization Requirements

DMARF categorises itself as an autonomic system which essentially covers the autonomic functioning of the distributed pattern-recognition pipeline and its optimization, especially in its Classification stage

The two most stressed functional requirements applicable to large DMARF installations related to self-optimization are [7]:

*Training set classification data replication*: DMARF- based system does lot of multimedia data processing and number crunching throughout the pipeline [7]. In DMARF pipeline, the classification and sample loading stages stores huge amount of information for I/O bound data processing. In addition feature extraction along with classification stages stresses to do heavy computations. It has been observed that among all the stages, classification stage holds large amount of data, which creates need for re-computation or replication of already computed data transformed on to another classification suite. Usually it adds additional over head on the communication nodes and would require a lot of computational effort for data replication.

*Dynamic communication protocol selection* - One of the most prominent feature of self-optimization is its automatic selection of most eligible protocol suite which runs in the current environment. Assume that DMARF initially starts by choosing a particular protocol for its communication and later can make an impression by changing its current environment to another suitable and capable communication protocol, further promoting flexibility and easiness.

ASSL SELF-OPTIMIZATION MODEL FOR DMARF

The model mainly emphasizes to be autonomic, thus striving to be ADMARF (Autonomic DMARF), which complements the whole architecture with its behaviour over the system, and employs self-management policies. The autonomic behaviour is encoded in a special ASSL construct denoted as SELF_OPTIMIZING policy [7]. The basic procedure starts when ADMARF enters into the classification stage, where self-optimization takes place. Prior to the initiation of the real Computation, each nodes initially tries to acquire capable communication protocol.

The ASSL construct is specified at two other levels apart from SELF_OPTIMIZING policy, they are AS-tier and AE-tier

AS Tier Specification - In this tier the actions and events complementing SELF-OPTIMIZING policy where used, where ASSL supports policy specifications with special constructs called fluents and mappings figure(6) [7].

```
SELF_OPTIMIZING {
  // DMARF enters in the Classification Stage
  FLUENT inClassificationStage  {
     INITIATED_BY { EVENTS.enteringClassificationStage }
     TERMINATED_BY { EVENTS.optimizationSucceeded,
                     EVENTS.optimizationNotSucceeded }
  }
  MAPPING {
     CONDITIONS { inClassificationStage }
     DO_ACTIONS { ACTIONS.runGlobalOptimization }
  }
}
```

**Figure 6: AS Tier SELF_OPTIMIZING policy [7]**

From figure 6 it is indicative that policy is triggered when DMARF enters the classification stage, and when the *FLUENT inClassificationStage* is initiated [7].

AE Tier Specification- This tier specifies, a unique node for each distinct AE. Further the communication protocol would likely to be adopted with single node of the specification stage which is quite similar to that of AS tier.

ASSL has a self-forensics autonomic property (SFAP) to enable generation of the Java-based Object- Oriented Intensional Programming (JOOIP) language code laced with traces of Forensic Lucid to encode contextual forensic evidence and other expressions [8].

The ASSL framework takes specification of properties from autonomic systems as input, does formal syntax and semantics checks. If the check passes, it generates a Java collection of classes and interfaces corresponding to the specification [8].

Subsequently, a developer needs to fill in some overridden interface methods corresponding to the desired autonomic policies as a proxy implementation within the generated Java skeleton application or map them to the existing legacy application [11, 10, and 9].

Self –Forensics autonomic property in ASSL toolset includes two steps:

- Adding the syntax and semantic support to the lexical analyzer, parser, and semantic checker of ASSL[8]
- Adding the appropriate code generator for JOOIP and Forensic Lucid to translate forensic events. The JOOIP code is mostly Java with embedded fragments of Forensic Lucid-encoded evidence [12, 13].

JOOIP code is generated by the ASSL toolset, next process involves sending the code to hybrid complier of GIPSY. Inside the GEE engine JOOIP and forensic Lucid specifications are linked together. The 3 choices of evaluation after the above process includes

- Traditional eduction model of GEE
- Aspect J-based eduction model
- Probabilistic model checking with the PRISM backend [8].

## B. GIPSY

General Intensional Programming System (GIPSY) is a framework for compilation and execution of Intensional Programming languages based on demand - driven architecture. It is a multitier complex system concentrating on multidimensional conceptual languages like LUCID with flexibility and adaptability. GIPSY also consists of a homogenous environment which is used to type check all hybrid and intensional programs.

Intensional programming, in the sense of Lucid, is a programming language paradigm based on the notion of declarative programming where the declarations are evaluated in an inherent multidimensional context space [14].Intensional Programming languages like LUCID deals with complex multidimensional concepts and also evolve at faster rate. Generally, GLU is the tool used which couldn't compensate with the evolving adaptability and flexibility of this language which lead to the introduction of GIPSY.

### GIPSY Architecture

General Intensional Programming System (GIPSY) consists of three subsystems, built to improve efficiency. If a system needs to replace any of the subsystem, the efficiency remains intact. The three subsystems are:

- GIPC (General Intentional Programming Language Compiler)
- GEE (General Eduction Engine)
- RIPE (Intensional Runtime Programming Environment)

**GIPC**

GIPSY programs consist of two parts, the lucid (Data dependencies) and sequential (compilation units) figure (7). GIPC converts any given program to 'c' and then compiles it.

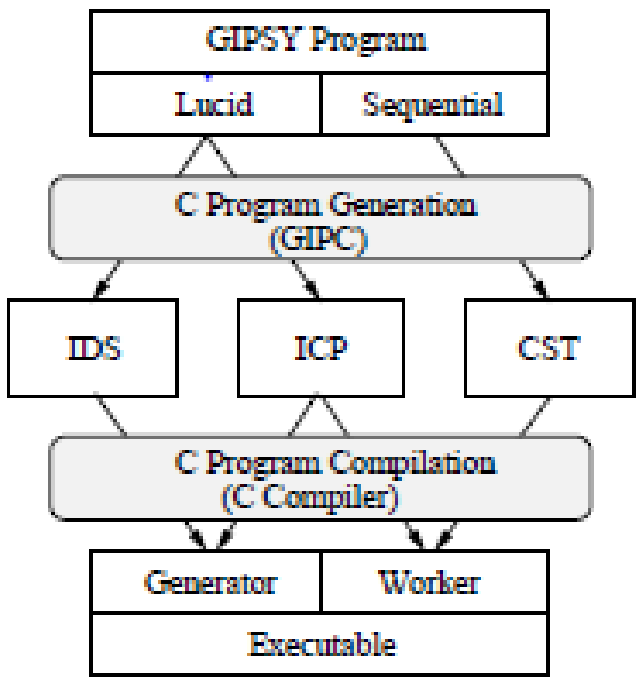

**Figure 7. GIPSY program compilation process [14]**

Conversion of program in c language indulges the creation of IDS (Intensional data dependency structure), ICP (Intensional communication procedure) and CST (C sequential threads) which deals with dependency, procedure calls and thread sequences respectively. Finally, a C compiler is used to form executable code from the program figure (8).

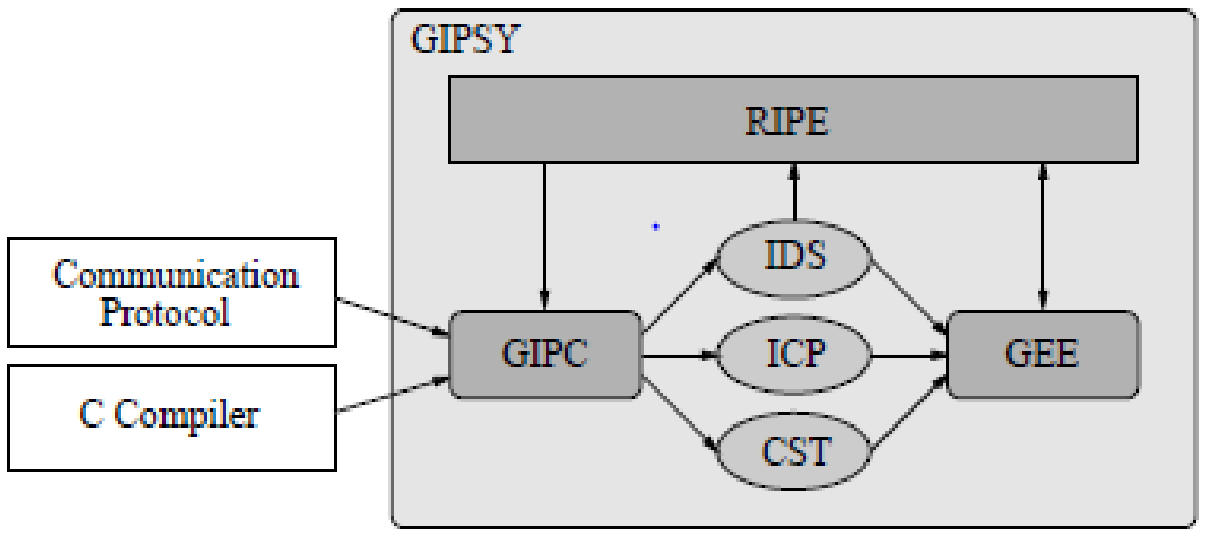

**Figure 8. GIPSY software Architecture [14]**

## GEE

GEE is an eduction engine made from the demand driven model using a generator worker architecture figure (9).
The engine receives procedure calls as demands, which it computes and stores in a cache named warehouse (IVW) .If the demand arrived is already computed then the result is extracted directly from the cache.

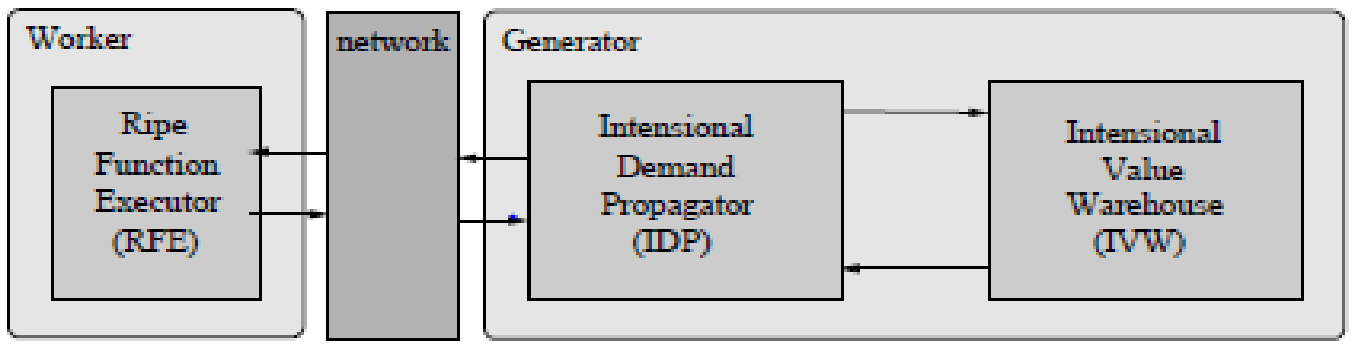

**Figure 9. Generator- worker execution architecture [14]**

Higher value procedure calls are evaluated in worker, and the lower value procedure calls are sent to the generator. All demands arrive in the queue and are computed on the basis of first come first serve.

## RIPE

RIPE is a visual run-time programming module. It translates the lucid program from graphical version to textual version and it compiles into operational version, also it can detect the data_flow during run-time of the system program.

The features that matter most for an architectural framework for runtime system which supports distributed execution through eductive model of computation are discussed.
Eduction can be described as "tagged-token demand driven dataflow" computing [15]. Core concept of this model of execution being generation, prorogation and consumption of demands.

The design implements a distributed multi-tier architecture where each tier can have any number of instances, where the execution is divided into three different tasks.
A GIPSY tier is an abstract and generic entity that represents a computational unit independent of other tiers [15].
A GIPSY node is a computer that has registered for the hosting of one or more GIPSY tiers.
A GIPSY Instance is a set of interconnected GIPSY Tiers deployed on GIPSY Nodes
Executing GIPSY programs
The Demand Generator Tier generates intensional demands and procedural demands according to an initial demand and the program declarations stored in the GEER generated for this GIPSY program.
The Demand Store Tier (DST) acts as a middleware between tiers in order to migrate demands between them.
The Demand Worker Tier is a tier that can process procedural demands. It consists of a Procedural Demand Processor that can process the value of any procedural demand corresponding to one of the elements of its Procedure Class Pool.
The GIPSY Instance Manager is a component that enables the registration of computational nodes to a GIPSY Instance and the allocation of various GIPSY tiers to these nodes, using a Nodes/Tiers Registrar.
Language independence [15]: Complier translates the system executed programs into generic language then provides a mechanism to wrap the functions into java classes. Lucid program and wrapper classes are integrated into GEER (Generic Eduction Engine Resources) which is language independent.
Scalability [15]: Scalability shares a major stake in successful implementation of a distributed system. The proposed Demand Store Tier (DST) solves the problem up to a great extent.
Flexibility of execution architecture [15]: The multi tire architecture incorporated in this paper makes huge Leaps
Opacity of run-time considerations [15]: The same GIPSY program can be executed in different execution topologies, which can be set prior to the starting of the program's execution, or even as the program is being executed [15].

Although a multi-threaded and distributed architecture using Java RMI has been initially designed, it was not fully integrated and many of the detailed working needed to be clarified. Meanwhile, two more separate branches of distributed computation for GIPSY emerged.
Creation of wrapper classes for each tier type-specifically DGT (Demand Generator Tier), DST (Demand Store Tier), DWT (Demand Worker Tier), and the GMT (General Manager Tier) is an evolution of the original architecture for the run-time system of the GIPSY.

It mentions four types of demands:

- Intensional demands

```
{GEERid, programId, context}
```

- Procedural demands

```
{GEERid, programId, Object params[], context, [code]}
```

- Resource demands

```
{resourceTypeId, resourceId}
```

- System demands

```
{destinationTierId, systemDemandTypeId, Object params[]}
```

Three classes which are essential for design and development are: EDMFImplemenation-enumerated type, Tier Factory-instance type, NodeController-abstaract type respectively.
The details of the structure are explained in the figure below

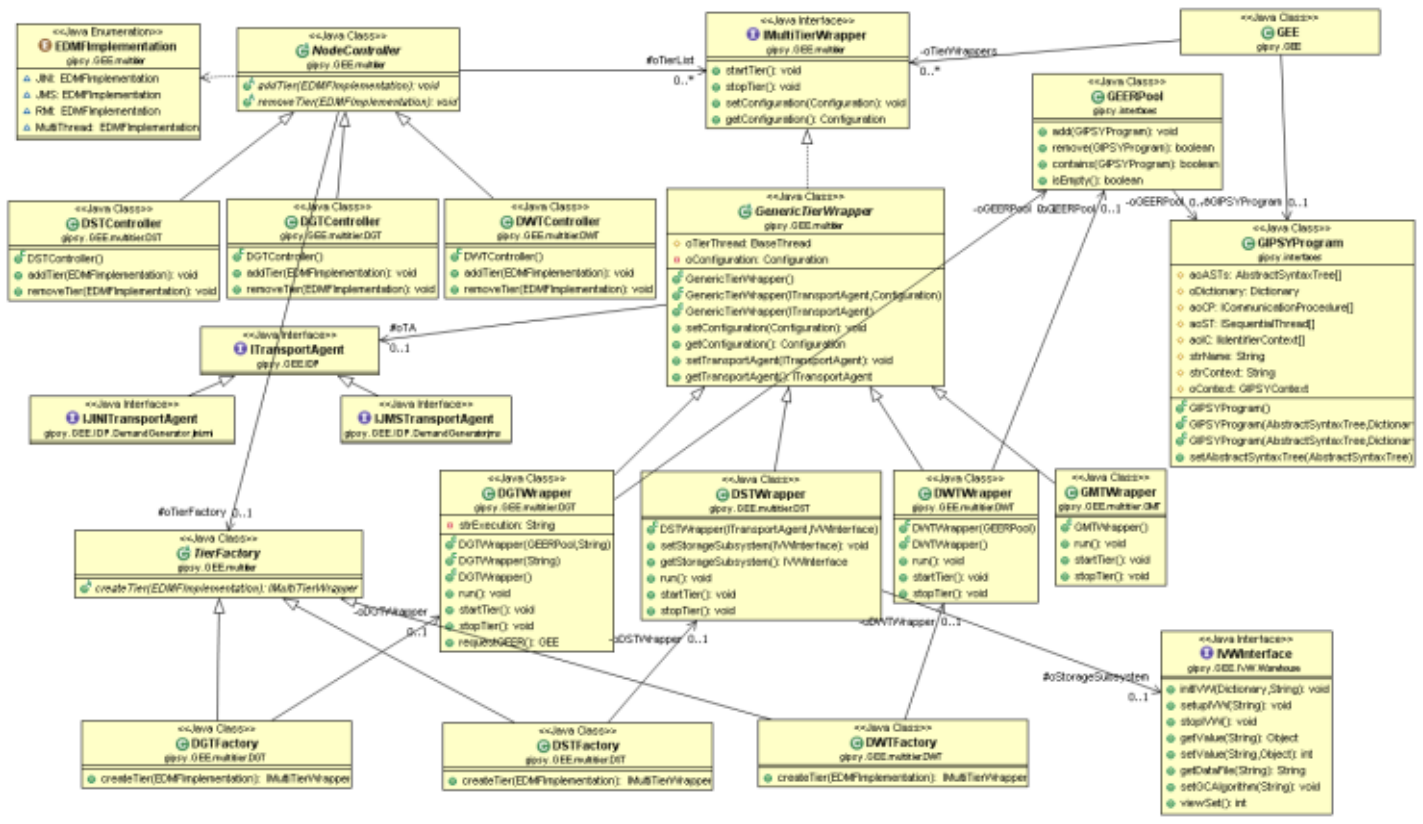

**Figure 10. Initial Multi-tier Architecture Design and Implementation [16]**

Though some extra layers of abstraction are present, the system remains extensible and flexible to accommodate any future changes to the design and implementation. Ongoing design and implementation presented in this work provides a feasible solution for the educative evaluation of hybrid intensional-imperative programs and tier management.
An interactive graph based GUI (graphical user interface) which allows the users to directly interact with GIPSY run time system is discussed. Prior to this, the GIPSY runtime system was totally managed by a command line interface. GUI provide the users - flexibility, usability in terms of managing the GIPSY network with minimum intervention. GUI translates easy and simple graphical interactions into complex message passing between various components, this allows the user to easily create, configure and control GIPSY network through the graph based interface. [17]

Design and implementation is based on representing GIPSY system as Graph based visualization. GIPSY tier networks are represented as nodes, each such node contains some data and properties associated with it. GIPSY configuration class is used to store configuration of various components of the system. GUI is implemented using JAVA/SWING library. JUNG library is used for modeling the data into network/nodes, it also provides many features regarding nodes like providing different color to differentiate among them.

The following are the features implemented by the GUI

1) Create a GIPSY network as a graph
2) Save/Load pre-configured GIPSY network
3) Start, Register and Stop the GIPSY nodes by using color differential list of nodes with their commands and properties
4) Allocate or Deallocate DST's (Demand Store Tier), DGT's (Demand Generator Tier), DWT's (Demand Worker Tier)
5) Start/Stop demand driven evaluation process on DGT through a contexted menu [17].

GMT operator view of the GUI allows a user to allocate and deallocate commands. It also provides drag and drop mechanism to change the connectivity among the tiers with ease, Furthermore users are allowed to start/stop nodes and to register them with GMT (GIPSY Manger Tier) with simple mouse clicks, Figure 11.When a new node is added to the network it is automatically pre-configured and associated/saved with users configuration file (Figure 11).Network graph editor allows user to create a GIPSY network or load an existing one[figure 11]Other run time system activities such as output of GMT,GIPSY nodes, tiers ,errors and log messages are displayed in separate view, this allows better failure traceability and better error troubleshooting. Set of JUNG interface classes are produced, this were used to manage, load, save GIPSY networks (Figure 12).Data structures are also detailed which were used to represent the network graphs and also to associate them to the appropriate GIPSY objects and action items(Figure 12).

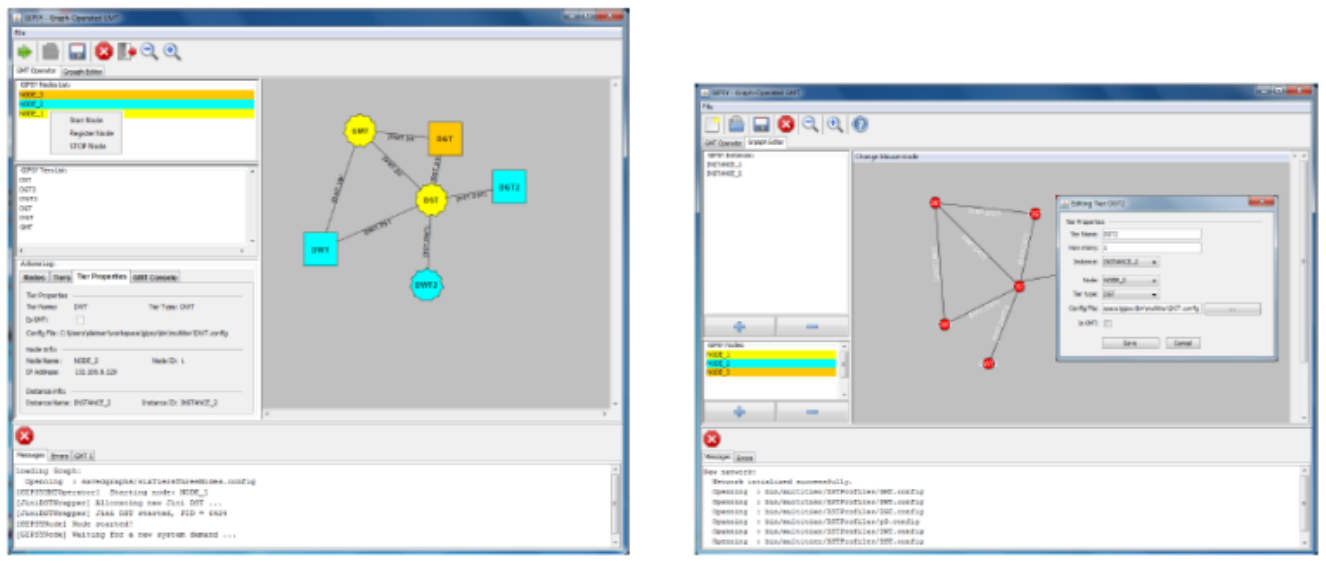

(a) GMT Operator View (b) Network Graph Editor View

**Figure 11. GMT operator view and Network graph editor view [17]**

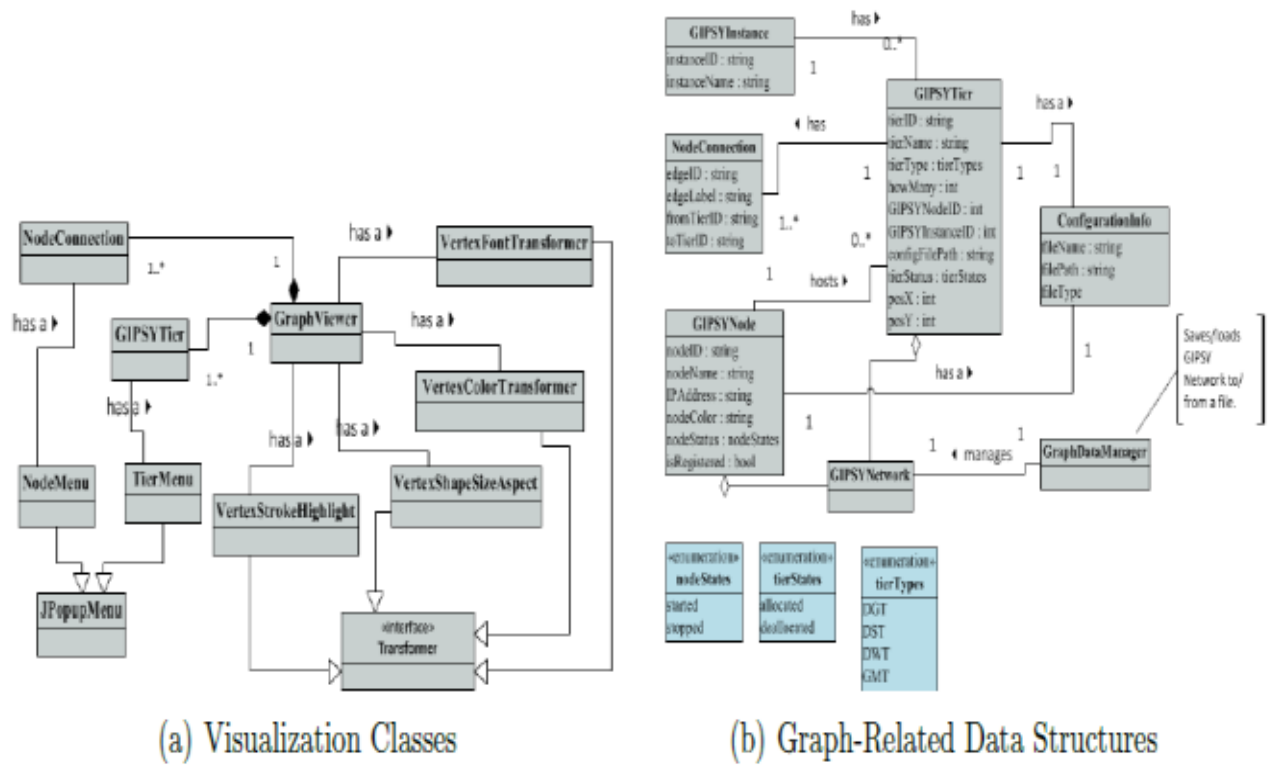

**Figure 12. Visualization and Graph related data structures [17]**

Overall this GUI is introduced to provide an effective solution for managing GIPSY run time system with ease. Future work on this GUI includes to allow a peer communication tool (Intra tool) to allow start up nodes not only on tiers but on remote computers too, extending the GUI to various platforms like mobile- Android and IOS [17].

A Modular intensional programming research system, GIPSY, to evaluate Higher-Order Intensional Logic (HOIL) expressions is discussed. The goal is to provide a flexible system for the investigation on programming languages of intensional nature, in order to prove the applicability of intensional programming to solve important problems. HOIL combines functional programming with various intensional logics to allow explicit context expression to be evaluated. The resulting contextual expression can be passed as parameters and returned as results of a function and constitutes a multi-dimensional constraint [18].
The overall architecture of GIPSY, is shown in Figure 13.For GIPC, the incoming GIPSY program's source code will be analyzed, divided into "chunks" preparing them to be fed to the respective concrete compilers for different languages.

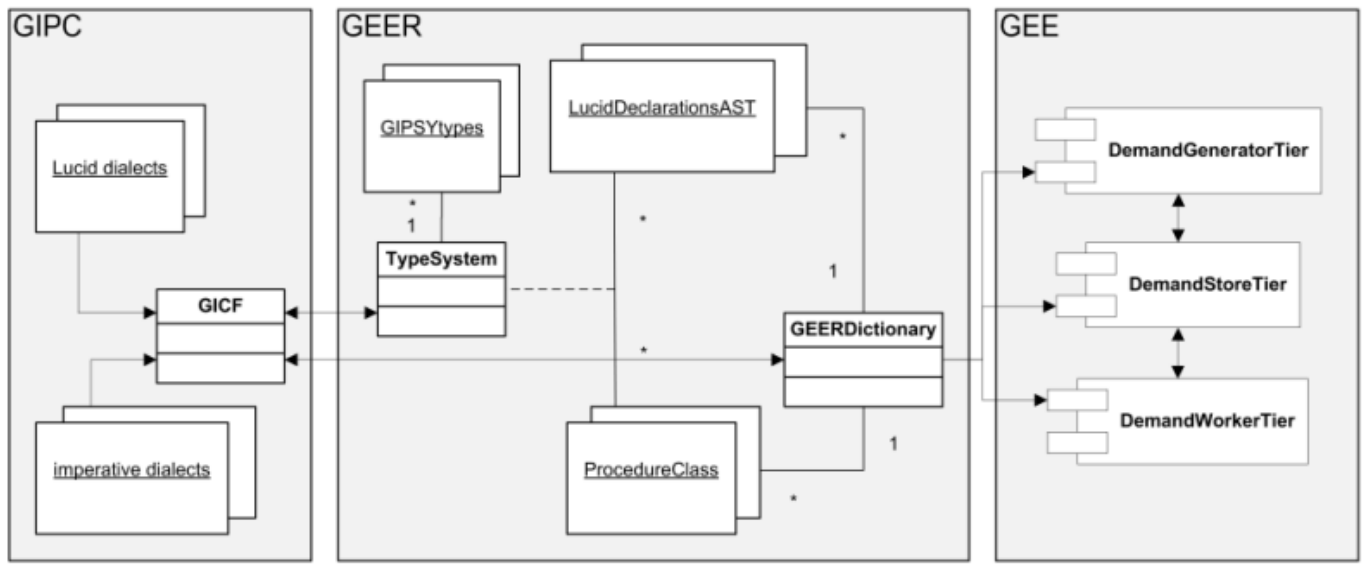

**Figure 13. GIPSY's GIPC-to-GEE GEER Flow Overview [18]**

Regarding General Education Engine (GEE), it has a distributed multi-tier architecture, where each tier can have any number of instances. It consists of Generic Education Engine Resources (GEER), GIPSY Tier, GIPSY Node, GIPSY Instance, Demand Generator Tier (DGT), Demand Store Tier (DST), Demand Worker Tier (DWT) and GIPSY Instance Manager (GIM).

For reasoning tasks of HOIL expressions, Higher Order Context (HOC) represents essentially nested contexts. The reasoning aspect of GIPSY is a particularity of a lucid dialect rather than the architecture.
In a nutshell, this paper presents GIPSY as a flexible, modular intensional programming research platform that can be used for reasoning tasks of HOIL expressions**.**

A multi-tier architecture that consists of [18]:

- Demand Generator Tire (DGT): It generates demands
- Demand Store Tire (DST): Stores and dispatch demands
- Demand Worker Tire (DWT): Computes demands
- General Manager Tire (GMT)

These tires are allocated in registered computers (GIPSY Nodes) and all of these tires and computers are managed by the general management tires figure 14.

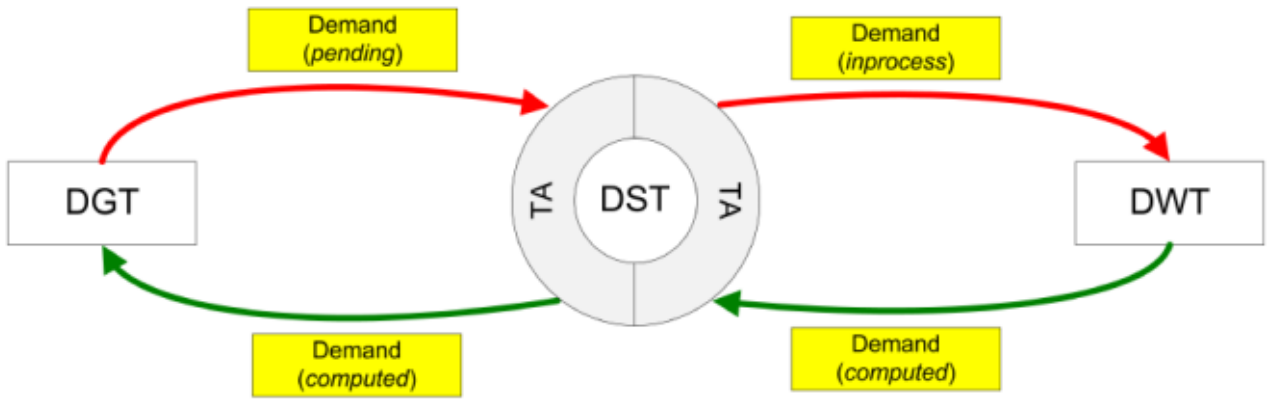

**Figure 14 Procedural demand migration among the DGT, the DST, and the DWT [18]**

Scalability:
The ability of a software system to handle increased workload and to achieve success on the long term while the system is facing growing demands can be achieved by adding resources to the system or by applying a cost-effective strategy in order to extend the system's capacity [18] [19].
In GIPSY, new tier implementations can be added without changing the source code of the existing system components. GIPSY system can deal with increasing workload and demand storage requirements by adding more nodes as registered computers, and allocating more GIPSY tiers in these nodes, therefore the GIPSY runtime system is scalable [18]. GIPSY runtime system has the ability to store more demands with acceptable memory usage (space scalability). It has also the ability maintain its performance (Space-time scalability). It has the ability of allocating more GIPSY tiers over more GIPSY nodes (Structural scalability). It has the ability to achieve anticipated demand processing quantity that is able to increase proportionally with the number of the software components that process the demands (Load scalability) [18].

With the existence of Autonomic GIPSY (AGIPSY) GIPSY is said to be self-manageable than what it is actually at present. Architecture of GIPSY using autonomic computing which often makes difficult computing systems easier and flexible to manage, automation also leads in reducing the overall complexity of maintainable system [20]. The emphasis here is to make the current GIPSY to be self-adaptive and autonomous for which an architecture was designed and modeled.

AGIPSY ARCHITECTURE

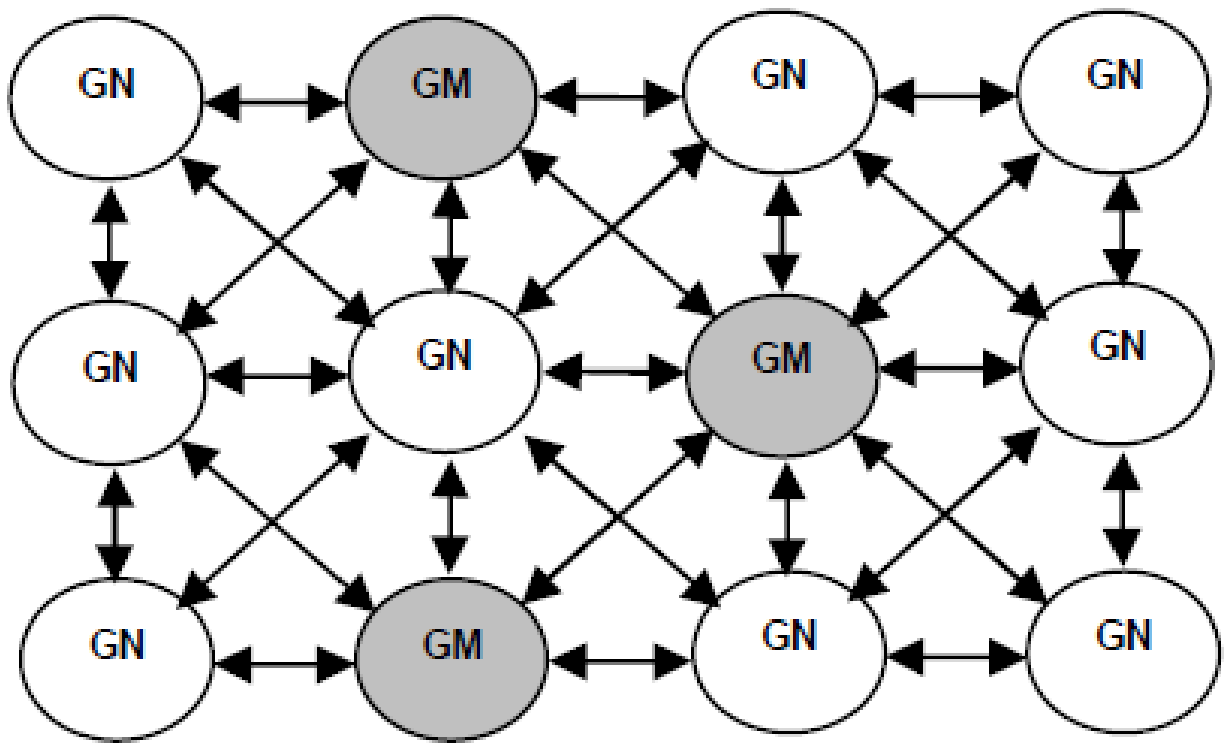

**Figure 15: AGIPSY Architecture [20]**

Mainly Node Manager (NM) is responsible for controlling the GIPSY nodes (GN's) which permits GN to comprehend its own thread of execution. Normally GN's are autonomous as they do not involve human intervention, while communication with external entities. Unlike the tier oriented architecture of GIPSY, AGIPSY holds all the prominent features of a multi agent distributed system [20]. From figure 15 it can be observed that the GN's are arranged as grid where they can share their instances to evaluate an intentional demand in context.

The salient features of AGIPSY are [20]:

Fault-Tolerance and Recovery -The main advantage of GN's is that they quickly recover from their past failures. GN's mainly uses ASSL protocol which saves information of GIPSY tiers after each information is transmitted or dispatched through a communication grid. At the point when a GN is begun, its state is restored from the recovery protocol data if accessible, on restoration all the tier components may continue their execution as there was not any intrusion.[20]

Self- Maintenance- This property entrails that for every distinct NM there is a corresponding GN which makes them more autonomous and self-maintainable.
Self-Optimization- In this aspect GM's are responsible for tracking the GN which eliminates the need for a GN to share information with its fellow GN's

Self-Healing- This property illustrates that, since every GN replicates its own essential states, the system may be easily recovered or healed when intended for.

Self-Protection. - The last and the important aspect is to restrain the GN from all the possible malicious attacks in order to reduce the incoming overhead of various threats.

A general architecture used for demand migration and evaluation of demands at runtime by the system is discussed. Here demand driven execution system is based on demand generators (DG) which controls the process by generating functional demands. If the workers are remote then the demands are migrated through a network from generator to worker. All these functional demands are independent [21].
Demand Migration System (DMS) which connects the execution nodes with using different middleware technologies. DMS is about process migration [21].

The following are main requirements for DMS is

1. Platform Interoperability.
2. Once delivery semantics
3. Asynchronous Communications
4. No demand discrimination
5. No worker discrimination
6. Secure communication
7. Fault tolerant demand migration
8. Distributed technologies independency
9. Hot plugging
10. Upgradability

DMS Architecture:

Demand dispatcher (DD) and Transport agents (TA) are two main subsystems for DMS architecture. They both run independently. Where DD acts as message storage mechanism and TA is to transport demands and results to DG's and workers. DD acts as a bridge between DG's and workers [21].

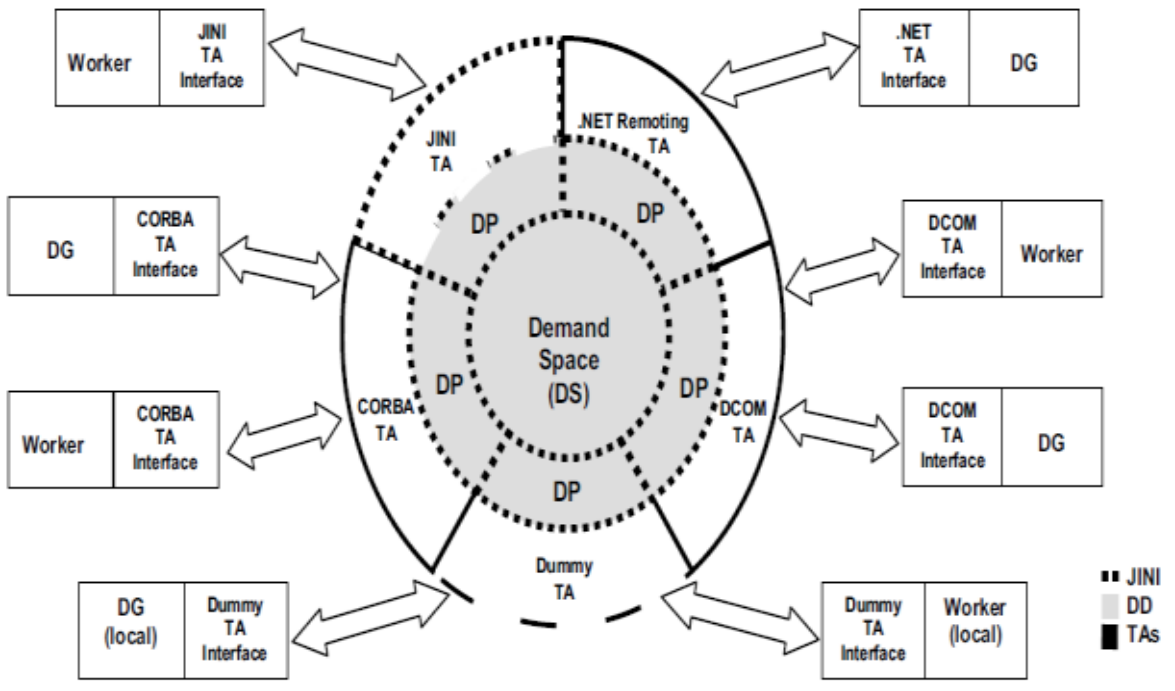

**Figure 16. GIPSY Demand Migration System [21]**

DD has two entities namely DS (Demand Space) and DP (Dispatcher Proxy). Where DS is internal object storage mechanism that stores demands and results. DP is entry point for DD, TA require DP to communicate with DD [21].
If DD and generator are placed locally then there is no need of TA and middleware technology. DG sticks to dummy TA interface. If DD and generators are placed remotely then we need TA in order to communicate between them. TA acts as GIPSY transport protocol [21].
DMS is depending on distributed technologies like JINI, COBRA etc. JINI is used for Multi-platform transportation.

The generators and workers communicate with demand space through DP to get and post demands. If a demand is stored in demand space then it should follow some rules:

-It must have a default no-argument constructor.

-All its instance variables must be public.

-All its instance variables must be serializable.

Dispatch process (Figure 17) depicts the demand which is dispatched between DG, DD, TA and workers.

TA is based on JINI. Which is a Java technology and JINI uses some internal protocols called discover, join and lookup.

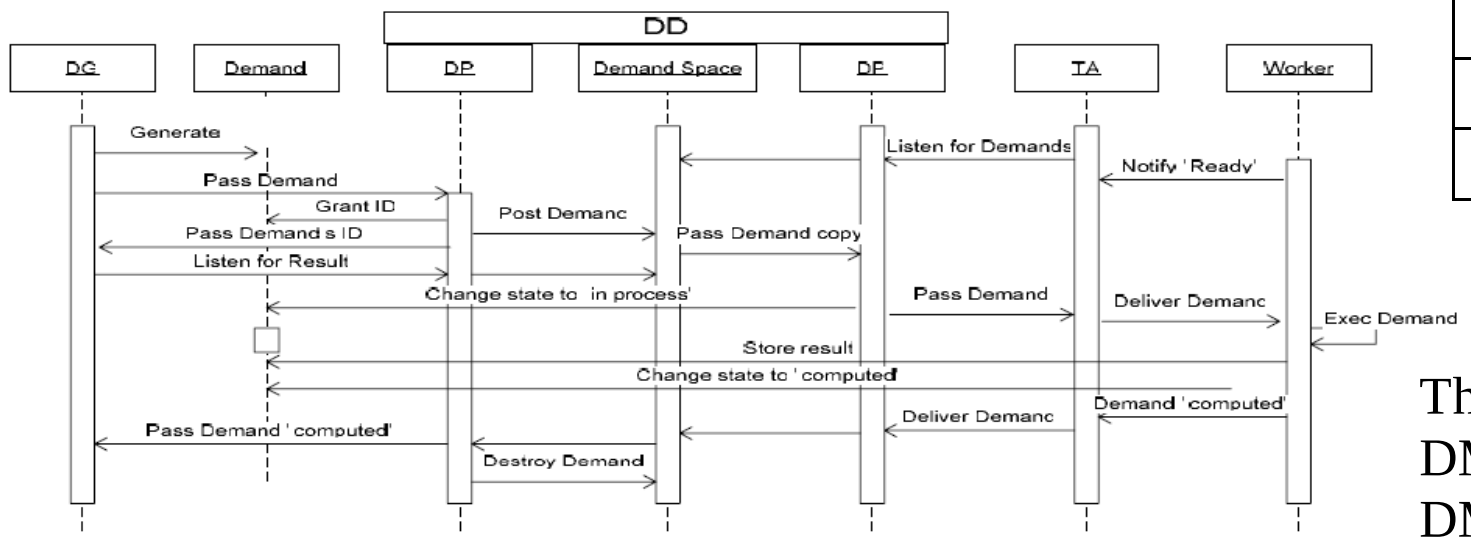

**Figure 17. Dispatch process [21]**

*C. Summary*

Distributed MARF (DMARF) is based on MARF platform, which supports processing of recorded audio, imagery or textual data for pattern recognition and bio metric applications as its domains. Since MARF is an open-source research platform, it has the ability of adding any module/algorithm implementation at any stage of the pipeline, DMARF has been introduced whose pipeline stages were made into distributed nodes. The communication among the nodes is done over Java RMI, CORBA], and XML-RPC web services. DMARF serves as a web application allowing NLP to be available on line. DMARF can be extended to robotic systems the software system should be able to manage itself with the dynamic requirements and threats just as the human body does. ASSL framework is used to develop and integrate self-management features to DMARF such as (Self-Configuration, Self-Healing, Self-Optimization, and Self-Protection).

GIPSY is considered as a more efficient and adaptable intentional tool due to its capability for any subsystem change and soothing of higher order functions. Furthermore it can also be considered as a flexible, modular intensional programming research platform that can be used for reasoning tasks of HOIL expressions. GIPSY also is scalable and supports fault tolerance and recovery. GIPSY framework is targeted to accommodate the feasibility of fluently developing components of complier for languages which are intensional in nature and to efficiently execute them on a self-reliant runtime system

OSS Case Study Estimations:

**SonarQube** is an open source platform used for continuous inspection of code quality which embeds with a tomcat server, and also integrates with Eclipse development environment. By using SonarQube we measured the number of classes, methods, files, lines of java code, for this we have installed SonarQube 3.7.4 and Sonar Runner 2.4 and set the corresponding sonar properties for project and then run it on the server to analyze the measures. The snapshots of the estimations for the case studies were included in (Appendix A).

| **Measurements** | **DMARF** | **GIPSY** |
|---|---|---|
| Java files | 1024 | 601 |
| Java Classes | 1054 | 665 |
| Methods | 7152 | 6261 |
| Lines of Java Code | 77297 | 104073 |

**Table 3: Case Study Measurements**

The total number of java files and java classes accounted for DMARF are twice more than that of GIPSY. Methods for DMARF are slightly more than GIPSY, but GIPSY is complex than DMARF when java lines of code is a measure.

## II. REQUIREMENTS AND DESIGN SPECIFICATIONS

### *A. Personas, Actors, and Stakeholders*

#### *1) DMARF*

*Actors*

Developer/Student

Is presently developing the application based on DMARF's framework. The student is supposed to develop a web application for forensic analysis, subject identification, and classification using a mobile equipment such as a laptop, or a cellphone.

Professor/Tester

Tests the functionality of the software. He uploads collected voice samples and tests how it matches the recorded voices in the corresponding data base.

*Stakeholders*

Students of Other Sections

Who can better learn from this case study and add more features which can be assigned based on their research or assigned by the professor.

University (Organization)

Can benefit from using this software for further developments and applications in the domain of forensic analysis. It provides all the possible means to help in evolving of DMARF.

*Persona*

| Persona | 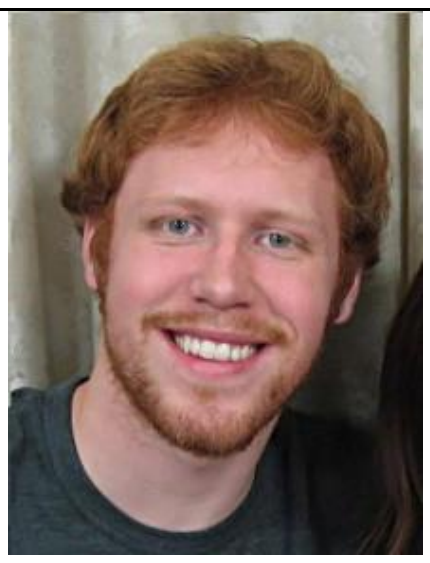<br>**Character Name:** Allen Armstrong |
|---|---|
| Job Title | Professor at Concordia University – ENCS Department |
| Experience | 11years of teaching and research |
| Skills | • Real-Time and Embedded Software Systems, Dependable parallel & distributed systems.<br>• Excellent communication skills and committed to team-building. |
| Goals | • Supervise a project of the *implementations of Distributed Modular Audio Recognition Framework* (DMARF) and its applications |
| Description | • Allen is 40 years old and received the PHD in Computer Science from the École *Polytechnique*, France, in 1999. He has been a professor at Concordia University since 2004.<br>• His research area is in Design patterns for parallel programming, Parallel architectural skeletons, and Dependable parallel & distributed systems. Ongoing projects include voice/sound/speech/text and natural language processing (NLP) algorithms, and machine intelligence, computer graphics systems such as MARF and its applications. |

2) *GIPSY*

*Actors*

*Student*

Is currently pursuing his PhD in computer Science and is keen to know the developments regarding intensional programming. He is the one who interacts directly with GIPSY GUI. Student has access to make modifications like adding a node to a GIPSY network. The main goal of user is to discover applicability of intensional programming by using the GIPSY GUI.

*Stakeholders*

*Developer*

Is the one who constructs the GIPSY GUI. Responsible for implementing the features according to the specifications. Also performs unit testing on each feature developed. Developer has access to make any in-depth level changes to the code and gets influenced when any major decisions about the system take place. The main goal of developer is to improve the overall usability of the GIPSY GUI.

*Architect*

Responsible for understanding the requirements and organizing them in accordance. Architect is the one who designs the overall system. Being an expertise he decides on the feasibility of the features and is the one whose decision affects the overall outcome of the project.

*Persona*

*Personal Profile*

Albin is from Toronto, Canada, graduated in computer science from University of Toronto. He is currently pursuing his PhD in computer science. He is very outgoing and has a lot of friends around him. Apart from his career and education Albin is also interested in social services. Every weekend he makes sure he does not miss his scheduled activities which involves visiting the local orphanage. From technical perspective as a part of his research work, Albin is currently working on intentional programming and its application.

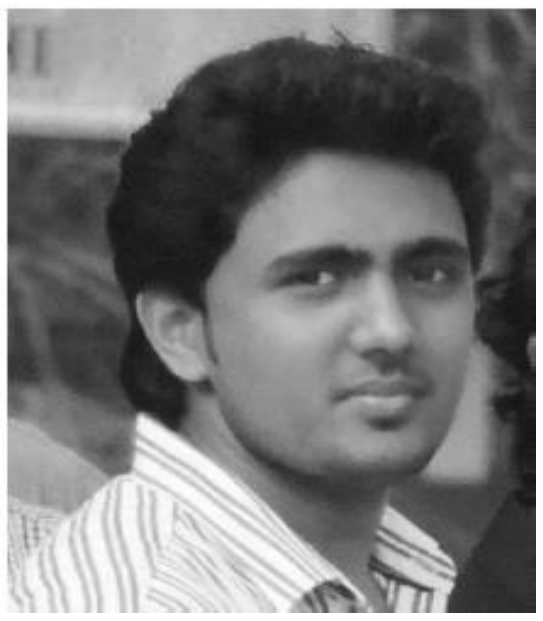

*Introduction*

*Persona type*- Final year PhD student

*Name*- Albin

*Age*- 25

*Location*- Toronto, Canada

*Job Title*- Student

*Back story*

- Born and raised in Canada

- Loves animals, swimming
- Graduated in computer science

*Characteristics*

- Quick Learner
- Logical Thinker
- Focused
- Tech Savvy

*Favorite Quote*

*"The artist is nothing without the gift, but the gift is nothing without work."*

*Ideal Experience*

- Worked on the applicability of intentional programming
- Easily Solve problems of intentional nature
- Analyze forensic investigations easily and effectively
- Investigate properties of programming language

*Info Sources*

- Competitor Websites
- University Articles
- Library References
- Previous Research Articles

*Scenario*

User main inclusion is to create a gipsy instance and start a node connection process and view the results in graphical interface managed by the network editor, and his experiences while communicating with GIPSY run-time system

*Needs*

GIPSY Node
GIPSY Tier
Node Connection

*B. Use Cases*

*1) DMARF*

The professor will be testing a DMARF application for forensic analysis, subject identification and classification using a mobile equipment such as a laptop, or a cellphone. He uploads collected voice samples and tests how it matches the recorded voices in the corresponding data base.

| | |
|---|---|
| Use Case ID | UC-1 |
| Use case name | Forensic analysis |
| Scope | Forensic analysis and subject identification and classification |
| Level | User Goal level |
| Primary Actor | Professor |
| Stakeholders and Interests | Student: developed the application and wants it to be tested by the professor.<br><br>Professor: Tests the functionality and performance of the application by uploading voice samples and receiving matched results. |
| Preconditions | Professor is authenticated by the system.<br><br>Sample file should be present in the Professor's system. |
| Post conditions | Results are saved for future references. |
| Main success scenario | 1. Professor <u>uploads</u> a voice sample.<br>2. Systems **Sample Loading** Service loads the uploaded audio file and converts for further preprocessing.<br>3. The Preprocessing Service accepts incoming voice.<br>4. Feature Extraction Service accepts data and <u>sends</u> data for classification.<br>5. **Classification** and Training Service accepts feature vectors and <u>updates</u> its database of training sets.<br>6. Professor uploads a voic e sample again to test the system.<br>7. Repeats steps 2, 3 and 5.<br>8. System shows the compa rative results between the sample files. |
| Extension | 2a. The system shows 'upload failed', if the "Sample Loading Service" does not support the voice sample format.<br>1. Professor uploads a correct audio format.<br>8a. The system does not find the matched voice in the training set.<br>1. The system asks the Professor to upload a different sample. |
| Special requirements | The system provide progress bar for showing upload.<br><br>The system also runs mobile equipment such as a laptop, or a cellphone.<br><br>The voice must upload in specially appointed format. |
| Technical and data variation list | The file format of the uploaded samples must be in WAV, MP3, SINE, SND, MIDI, AU, AIFF, or AIFFC format. |
| Frequency of occurrence | On Demand |
| Miscellaneous | System needs a good recovery support, in case of system crashes while uploading an audio file or in case of showing the results. |

*2) GIPSY*

| | |
|---|---|
| Use Case ID | UC-2 |
| Use case name | Running GIPSY GUI |
| Scope | Analysis and Evaluation of Intensional Programming |
| Level | User Goal Level |
| Primary Actor | Student |
| Stakeholders and Interests | ● Student: Wants effective analysis of intensional programming.<br>● Developer: Wants to increase the usability of GIPSY. |
| Preconditions | ● The GUI application should be installed in the Student's system. |
| Post conditions | ● The graph related to GIPSY network is displayed. |
| Main success scenario | **1.** Student creates a **GIPSY instance.**<br><br>2. Student creates **GIPSY node** and assigns properties such as node name, IP address and color.<br>3. Using the generated **GIPSYInstance** and node Student creates **GIPSY tier** and allocates properties like name, number of instances.<br>4. Student saves the generated **GIPSY network** as a graph.<br>5. Student can start/stop/register the nodes by maintaining a color differentiated list of nodes.<br>6. The system displays actions taken as log messages.<br>7. Student can allocate and de-allocate DST's (Demand Store Tier), DGT's (Demand Generator Tier), and DWT's (Demand Worker Tier.<br>8. Student can start/stop the demand driven evaluation process on a DGT.<br>9. Resulted GIPSY network graph is displayed. |
| Extension | 2a. Student loads an existing pre-configured GIPSY network file in the system.<br>1. System accepts the preconfigured file and displays the GIPSY network as a graph<br><br>4a. Student adds a node to existing Gipsy network<br>1. Student enters properties associated with the node.<br>2. System automatically pre configures and associates the node with the configuration file. |
| Special requirements | ● The system provides flexibility by providing drag and drop node mechanism.<br>● The system provides log and error messages which contributes understandability |
| Technical and data variation list | ● Configuration file should be with .config extension<br>● Number of instances, maximum demands are the parameters defined by the configuration file. |
| Frequency of occurrence | On Demand |
| Miscellaneous | ● System is not portable through all the platforms, it only supports few of them.<br>● System does not support more problem-specific tiers like MARFCAT.<br>● System is not distributed in nature, no peer communication is possible. |

*C. Domain Model UML Diagrams*

*1) DMARF*

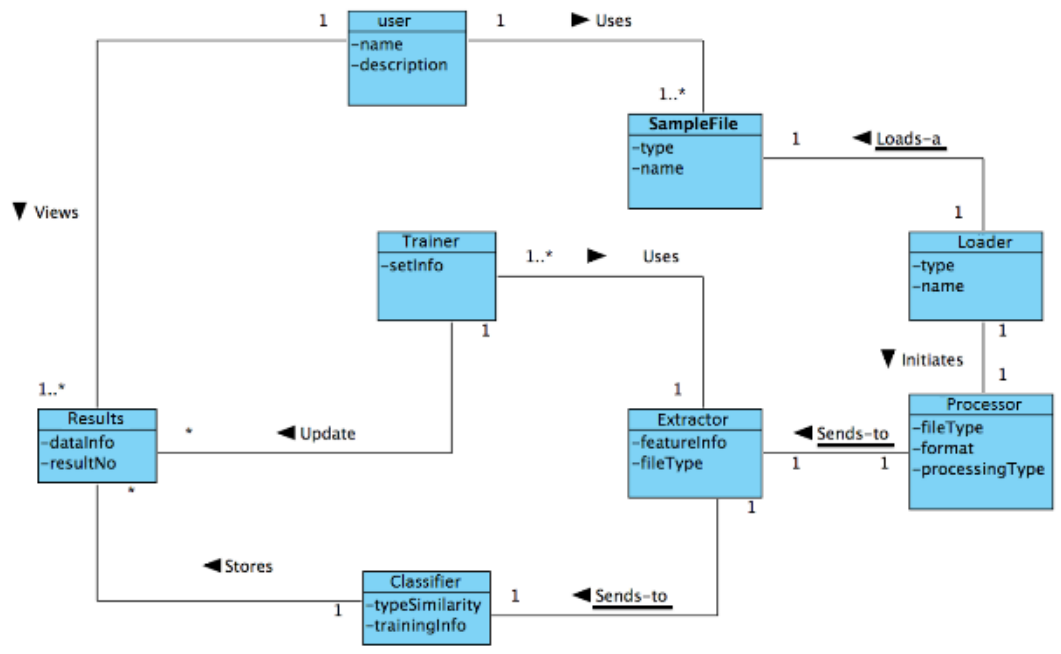

**Figures 18: DMARF Domain Diagram (see Appendix B for clear view)**

For readable view (refer to **Appendix B DMARF Domain Diagram**)

DMARF pipeline is composed of 4 stages Figure (1):

- sample loading services
- preprocessing,
- feature extraction
- training/classification

The user uploads a sample audio file, and the Sample Loading Services (loader) loads the sample file to the system and converts it for further preprocessing. The Preprocessing Service accepts the incoming audio file sample, and does the required filtering. All the features of preprocessed file will be extracted by the Feature Extraction Service. Classification and Training Service accepts feature vectors and updates its database of training sets or perform classifications in the training data base. User uploads a voice sample again to test the system [1].

The noun phrases and verb phrases are identified in the domain diagram (**see Appendix B for clear view**)

*2) GIPSY*

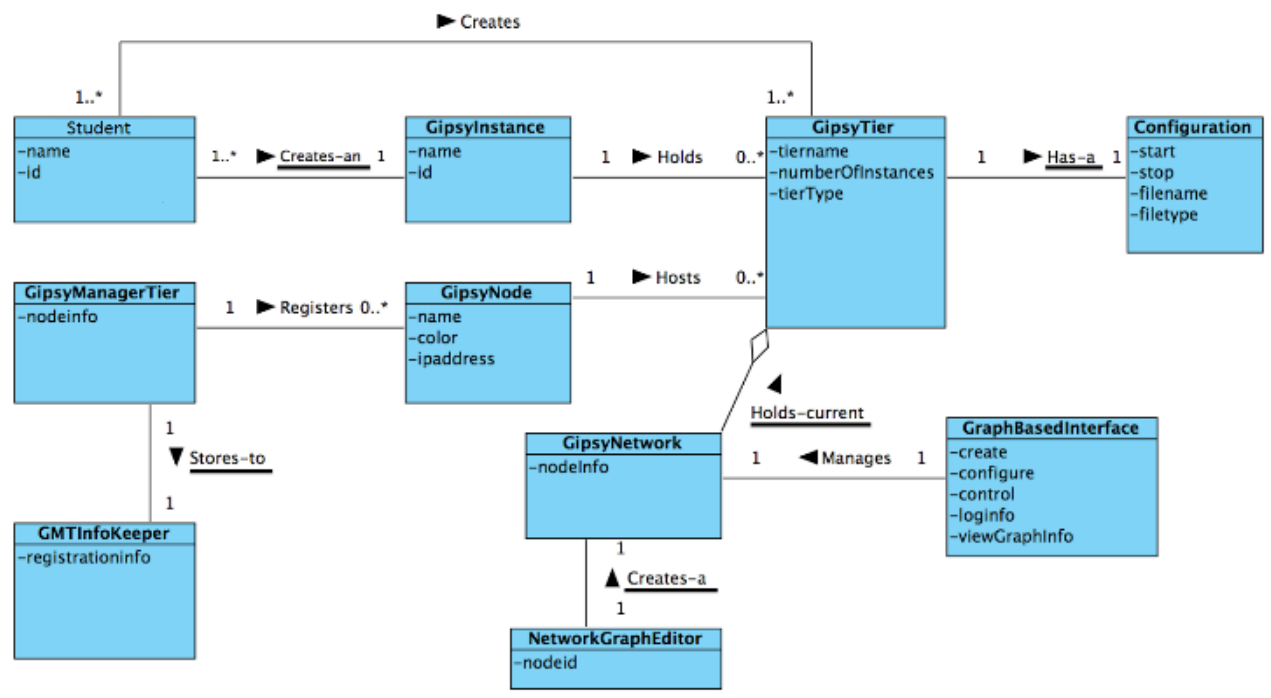

**Figures 19: GIPSY Domain Diagram (see Appendix B for clear view)**

For readable view (refer to **Appendix B GIPSY Domain Diagram**)

The above domain modeling diagram illustrates the interactions of typical user (let's say Student) while communicating with Gipsy Run time-system. As Gipsy being a Demand Driven process it allows users to store demands in a coherent flow manner. Initially the user initiates the process for creation of Gipsy Instance by providing the Instance Name and Instance Id and then creates a Gipsy Tier which is hosted by the Gipsy Node where node registration should be done prior using Gipsy Manager Tier which is responsible for registering and storing the registration information of the particular node which is under discussion. As Gipsy being a Multi-Tier Architecture it holds a unique details, similarly the current tier has configuration setup which allows users to activate or deactivate the node processing.
The noun phrases and verb phrases are identified in the domain diagram (**see Appendix B for clear view**)

*3) Fused DMARF-Over-GIPSY Run-time Architecture (DoGRTA)*

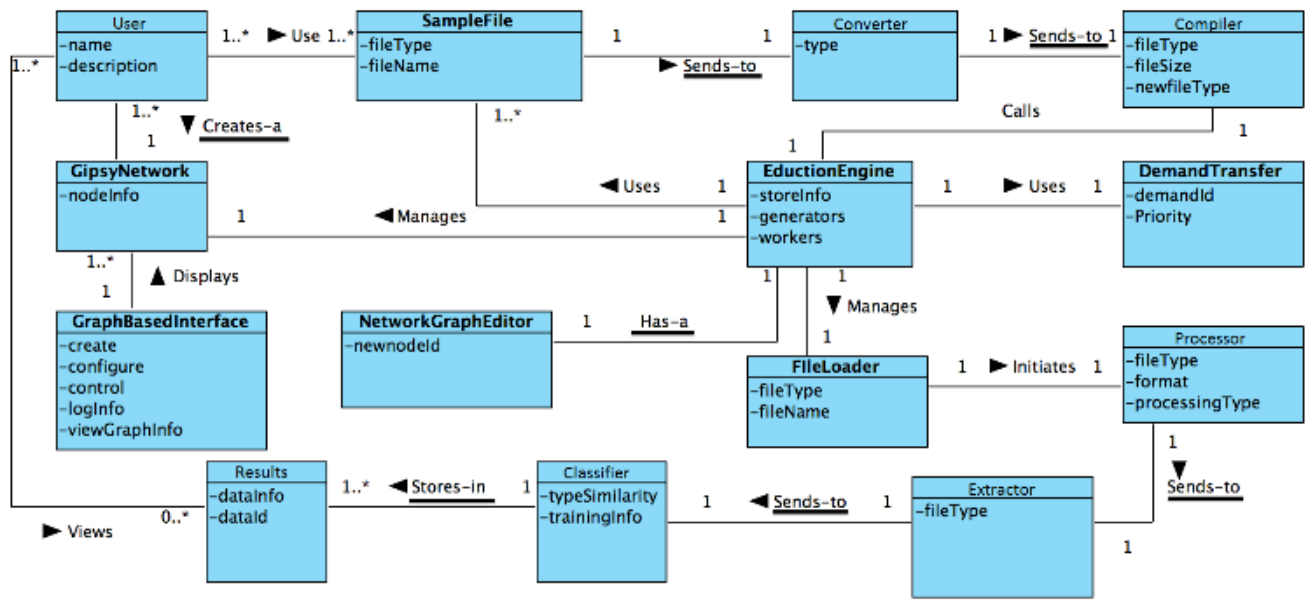

**Figures 20: FUSED DMARF- over- GIPSY Domain Diagram**

**(See Appendix B for clear view)**

For readable view (refer to **Appendix B FUSED DMARF-over- GIPSY Domain Diagram**)

GIPSY uses demand driven architecture on the other hand DMARF uses pipelined architecture. GIPSY uses its demand driven eductive execution model called GEE (General Eduction Engine).GEE assess intentional expressions for which a demand is generated, this demand is generated, delivered to a networked demand store. This demand can be picked up by a worker who is observing on a corresponding node, which computes the result and places into the warehouse to be picked by generator and return back to the executing program. DMS (Demand Migration system) is responsible for this distributed asynchronous communication. GIPC is responsible for compilation and acts as a network protocol and sends the process information to the GEE where the demand driven process initiates.
DMARF adopts synchronous communication. While there is coordination among the processes in the pipeline, the path may be different for each subject or sample.
In the above domain diagram, the sample file from the DMARF is converted to a LUCID form using a converter and sent to the GEE (General Eduction Engine) via GIPC. GEE uses DMS for migrating the demands received. The Demands from DMARF phases are generated and the corresponding worker pick them up for processing. Workers stores the result in the GEE‘s warehouse, the results are picked by generators to respective processes. The above process in the domain model explains the use of demand driven procedure with DMARF rather than the regular pipeline procedure.

The noun phrases and verb phrases are identified in the domain diagram (**see Appendix B for clear view**)

### *D. Actual Architecture of UML Diagrams*

#### *1) DMARF*

In the conceptual class diagram, real-world components are represented, not a software itself. In other words, it clarifies meaningful concepts in a problem domain. For example, the "User" conceptual class was represented in the domain diagram, and it has an association called "use" which explains that the user will use an audio sample file, and this sample is represented as a conceptual class. The "sample" audio file can be uploaded to the "loader" conceptual class using the association "load". This concept can be done in the Class Diagram: In this diagram, the **classes** of the system are presented, and their inheritance, aggregation, association, and the operations and attributes.

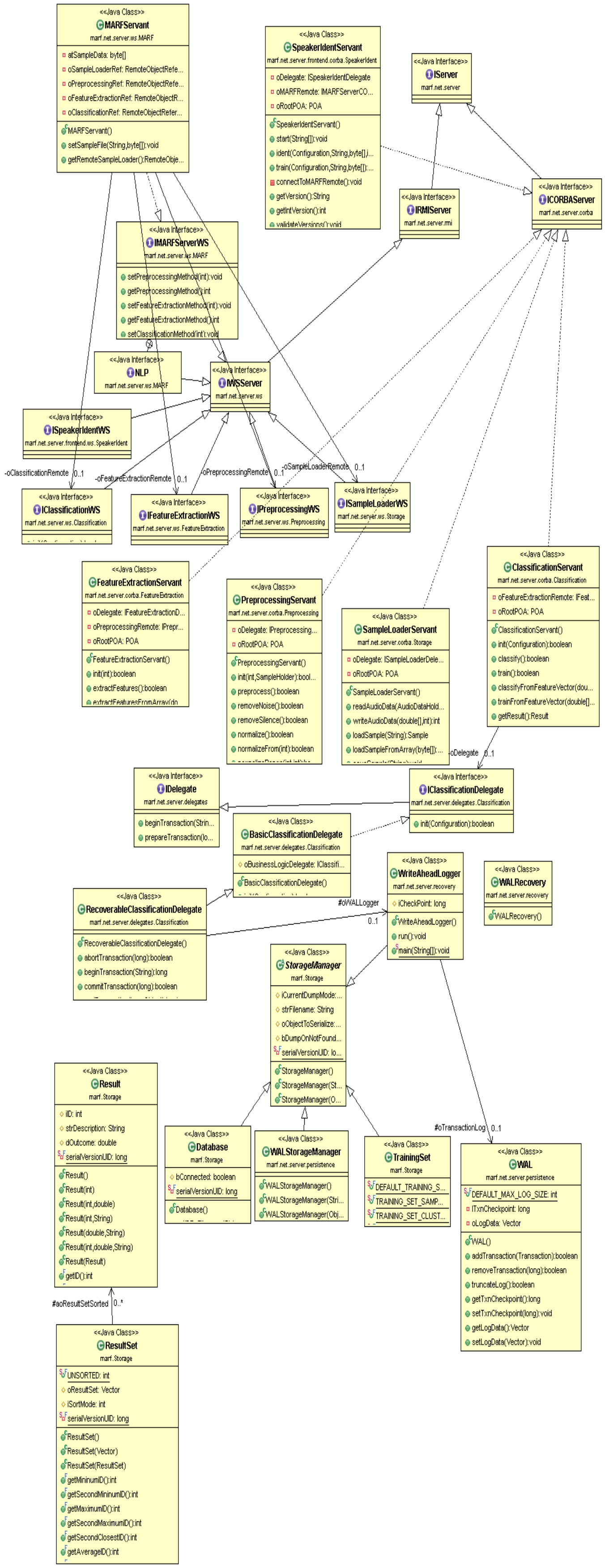

**Figure 21: Class diagram for DMARF (See Appendix B for Clear View)**

According to the scenario mentioned above, this table (4) represents the corresponding conceptual classes and the system class

| Actual Classes | Conceptual Classes |
|---|---|
| SampleLoaderServant | Sample File |
| PreprocessingServant | Processor |
| FeatureExtractionServant | Extractor |
| ClassificationServant | Classifier |
| TrainingSet | Trainer |
| Result | Results |

**Table 4: Mapping of actual classes to conceptual classes in DMARF**

In fact, they do not match 100%. Most of the conceptual classes exist in the class diagram. The domain model diagram usually clarifies meaningful concepts in a problem domain; according to a specific scenario.

In class diagram a description of the system design must be illustrated in details. All **classes** of the system are presented, and their inheritance, aggregation, association, and the operations and attributes.

To have a consistency between the conceptual and the actual classes, it means that the solution of the problem space has been met using the system. In other words, it reflects that the users' needs can be met by the output of the functionality of this system.

The relationship between the **class** SampleLoaderServant, and the **class** PreprocessingServant is done through CORBA server for methods' calls. The **class** SampleLoaderServant and the **class** PreprocessingServant extend ISampleLoaderCORBAPOA which implanted in ICORBAServer in order to communicate. The DMARF is uses delegate implementation as long as IDelegate in marf.net.server.delegates is implemented. Delegate implementations allow sharing all of transactions, and communication needed.

**class** MARFServant recognize an audio sample file through recognize(), and then it starts the recognition pipe line through startRecognitionPipeline(). Afterwards, the **class** SampleLoaderServant loads a sample file from **Class** MARFServant using the Sample loadSample(), and then this will be loaded to the **class** PreprocessingServant through the class sample using getSample() and does normalization using normalize() method, and process it to the **class** FeatureExtractionServant [22].

**SampleLoaderServant Class:**

```
public class SampleLoaderServant
extends ISampleLoaderCORBAPOA
implements ICORBAServer
{
	private ISampleLoaderDelegate oDelegate = null;

	public SampleLoaderServant()
	throws InvalidSampleFormatException, Exception
	{
		super();
		new Logger("sampleloader.corba.log");
		this.oDelegate = new BasicSampleLoaderDelegate();
	}

	public Sample loadSample(String pstrFilename)
	throws CORBACommunicationException
	{
		try
		{
			return MARFObjectAdapter.getCORBASample(this.oDelegate.loadSample(pstrFilename));
		}
		catch(StorageException e)
		{
			throw MARFObjectAdapter.getCORBACommunicationException(new CommunicationException(e));
		}
	}
}
```

**Preprocessing Servant Class:**

```
public class PreprocessingServant
extends IPreprocessingCORBAPOA
implements ICORBAServer
{

	private IPreprocessingDelegate oDelegate = null;

	public Sample getSample()
	{
		return MARFObjectAdapter.getCORBASample(this.oDelegate.getSample());
	}

  public boolean normalize()
	throws CORBACommunicationException
	{
		try
		{
			return this.oDelegate.normalize();
		}
		catch(PreprocessingException e)
		{
	throw MARFObjectAdapter.getCORBACommunicationException(new CommunicationException(e));
		}
	}
```

```
public void setSample(Sample poSample)
	{
		try
		{

	this.oDelegate.setSample(MARFObjectAdapter.getMARFSample(poSample));
		}
		catch(InvalidSampleFormatException e)
		{
			e.printStackTrace(System.err);
			throw new RuntimeException(e);
		}
	}
}
```

### 2) *GIPSY*

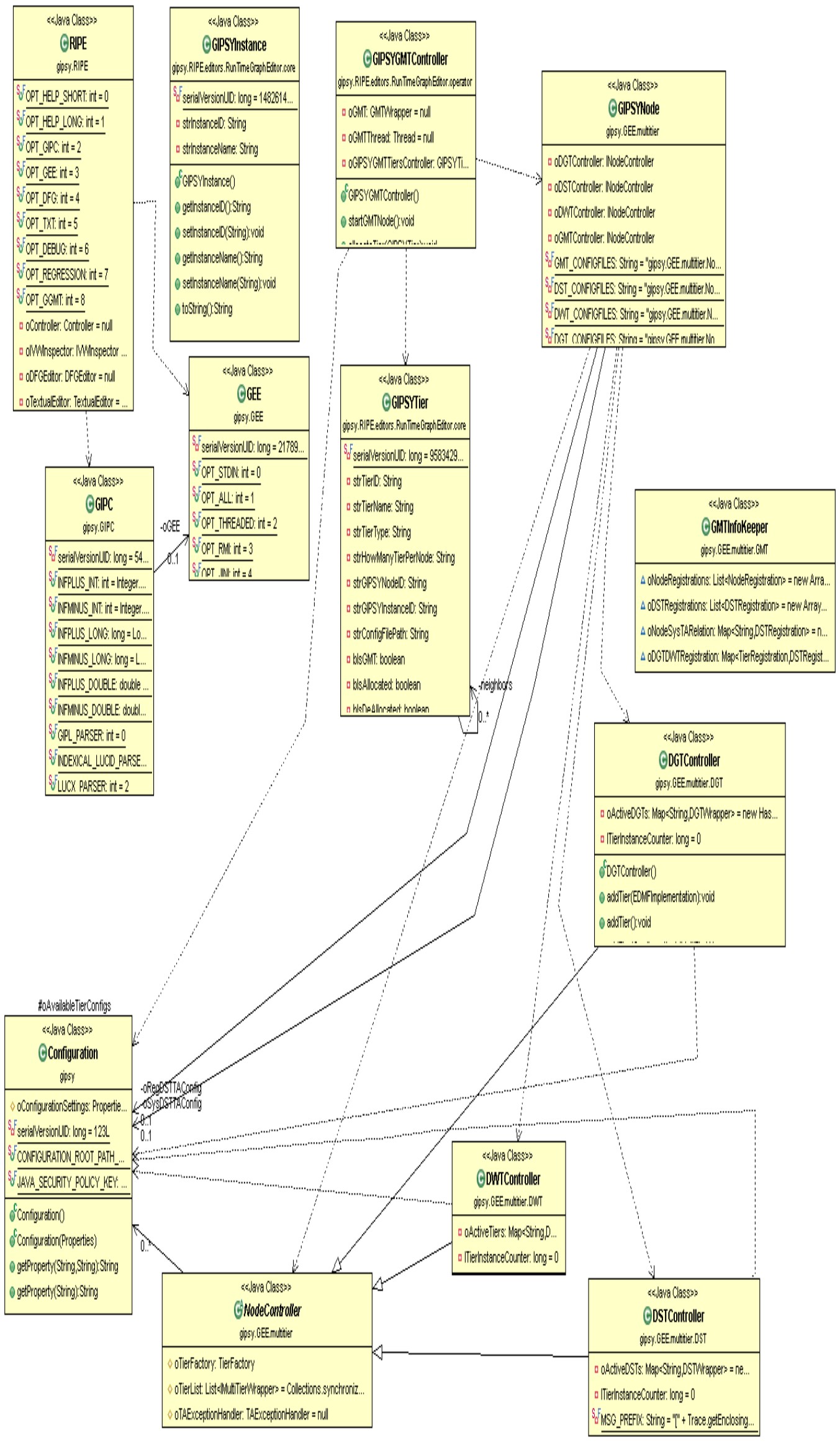

**Figure 22: class diagram for GIPSY (See Appendix B for Clear View)**

The above diagram illustrates that GIPSY comprises of three important modules for its working they are: RIPE (Intensional run time programming environment), GIPC (General Intensional programming language compiler) and GEE (General Eduction Engine).

RIPE which is a run time programming environment provides users with visualization of data flow diagrams related to the lucid part of GIPSY programming. The user can interact with the RIPE by changing the input/output channels, changing communication protocols or either changing the parts of GIPSY itself like garbage collector. For this runtime environment, a compiler called GIPC is used to compile the programs of intentional nature. Therefore a connection lies between **RIPE class and GIPC class** in the diagram, GIPC also acts as a communication protocol.

GIPSY uses Eduction process, **GEE class** is responsible for implementing a demand driven model of computation. Here each demand generates a procedure call which computed locally or remotely. For every computed process the value is placed in the warehouse, from which the values are taken by the respective expressions. **DemandGenerator class** in the diagram is responsible for generating demands and **DemandWorker class** is responsible for processing any of the required demands. DemandGenerator calls **DemandDispatcher class** to read, write or cancel demands and also allows to view the results. **Cache class** acts as the warehouse to store the results.

GIPSYGMTController class is responsible for starting GMT node and is connected to **GIPSYTier class** for allocation and de-allocation of tiers. GMTController is also connected to **Configuration class** which holds properties for each node and it is connected to a **GIPSYNode class** which processes adding and removing of GIPSY Tiers.

**NodeController Class** acts as the controller for the GIPSY node which allows to set configuration properties for each of the GIPSY node. GMTInfoKeeper class keeps the registered nodes information.

The conceptual classes describe GIPSY GUI application as a domain model, whereas the actual architecture of GIPSY represents software components interacting with each other. Conceptual classes are based on a scenario, on the other hand actual architecture provides overall layout of the software system.

GIPSY instance, GIPSY tier, GIPSY node, GMT (GIPSY manager tier), GMT infokeeper, configuration are the conceptual classes which map to actual classes.

Yes there exists discrepancy between concepts and the actual classes. Concept is a broad abstract idea or a general guiding principle of representing system's artefacts. While the software architecture is a high level structure of a software system and its documentation.

| Conceptual Classes | Actual Classes | Comments |
|---|---|---|
| GipsyInstance | GIPSYInstance | Gets the information regarding basic parameters like instance id instance name. |
| GipsyNode | GipsyNode | Gipsy node is associated with the gipsy networks configuration |
| GipsyManagerTier | GIPSYGMTController | Initializes the node process and also responsible for allocating the tier |
| GMTInfoKeeper | GMTInfoKeeper | Responsible for node registration and saving information regarding the node and can perform modification corresponding to the node. |
| Configuration | Configuration | Deals with the corresponding to the node |

**Table 5: Mapping of actual classes to conceptual classes in GIPSY**

**GIPC Class**:
**Package** gipsy.GIPC;

```
import gipsy.GIPC.DFG.DFGGenerator.DFGCodeGenerator;
import gipsy.GIPC.imperative.EImperativeLanguages;
import gipsy.GIPC.intensional.EIntensionalLanguages;
import gipsy.GIPC.intensional.IIntensionalCompiler;
import gipsy.GIPC.intensional.IntensionalCompiler;
import gipsy.GIPC.intensional.IntensionalCompilerException;
import gipsy.GIPC.intensional.SimpleNode;
import gipsy.GIPC.intensional.GIPL.GIPLCompiler;
import gipsy.interfaces.AbstractSyntaxTree;
import gipsy.interfaces.GIPSYProgram;
import gipsy.storage.Dictionary;

import java.io.InputStream;

import marf.util.Debug;
import marf.util.OptionProcessor;

public class GIPC
extends IntensionalCompiler
{
public static final int GIPL_PARSER = 0;
public static final int OPT_STDIN = 1;
public static final int OPT_GIPL = 2;
public static final int OPT_GIPL_SHORT = 3;
public static final int OPT_TRANSLATE = 12;
public static final int OPT_TRANSLATE_SHORT = 13;
public static final int OPT_DFG = 17;
public static final int OPT_GIPC = 25;
private Dictionary oDictionary = null;
private OptionProcessor oOptionProcessor = new OptionProcessor();

private IIntensionalCompiler[] aoIntensionalCompilers = null;

private GIPSYProgram oGIPSYProgram = null;

public static int siPrimaryParserType;

public GIPC()
throws GIPCException
{
super();
setupDefaultConfig();
this.oObjectToSerialize = this.oGIPSYProgram;
}
public GIPC(String[] argv)
throws GIPCException
{
setupConfig(argv);
}

protected void setupConfig(String[] argv)
throws GIPCException
{
try
{
this.oOptionProcessor.addValidOption(OPT_STDIN, "--stdin");

this.oOptionProcessor.addValidOption(OPT_GIPL, "--gipl");
this.oOptionProcessor.addValidOption(OPT_GIPL_SHORT, "-G");

}
catch(Exception e)
{
throw new GIPCException(e);
}
}

public GIPC(InputStream poInputStream)
throws GIPCException
{
super(poInputStream);
setupDefaultConfig();
}


protected void setupDefaultConfig()
{
this.oOptionProcessor.addActiveOption(OPT_STDIN, "--stdin");
this.oOptionProcessor.addActiveOption(OPT_TRANSLATE, "--translate");
this.oOptionProcessor.addActiveOption(OPT_TRANSLATE_SHORT, "-T");
}

public GIPSYProgram process()
throws GIPCException
{
String strPhase = "process() begun";

try
{
if(this.oOptionProcessor.isActiveOption(OPT_GIPL))
{
Debug.debug("GIPL-only processing");
strPhase = "GIPL";
```

```
siPrimaryParserType = GIPL_PARSER;
GIPLCompiler oGIPLCompiler = new GIPLCompiler(this.oSourceCodeStream);
this.oAST = oGIPLCompiler.compile();

if(this.oOptionProcessor.isActiveOption(OPT_DFG))
{
DFGCodeGenerator oDFGCodeGenerator = new DFGCodeGenerator();
oDFGCodeGenerator.generateDFG((SimpleNode)this.oAST.getRoot(),
GIPL_PARSER, null);
}
}
return this.oGIPSYProgram;
}

catch(Exception e)
{
Debug.debug("GIPC foobared: " + e);
e.printStackTrace(System.err);
throw new GIPCException("Phase: " + strPhase + ", " + e.getMessage() + e,
e);
}
}

public void init()
throws GIPCException
{
this.oDictionary = new Dictionary();
}

public AbstractSyntaxTree parse()
throws GIPCException
{
return this.oAST;
}

public AbstractSyntaxTree translate()
throws IntensionalCompilerException
{
for(int i = 0; i < this.aoIntensionalCompilers.length; i++)
{
AbstractSyntaxTree oCurrentIntensionalAST =
this.aoIntensionalCompilers[i].translate();
}

return this.oAST;
}

public String lookupCompiler(String pstrLanguageName)
{
int i;
for(i = 0; i < EIntensionalLanguages.INTENSIONAL_LANGUAGES.length;
i++)
{
if(pstrLanguageName.equals(EIntensionalLanguages.INTENSIONAL_LAN
GUAGES[i]))
{
return EIntensionalLanguages.INTENSIONAL_COMPILERS[i];
}
}

for(i = 0; i < EImperativeLanguages.IMPERATIVE_LANGUAGES.length;
i++)
{
if(pstrLanguageName.equals(EImperativeLanguages.IMPERATIVE_LANG
UAGES[i]))
{
return EImperativeLanguages.IMPERATIVE_COMPILERS[i];
}
}
return null;
}

public Dictionary getDictionary()
{
return this.oDictionary;
}

public AbstractSyntaxTree compile(Object poExtraArgs)
throws GIPCException
{
init();
process();

return this.oAST;
}

public GIPSYProgram getGIPSYProgram()
{
return this.oGIPSYProgram;
}

public GIPSYProgram getGEER()
{
return getGIPSYProgram();
}
}
```

GIPLCompiler:

```
package gipsy.GIPC.intensional.GIPL;

import java.io.InputStream;

import gipsy.GIPC.GIPCException;
import gipsy.GIPC.intensional.IntensionalCompiler;
import gipsy.interfaces.AbstractSyntaxTree;

public class GIPLCompiler
extends IntensionalCompiler
{

private GIPLParser oParser;

public GIPLCompiler()
throws GIPCException
{
super();
}
public GIPLCompiler(InputStream poInputStream)
throws GIPCException
{
super(poInputStream);
}

public GIPLCompiler(String pstrFilename)
throws GIPCException
{
super(pstrFilename);
}

public void init()
throws GIPCException
{
this.oParser = new GIPLParser(this.oSourceCodeStream);
}

public AbstractSyntaxTree parse()
throws GIPCException
{
return this.oParser.parse();
}
}
```

*Tool Support:*

We use ObjectAid UML plug-in for reverse engineering. It is a graphical representation of code in form of UML diagrams. The tool is agile and lightweight for Eclipse. It is easy to use just drag and drop java classes from package explorer to ObjectAid view. It automatically shows the relationship between the classes. If there is any change in the code then automatically shows the changes in the class diagrams The ObjectAid UML Explorer is an Eclipse plug-in. When there is an update/refactor in the source code, then it reflect the changes in the diagrams.it allows to Save UML diagram as a GIF, PNG or JPEG file.

## III. METHODOLOGY

*Refactoring*

*Identification of code smells and system level refactorings*

*a) DMARF*

| **Code Smell** | **Refactoring Method** |
|---|---|
| Feature Envy | Move method |
| God Class | Extract class |
| Switch Statement | Replace Type Code with State/Strategy |
| Obscured Intent | Replace Magic Number with Symbolic Constant |
| Long Method | Extract Method |
| Dead Code | Delete the Code |

**Table 6: mapping between source code smells and refactoring methods**

**Feature Envy smell:** When a method wants to be somewhere else, Move Method will be the best solution in this case. If one part of the method has this problem (smell), then extract Method on the jealous part and Move Method to the suitable class.

In figure 24 and figure 25, we can see that feature envy smell has been detected by JDeodorant. The fillInTransitionTable method was found in GrammarCompiler class in marf.nlp.Parsing.GrammarCompiler package. In order to resolve the code smell, use refactoring method which is called move method to move this method to TransitionTable class in marf.nlp.Parsing package.

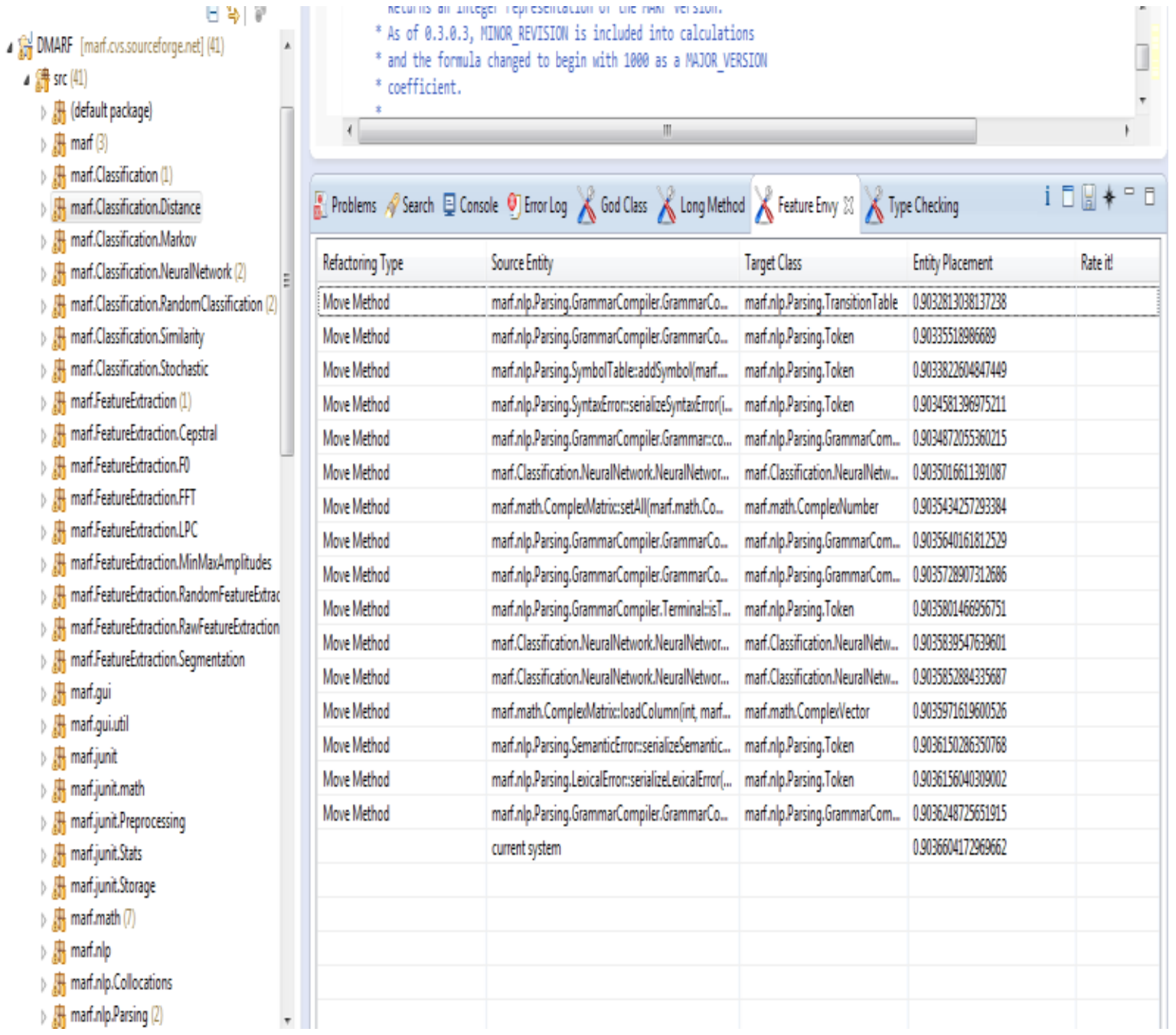

**Figure 23. Feature Envy smell for DMARF**

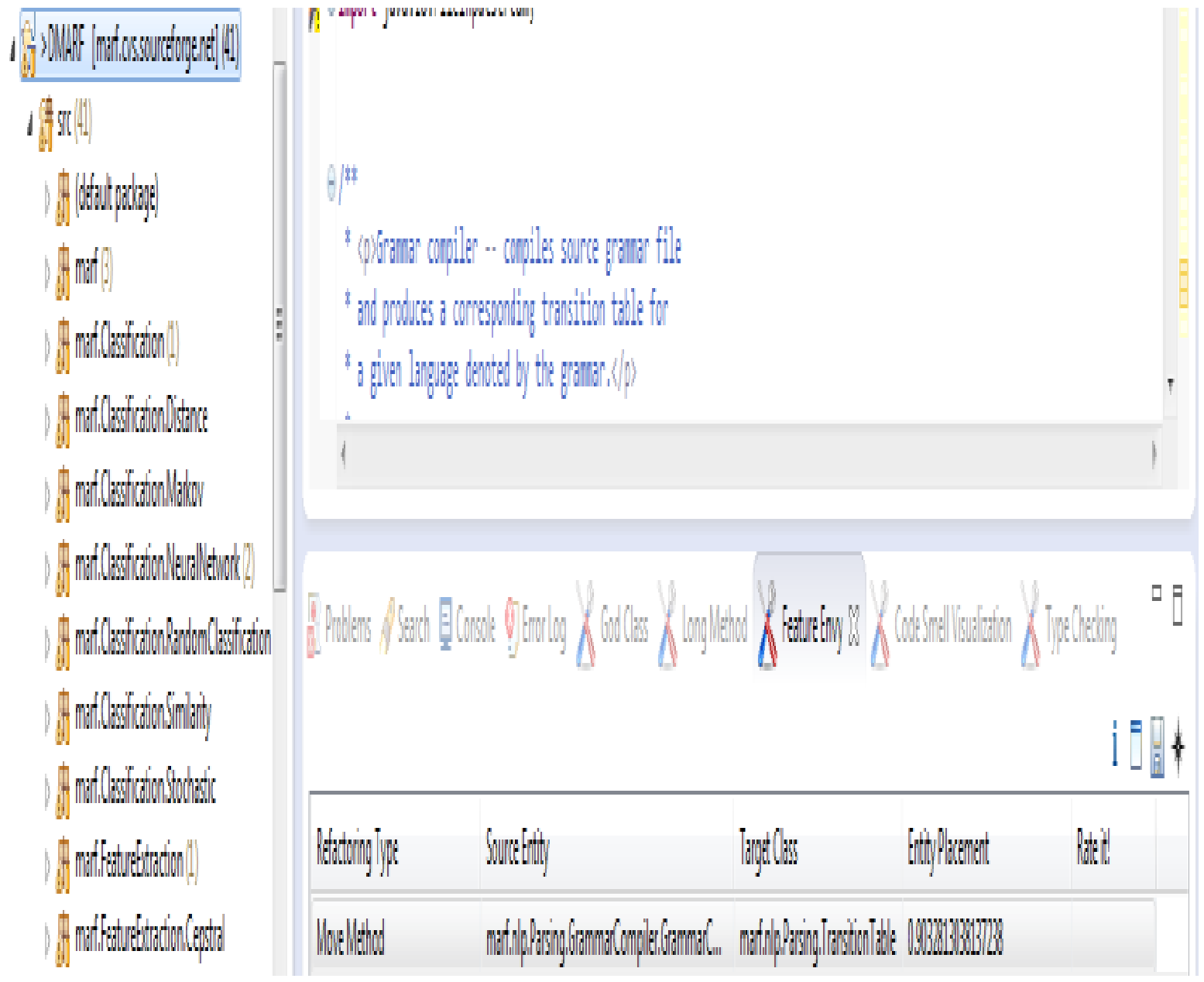

**Figure 24. Feature Envy smell for DMARF**

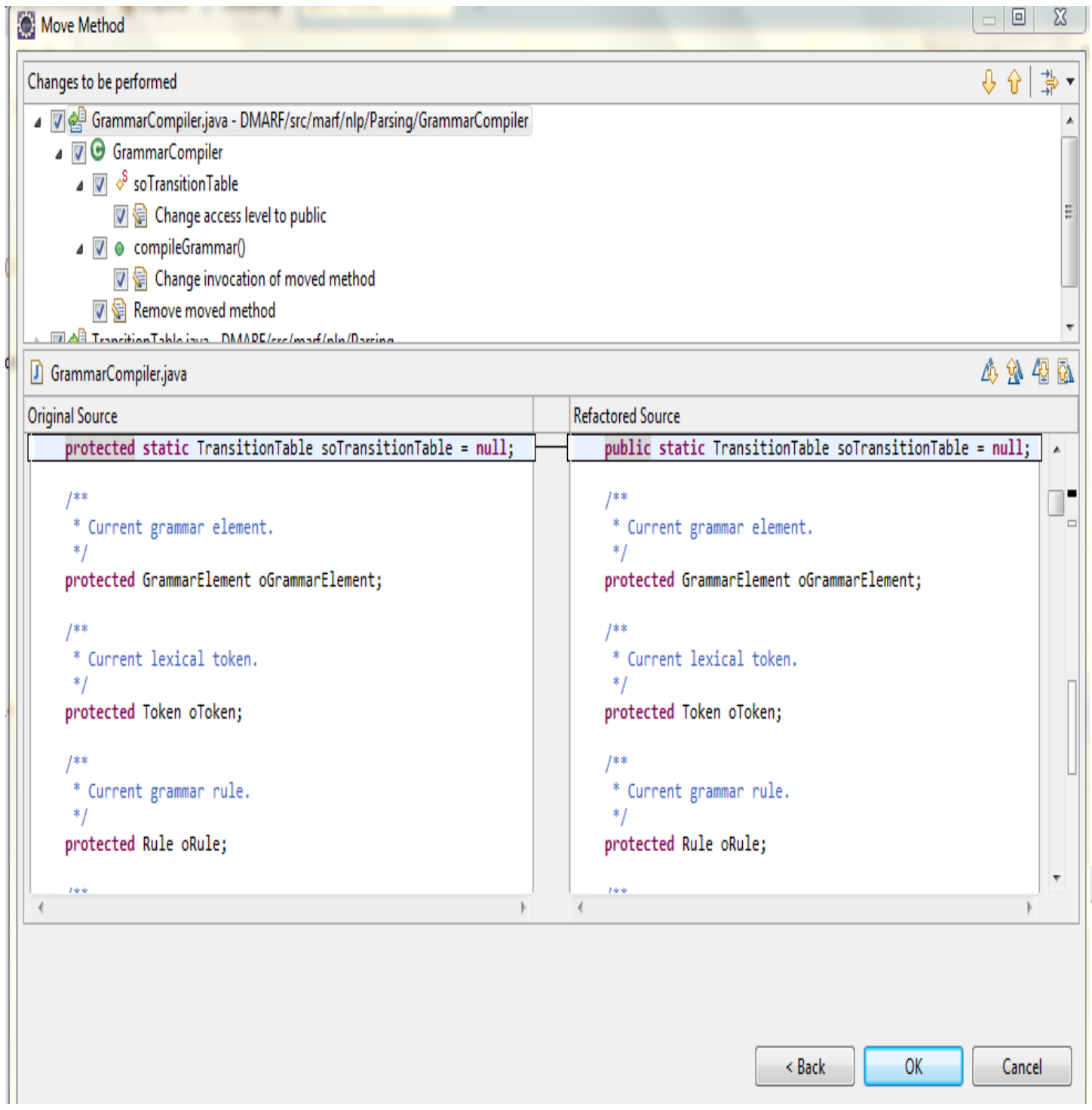

Figure 25. Feature Envy smell for DMARF

**2. God Class smell:** a class that has grown too large.

In figure 23, we can see that god class smell has been detected by JDeodorant. The smell is located at DMARF.src.marf.util.OptionFileLoader.java. In order to resolve the code smell, use refactoring method which is called Extract class to create a new class DMARF.src.marf.util.OptionFileLoaderProduct.java ( From figure 28) and move the relevant fields getKey( ), getValue(), removeComment() and methods loadConfiguration() from the old class DMARF.src.marf.util.OptionFileLoader.java into the new class DMARF.src.marf.util.OptionFileLoaderProduct.java.

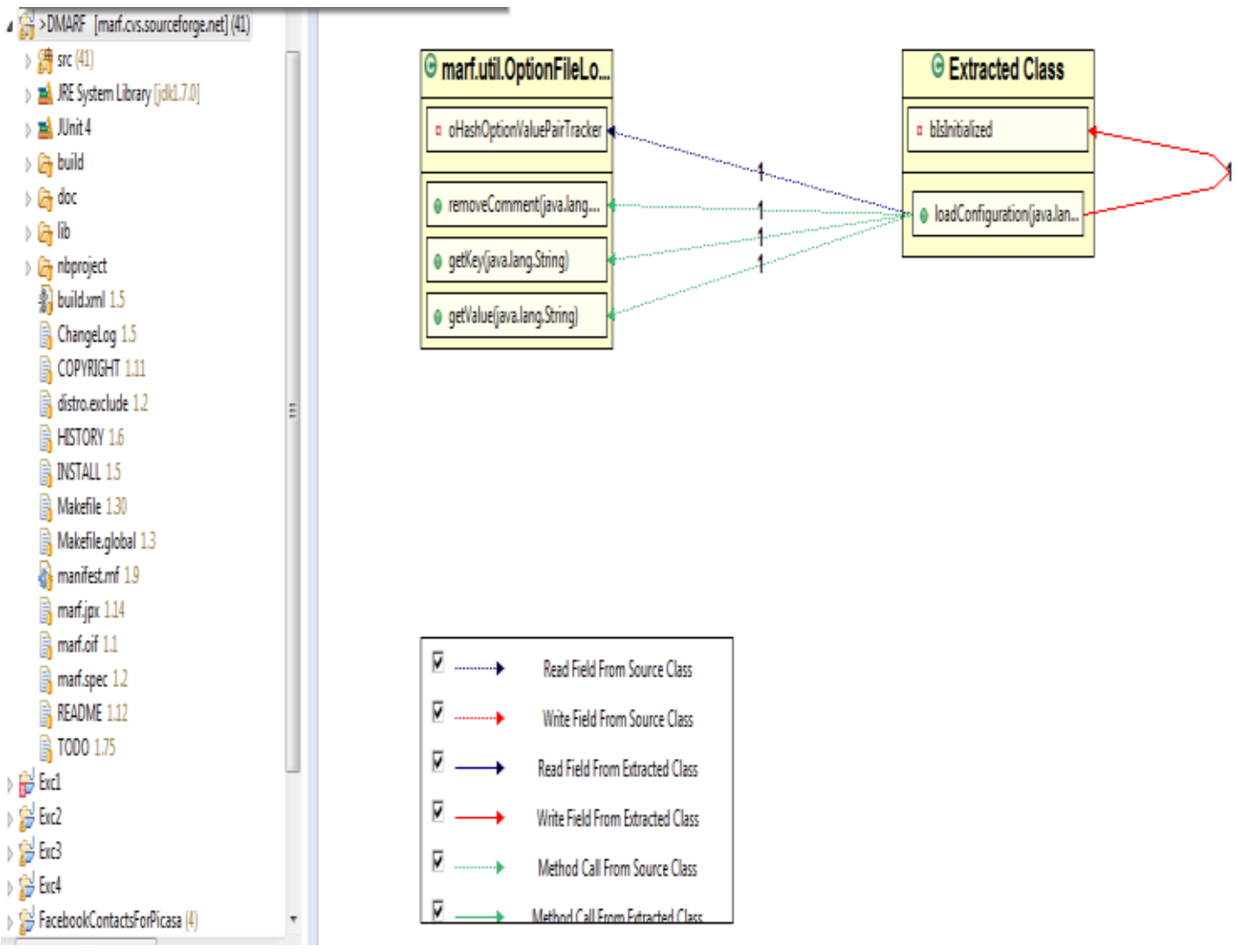

Figure 26. Restructure diagram God Class smell for DMARF

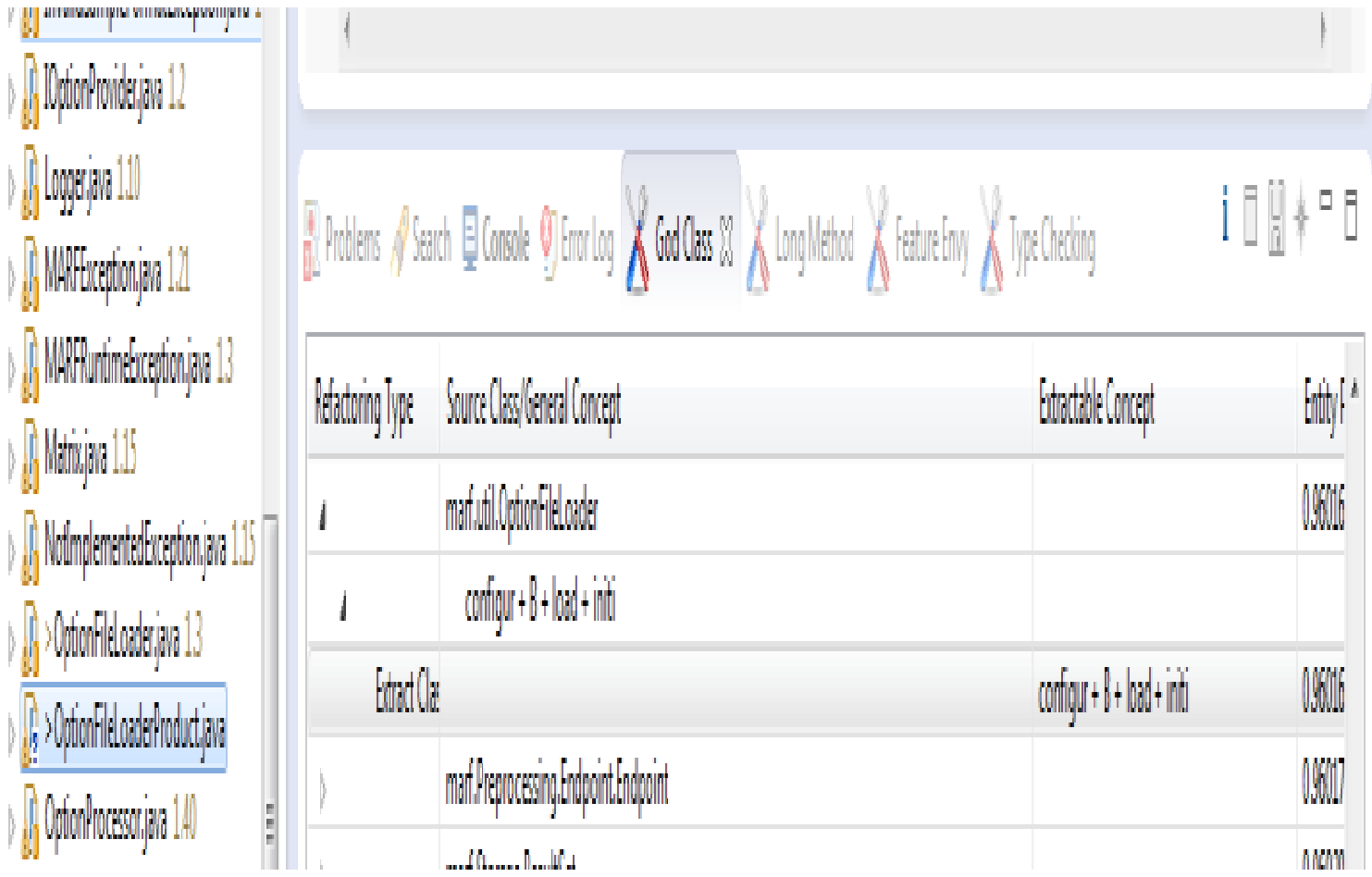

Figure 27. God Class smell for DMARF

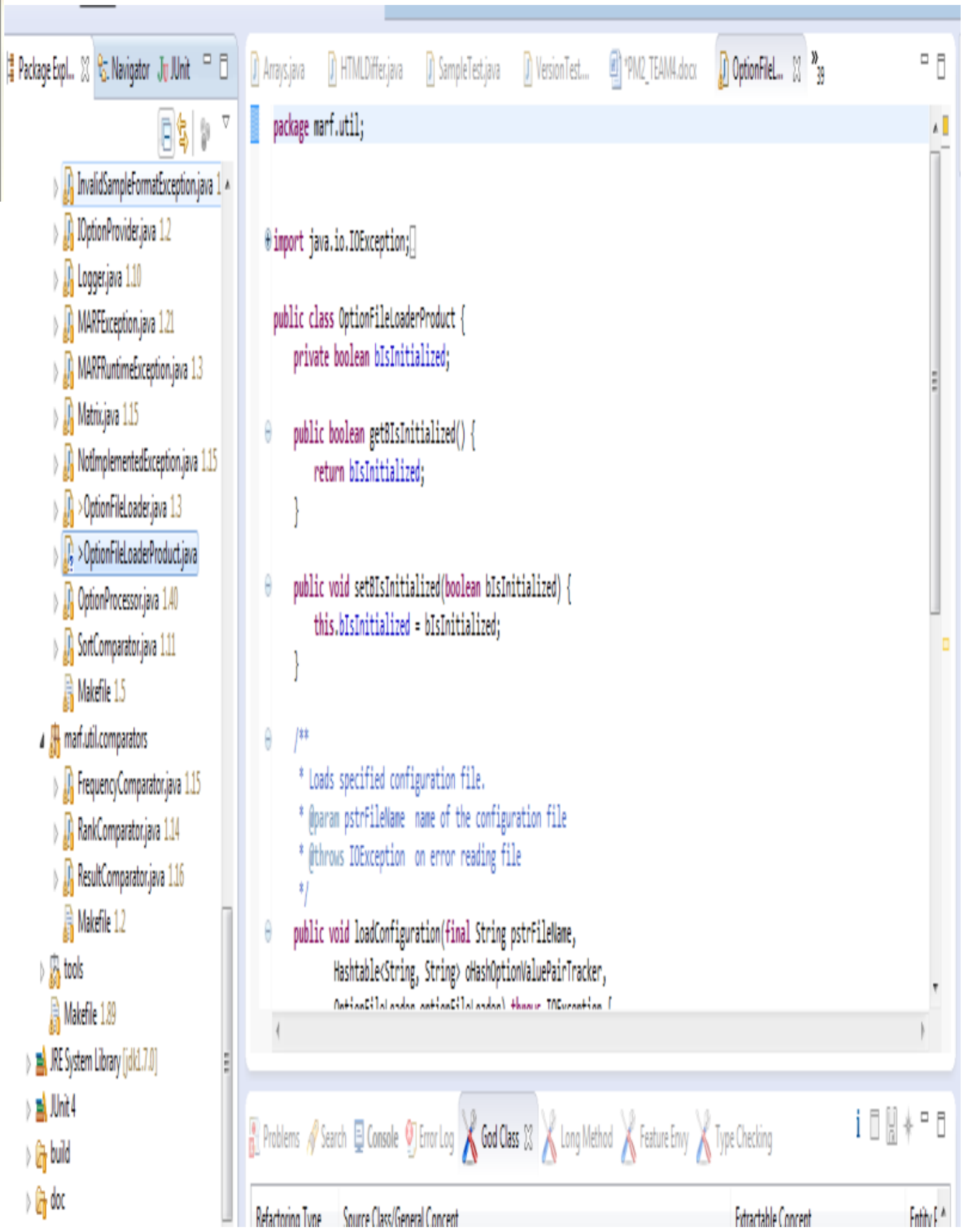

Figure 28. God class smell for DMARF (After refactoring)

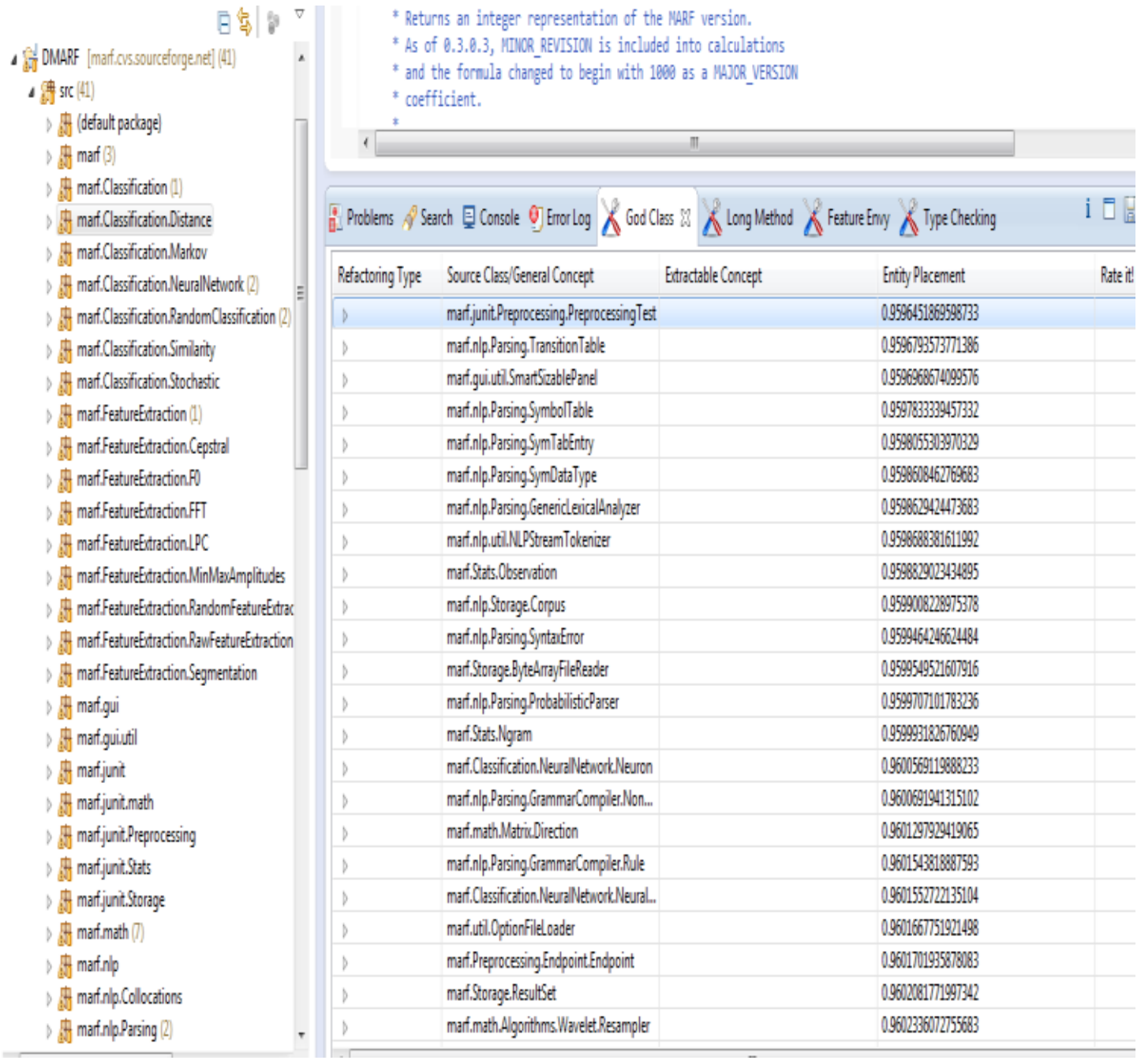

**Figure 29. God Class smell for DMARF**

### 3. Switch statement smell:

Switch statement means to consider polymorphism. The problem is where the polymorphism should happen. Usually, the switch statement switches on a type code. Extract Method to extract the switch statement and then Move Method to get it onto the class where the polymorphism is needed. A decision must be taken to Replace Type Code with Subclasses or Replace Type Code with State/Strategy. When the inheritance structure is set, the Replace Conditional can be used with Polymorphism.

In figure 30, it can be seen that switch statement smell has been detected by JDeodorant. Switch statement smell located at ModuleParams.java from package DMARF.src.marf.storage From figure 30, to refactor comment smell, we use method called Replace Type Code with State/Strategy. Replace the typecode which in green from figure 6 with a state object Sem.

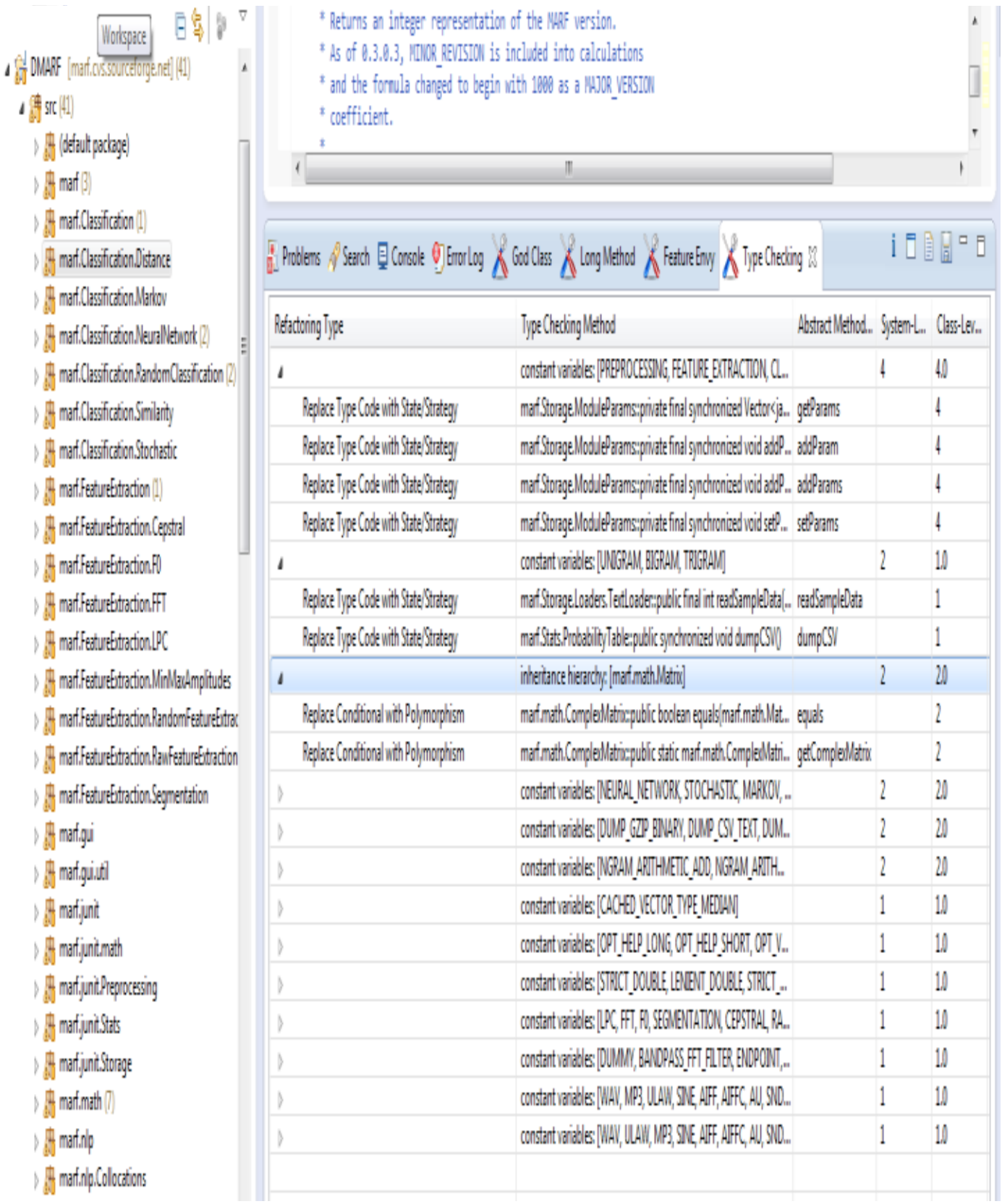

**Figure 30. Switch statement smell for DMARF**

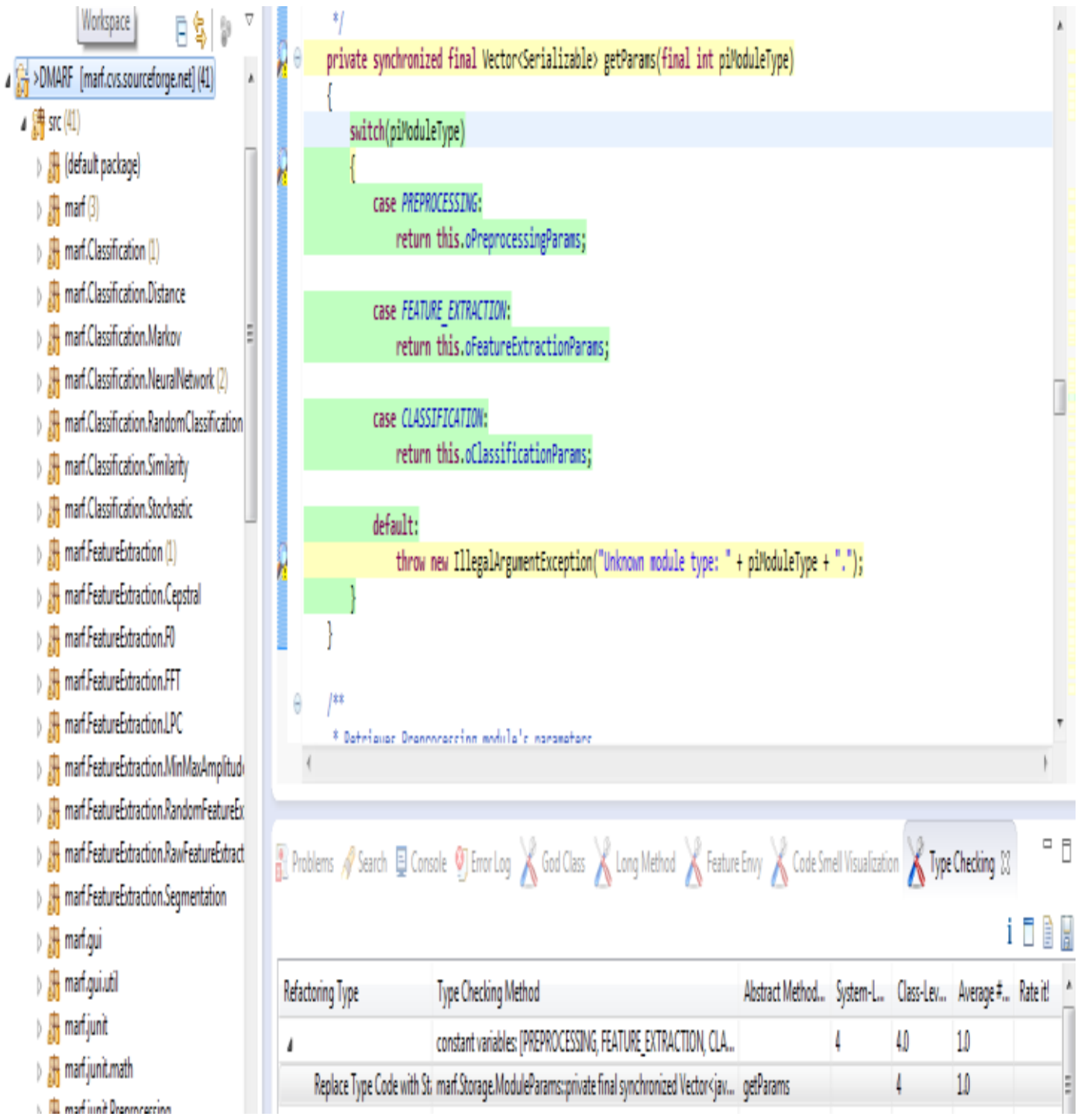

**Figure 31. Switch statement smell for DMARF**

**4. Obscured Intent:** The code is not clear and expressive enough [24].

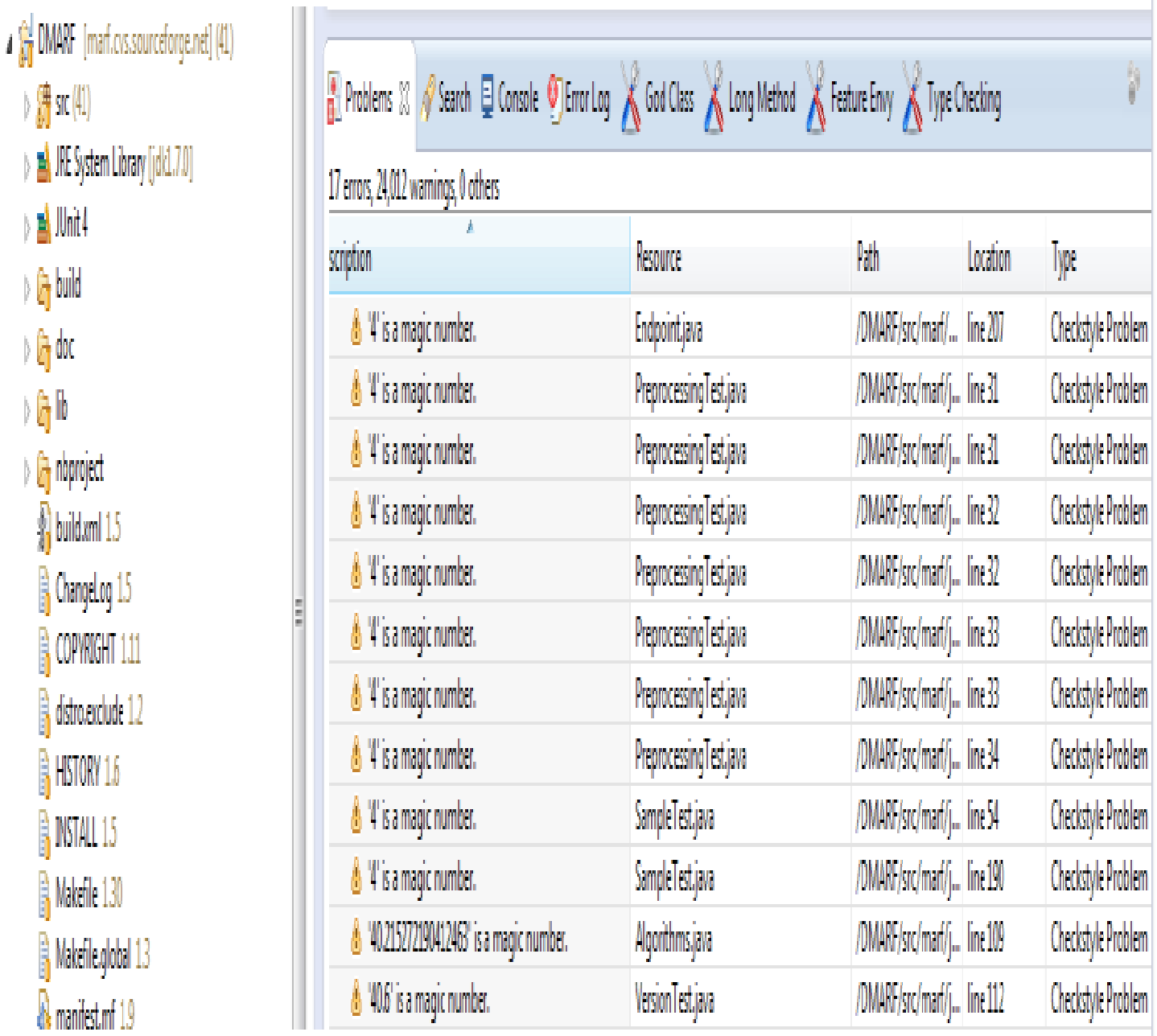

**Figure 32. Obscured Intent smell for DMARF**

Replace Magic Number with Symbolic Constant

static final int SAMPLEVALUE = 10;
static final int D_VERSION = 40.6;

**5. Dead code smell**: The code is not executed, and is not updated when the designs change. In fact, it compiles, but it does not follow newer rules, because it was written at a time when the system was different [24].

In SampleRecorder.java, the variable bais is not used
In LowPassFilter.java, the variable bcd is not used
In CFEFilter.java dLowerBound is not used
In CFEFilter.java dStep is not used
In CFEFilter.java dUpperBound is not used
In Corpus.java oCorpusToCompare is not used
In CFEFilter.java oWVector is not used.

The above code (variable) are not executed, and need to be removed as a refactoring strategy.

**6. Long method**: The longer a procedure is, the more difficult it is to understand, best refactoring strategy is to shorten a method is Extract Method. If a method has lots of parameters and temporary variables, these elements get in the way of extracting methods

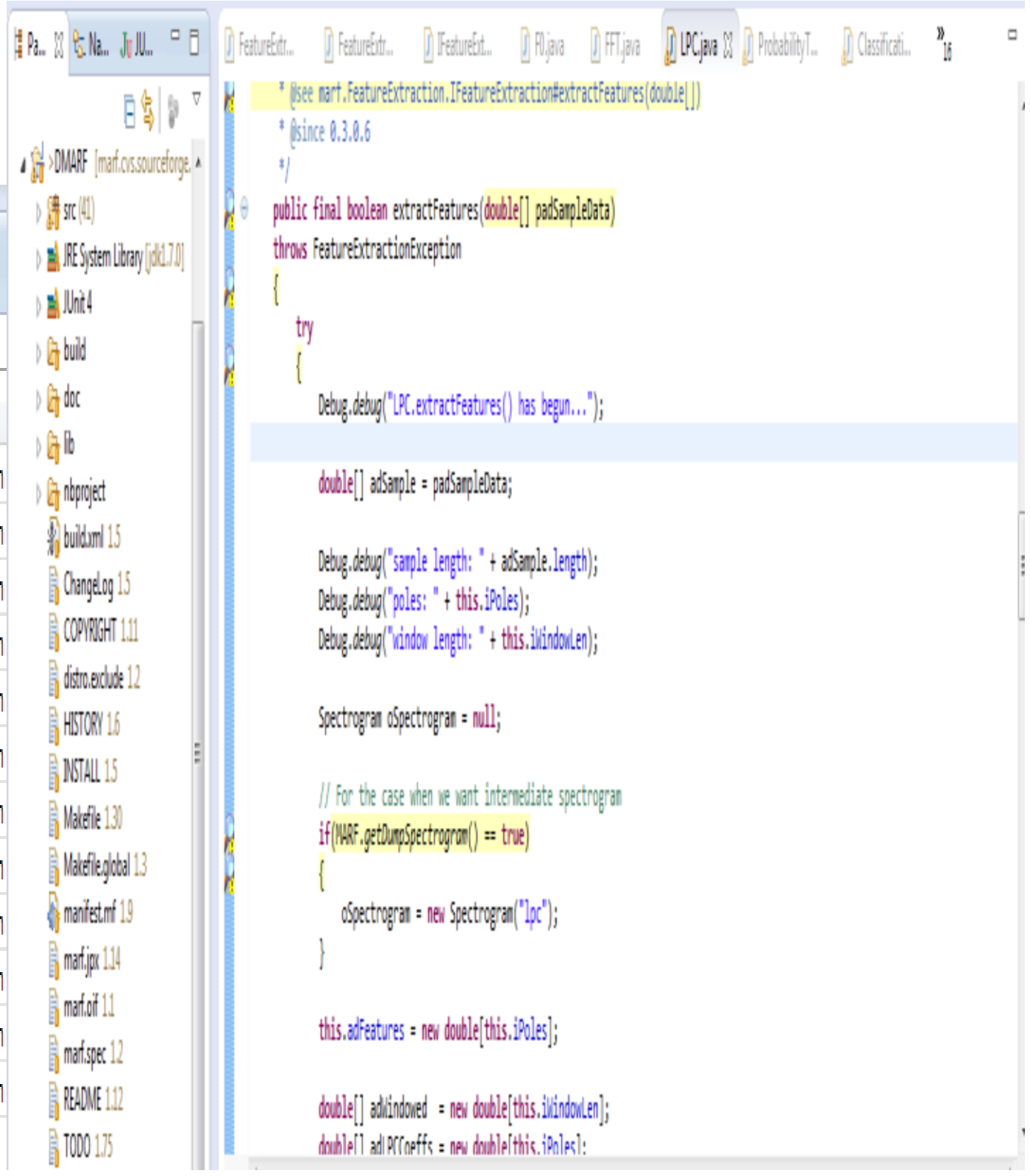

**Figure 33. Long Method for DMARF**

Restructure the system design:

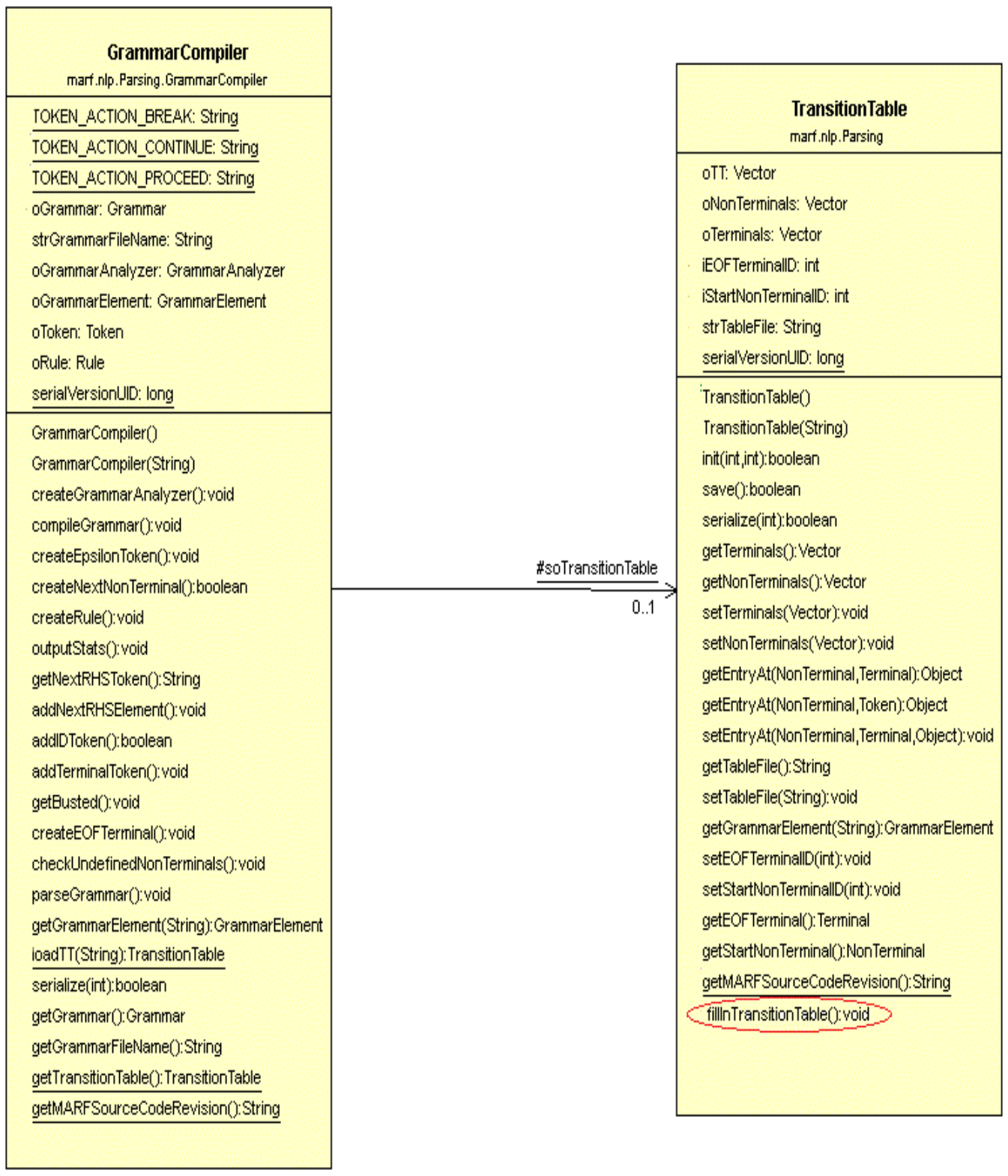

**Figure 34. Feature Envy for DMARF (after Refactoring)**

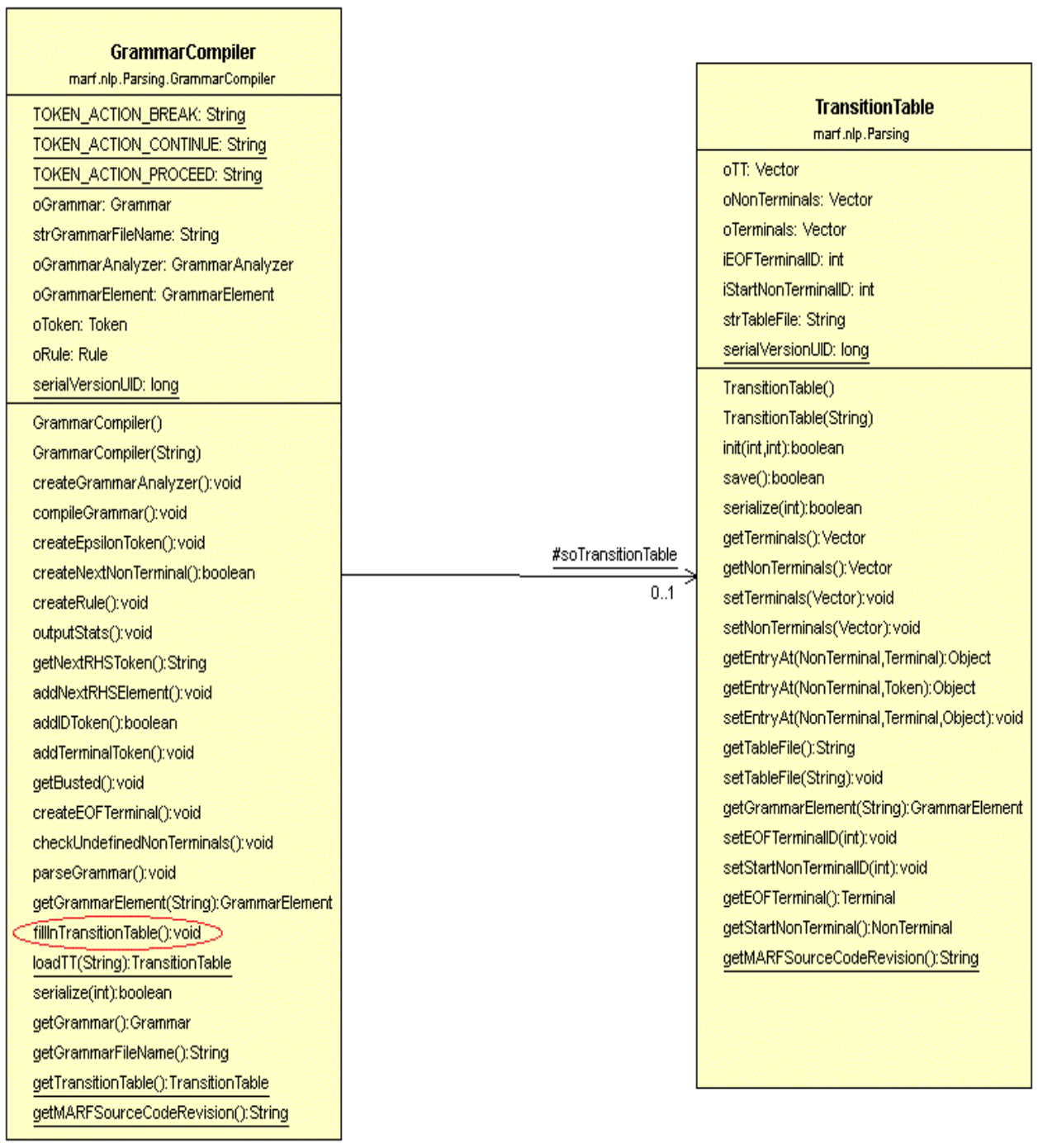

**Figure 35. Feature Envy code smell before refactoring for DMARF**

*b) GIPSY*

| Code smells | Refactoring Type |
|---|---|
| Long Parameter List | Introduce Parameter Object |
| God Class | Extract Class |
| Switch Statement | Replace Type code with status strategy |
| Dead Code | Delete the code or fill the code |
| Long Method | Extract Method |
| Feature Envy | Move method |
| Duplicate Code | Remove or replace Duplication |

**Table 7: List of code smells with respective Refactoring types**

Long Parameter List Are difficult to interpret and become inconsistent, difficult to use. It has been observed that the following class has too many parameters declared way more than specified. To fix this smell we use Introduce Parameter Object as a refactoring type. The figure below depicts that there are 8 parameters declared for the method JOOIPToJavaTranslationItem() violating the threshold, it can be overcome by creating a new object to bind all the parameters.

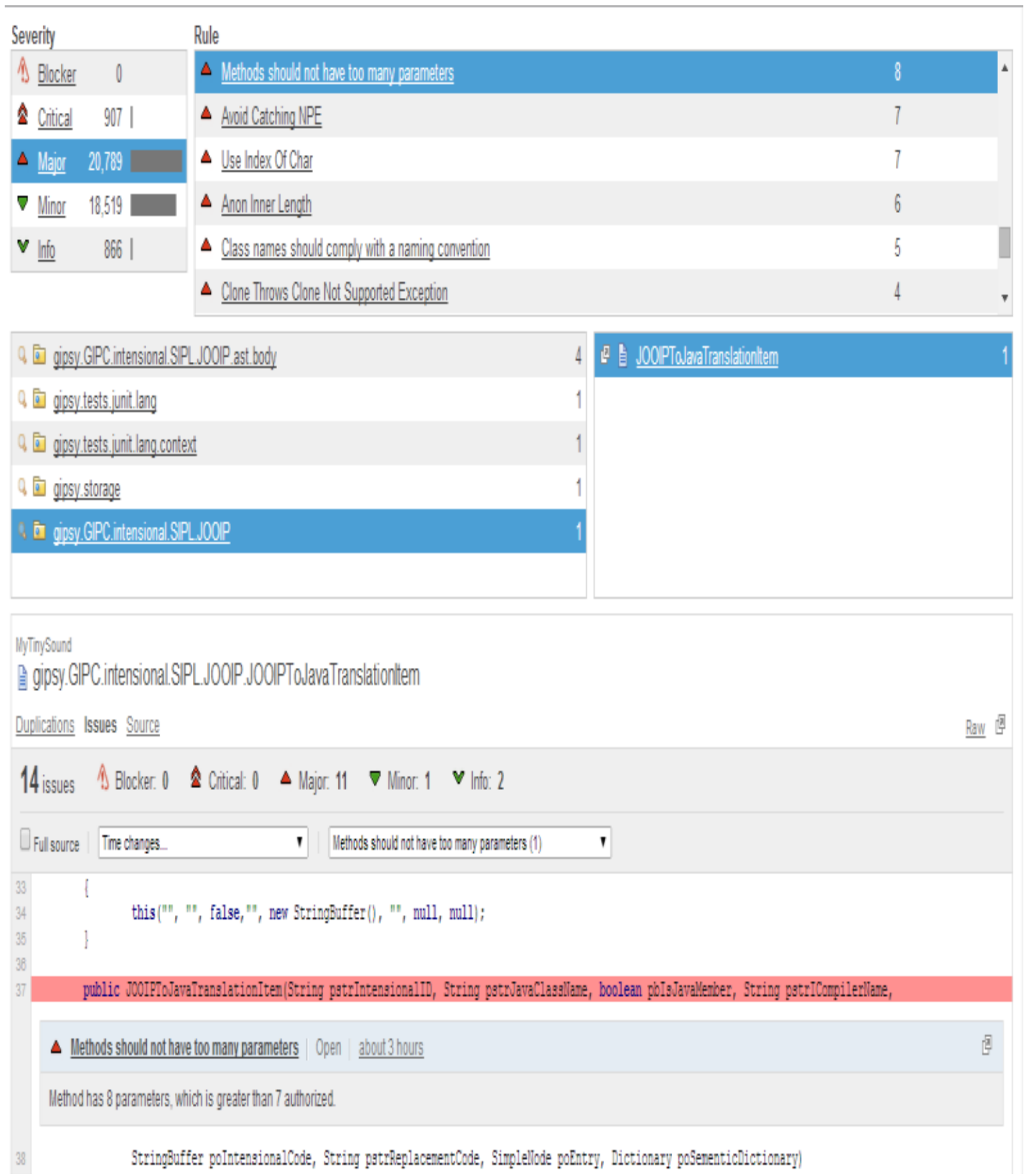

**Figure 36. Long parameter list for GIPSY**

**God Class:** By definition god class is about handling too many responsibilities that a particular class holds, or can be identified by looking at how many instance variables it has. In the GMTInfoKeeper class there are too many methods like (saveNodeRegistration(),saveTierRegistration(),updateSysDSTRegistration(),removeRegistration(),getNodeSysDST(),getNodeRegistartionsSize()getDSTRegistration(),removeDGTDWTRegistration() ) Using many attributes from external classes directly or by using accessor methods

**Ex:** List<DSTRegistration> oDSTRegistrations= new ArrayList<DSTRegistration>();

The refactoring type is **Extract Class.**

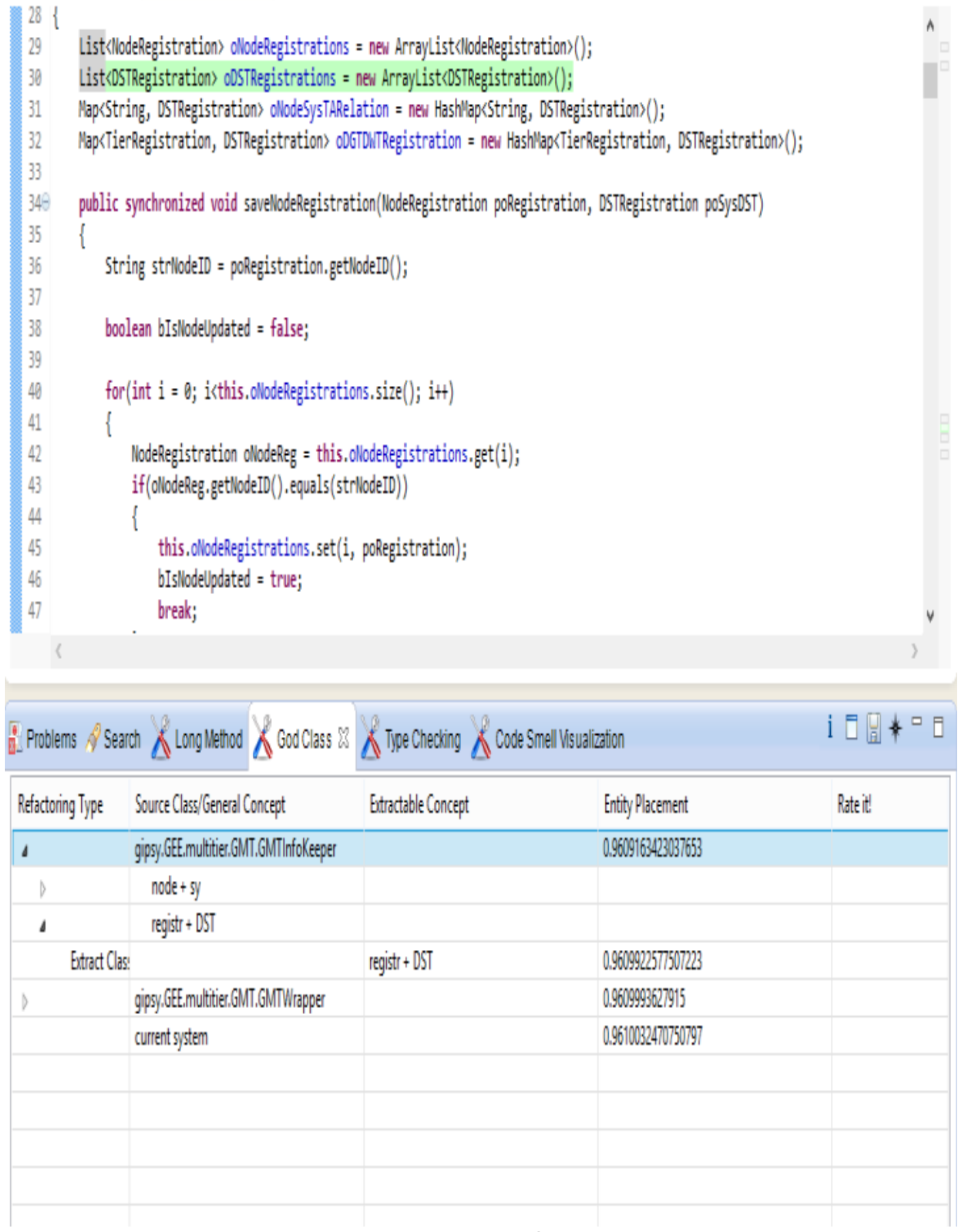

**Figure 37. God Class for GIPSY**

## UML diagram explaining the refactoring of god class

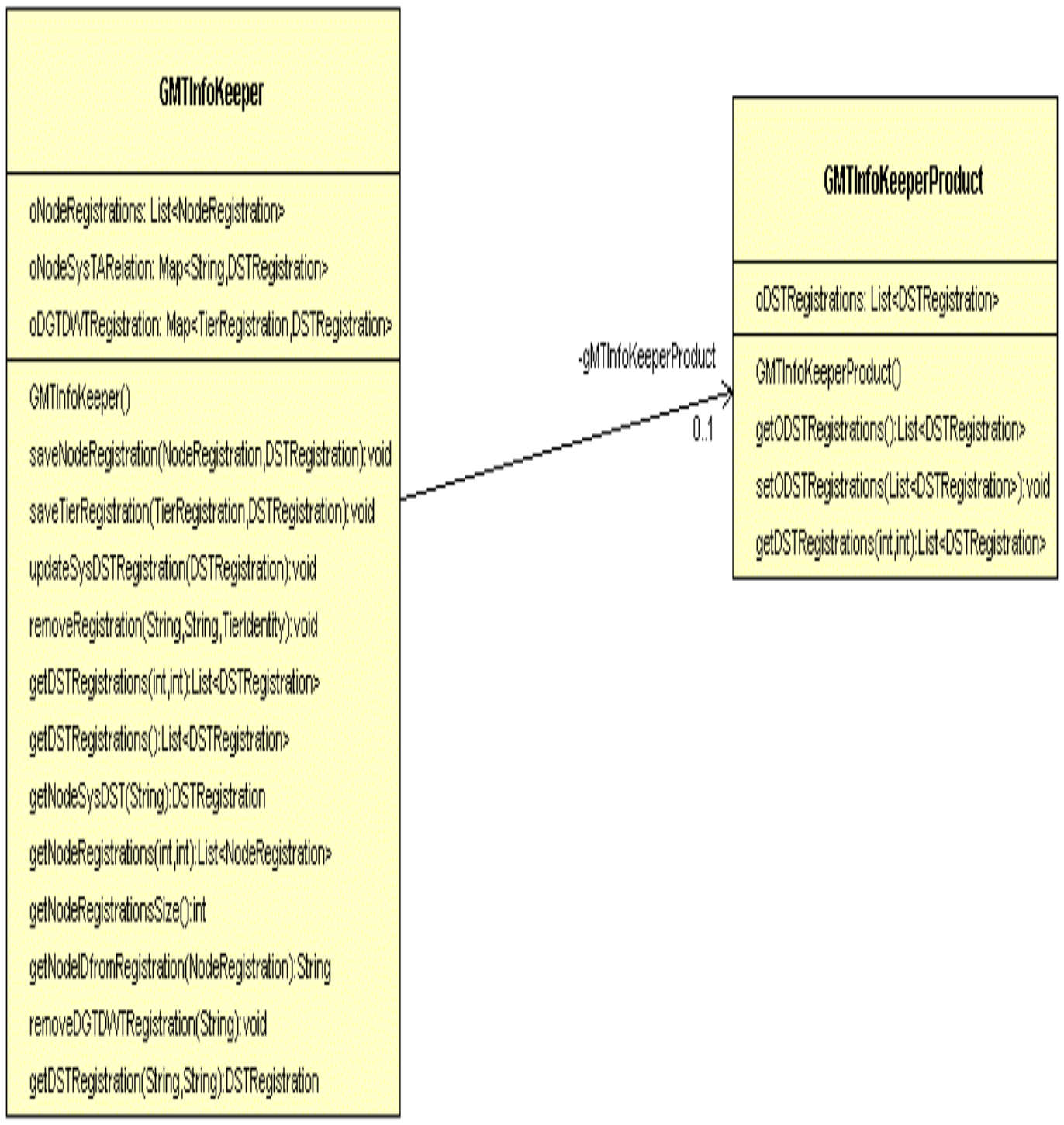

**Figure 38. UML diagram explaining the refactoring of god class**

**Process: GMTInfokeeper class** initially holds many methods in it which results in making the class a god class. We create an extract class **GMTInfoKeeperProduct** and move some of the methods from the god class to the extract class to decrease the complexity of the god class.

**Switch Statement** Switch statements often consist duplicated code, as the same code repeats in different case. And similar switch statement exist all over the code in different parts.

Refactoring technique **Replace Type code with status strategy**

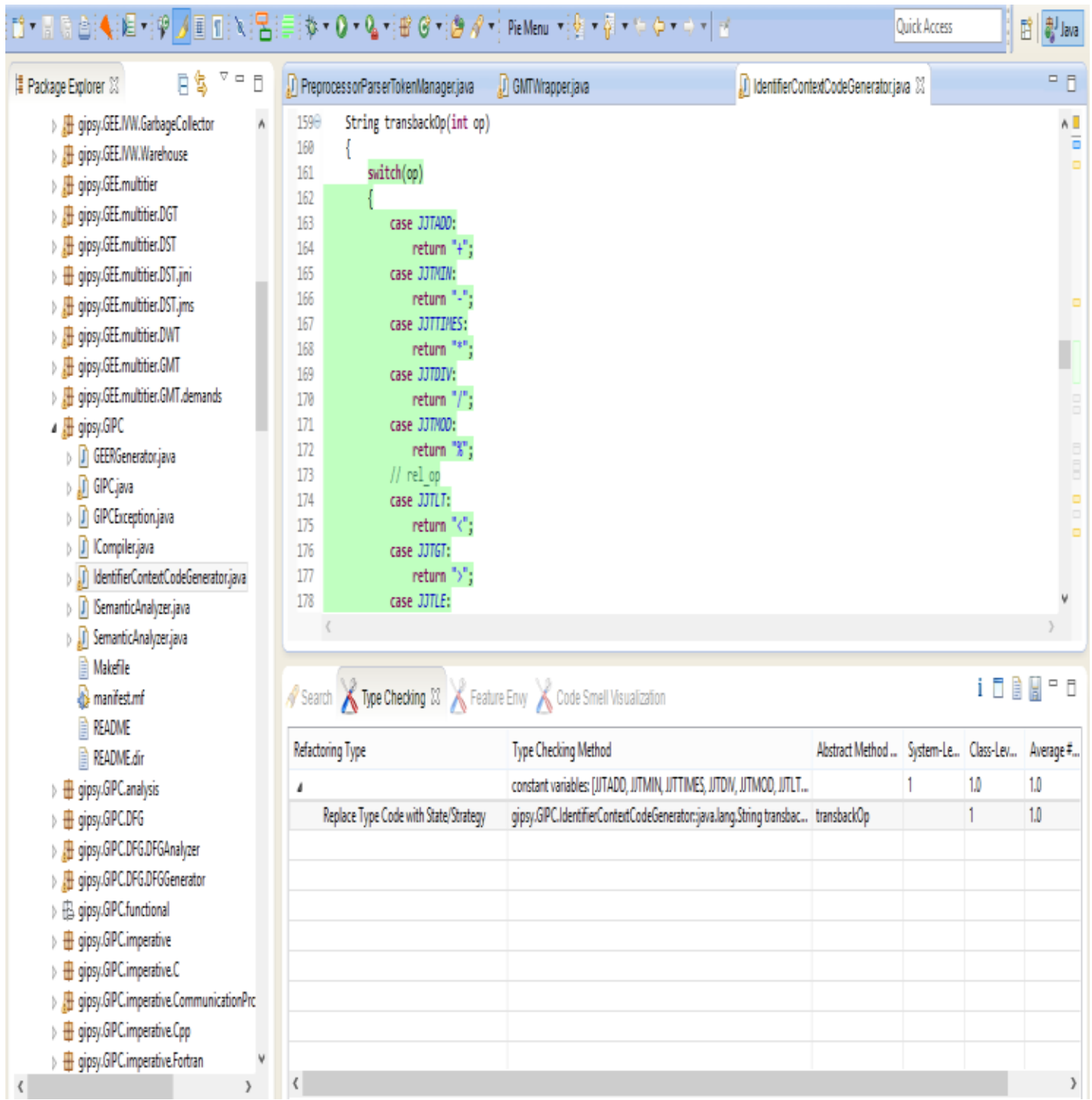

**Figure 39. Switch statement**

## Dead Code

A variable, parameter, method, code fragment, class, etc. Is not used anywhere and therefore does not favor the functionality of the code can be deleted or the corresponding block can be accommodated with lines of code. The corresponding statement as shown in the figure can be deleted or filled.

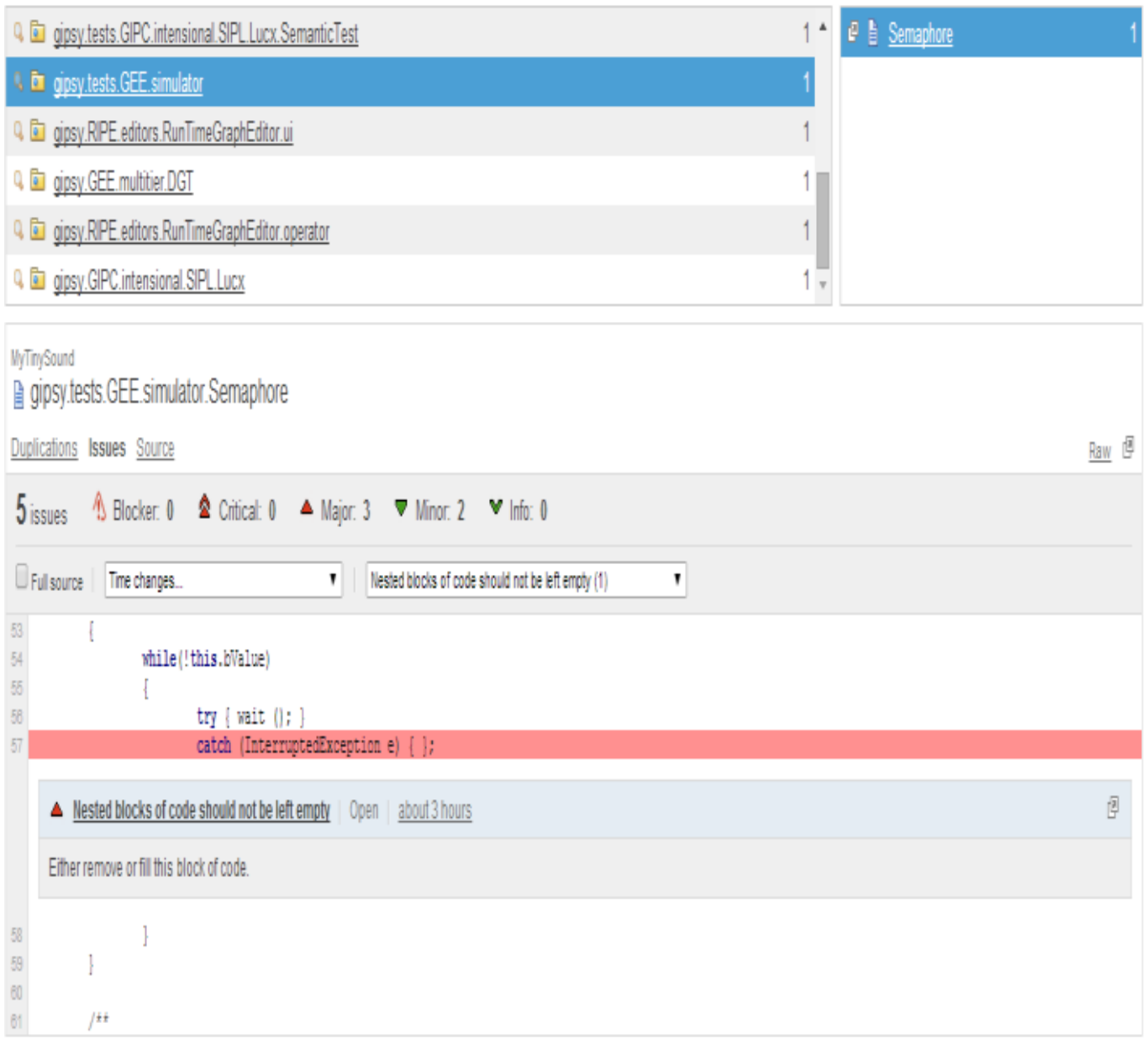

**Figure 40. Dead code**

**Long Method:** The longer the method the harder it is to see what it is doing. As suggested by the tool the Cyclomatic Complexity of the method setupConfig() is 18 which is greater than 10 authorized. We have identified that the particular method is more than 100 lines of code which is difficult to understand with lot of duplication which can be eliminated or we can fix it by using **Extract Method** refactoring type where it can be fixed by grouping code which goes together seamlessly and creates a new method.

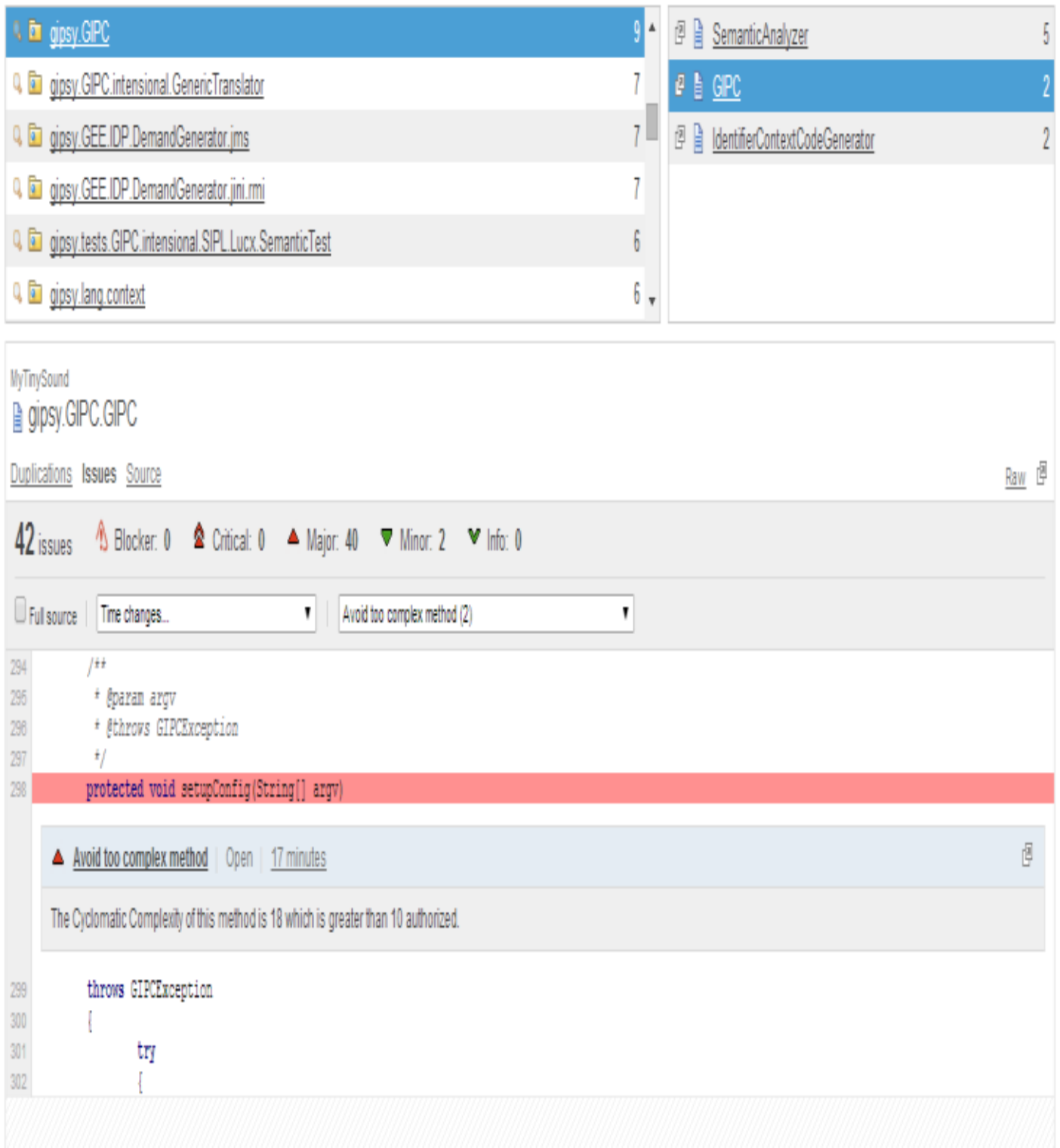

**Figure 41. Long method**

**Duplicate Code:** Reusing existing code in different locations is the simplest form of reuse mechanism in the development process, which results in duplicate code. As shown in the figure the following statement is found repeated several times in the GMTWrapper class, this can be eliminated by removing Duplication or replacing it.

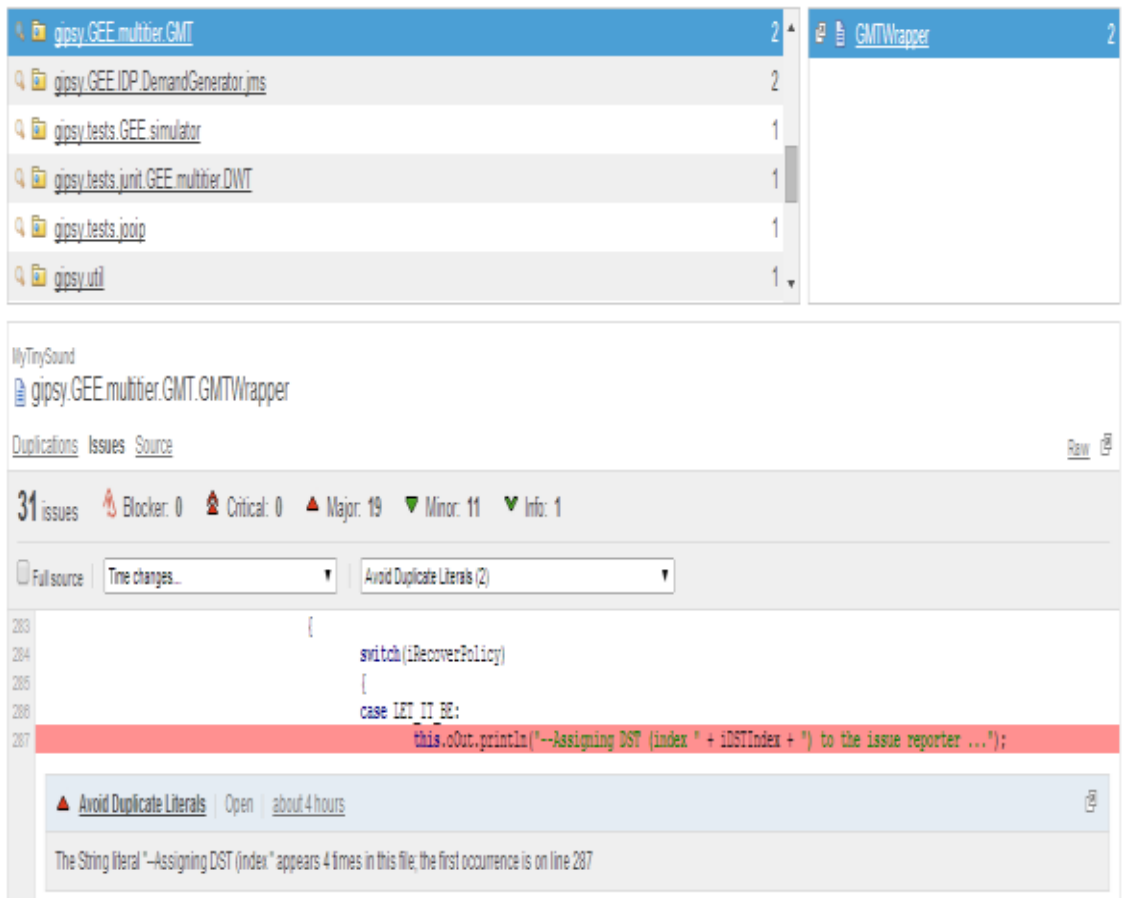

**Figure 42. Duplicate code**

*Tool Support*

*JDeodorant*

JDeodorant is an Eclipse plug-in that can be used to identify bad smells. It is able to identify god classes, long method, feature envy and type checking code smells. Following the identification of the bad smell it recommends and is able to implement a refactoring to resolve the smells in the project.

*SonarQube*

SonarQube is a quality tool to analyze a project's source code. It broadly classifies the issues into critical, major, minor categories where each of them contains series of suggestions for code improvement. Based on the results of this tool we were able to identify areas of the project source code which were of concern. This tool was also ran throughout the course of the project to validate the refactoring methods which were applied to correct code smells.

*Specific Refactoring's that will be Implemented*

*DMARF and GIPSY*

Test cases for GIPSY and DMARF already covers all the aspects. For GIPSY a package name gipsy.test consists of all the relevant test cases to ensure the external behavior of the system remains intact.

| Class Component | Corresponding Package Name (JUnit) |
|---|---|
| Demand Generator Tier | Gipsy.tests.junit.GEE.multitier.DGT |
| Demand Store Tier | Gipsy.tests.junit.GEE.multitier.DST |
| Demand Worker Tier | Gipsy.tests.junit.GEE.multitier.DWT |
| Demand Store | Gipsy.test.junit.interfaces |

For DMARF import a package from CVS named Apps which consists of all the test cases for testing DMARF system.

| Feature | Corresponding Folder name (Test App) |
|---|---|
| Speaker Identification | SpeakerIdentApp |
| Loader | TestLoaders |
| Filter | TestFilters |

The above mentioned test cases covers the complete system. We don't need to create any additional test cases for testing the systems.

The following refactoring's will be applied in next sections.

1. Fixing the long method:

Method called setupConfig() consists of several lines of code which makes it less understandable. The following as the type of refactoring which could be applied to fix this issue:

- Extract method: Split the large code method to simpler and smaller methods with appropriate names.

2. Fixing duplicated code:
GMTWrapper class consists a large amount of code which is duplicated. The following is the refactoring which is relevant to remove this type of bad code smell:

- Remove duplicated code or replace the code.

3. Fix the switch statement code smell:
IdentifierContextCodeGenerator is the class which consists switch statement which in turn has some duplicated code. The following refactoring can be applied to fix this

- Replace Type Code with State/Strategy.
- Replace Conditional with Polymorphism.

Present problematic Code for identifierContextCodeGenerator:

```
String transbackOp(int op)
	{
		switch(op)
		{
			case JJTADD:
				return "+";
			case JJTMIN:
				return "-";
			case JJTTIMES:
				return "*";
			case JJTDIV:
				return "/";
			case JJTMOD:
				return "%";
			// rel_op
			case JJTLT:
				return "<";
			case JJTGT:
				return ">";
			case JJTLE:
				return "<=";
			case JJTGE:
				return ">=";
			case JJTEQ:
				return "==";
			case JJTNE:
				return "!=";
			// log_op
			case JJTAND:
				return "&&";
			case JJTOR:
				return "||";
			default:
				return (" bad Operator");
		}
	}
```

4. Fixing the feature envy code smell:
GrammerComplier class consists of feature envy smell. The following refactoring could be applied to fix the issue:

- Move method is applied to move a method to a different class.

5. Fixing god class smell:
OpenFileLoader consists of god class smell. The following is the type of refactoring which could be applied to fix the issue:

- Extract class- create a new class and move relevant fields and methods from old class to the new.

6. High Coupling:
From the re-engineering tools used (ObjectAid) for visualization of class diagrams, there exist a lot of method calls from a class to the other classes in the systems design, this gives a hint of high coupling smell which indicates that one module relies on one or more modules (dependency).High coupling indicates the class is less reusable and difficult to maintain.

7. Less Cohesion:
File called objectiveIndexicalLucidParser in GIPSY has LCOM (Lack of cohesion) value of 13422. Cohesion is defined as how strongly components in a module are connected to each other, the high value indicates lack of cohesion which is considered as a code smell. High LCOM value indicates that the class can be split into sub classes of high cohesion.

*Identification of Design Patterns*

a) *DMARF*

**State**

Let an object to modify its behavior when it’s internal state changes. The object will appear to change its class. In other words, an object's behavior depends on its state, and it should change its behavior depending on that state at run-time. Operations have large conditional statements which depend on the object's state. The State pattern places each branch of the conditional in a separate class. Puts all behavior associated with a state into one object [25].

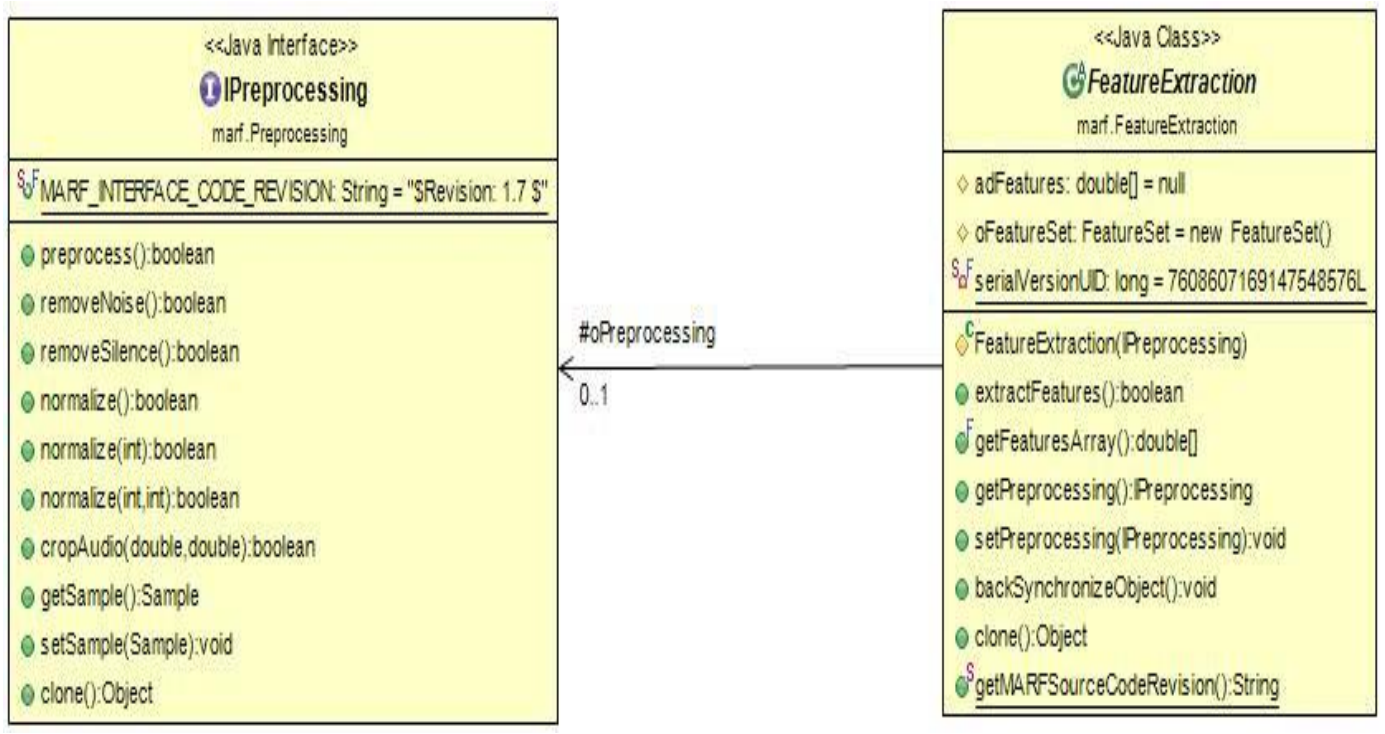

**Figure 43. UML Class Diagram for State Design Pattern**

```
package marf.FeatureExtraction;

import marf.Preprocessing.IPreprocessing;

public interface IFeatureExtraction
{

	String MARF_INTERFACE_CODE_REVISION = "$Revision: 1.1 $";

	boolean extractFeatures()
	throws FeatureExtractionException;

	 */
	boolean extractFeatures(double[] padSampleData)
	throws FeatureExtractionException;

	double[] getFeaturesArray();
	IPreprocessing getPreprocessing();


	void setPreprocessing(IPreprocessing poPreprocessing);
}
```

Context:marf.FeatureExtraction. FeatureExtraction
State/strategy:marf.Preprocessing.IPreprocessing

In State pattern a class behavior changes based on its state. This type of design pattern comes under behavior pattern.

In State pattern, we create objects which represent various states and a context object whose behavior varies as its state object changes.

We're going to create a IPreprocessing interface defining these actions and concrete state classes implementing the IPreprocessing interface. Context is a class which carries a State.

**Singleton**

The intent of the Singleton pattern as defined in *Design Patterns* is to "ensure a class has only one instance, and provide a global point of access to it".
Singleton controls how class instances are created and then ensures that only one instance gets created at any given time. This ensures exactly the behavior that is required, and releases a client from having to know any class details [26].
Singleton design patterns are should be used moderately, the singleton's instance variable is static, which means that all derived classes will share a single copy of it [27].

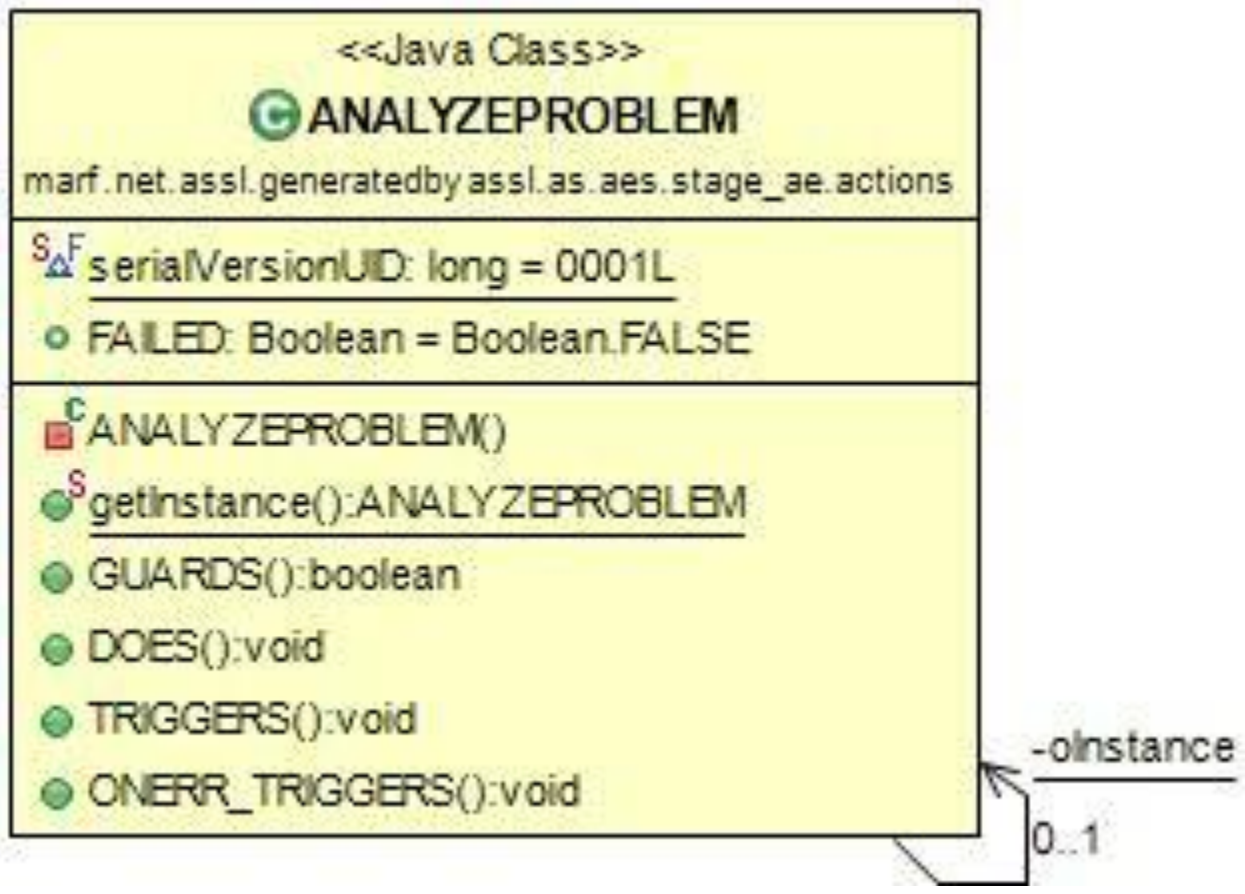

**Figure 44. UML Class Diagram for Singleton Design Pattern**

```
public class ANALYZEPROBLEM
        extends ASSLACTION
        implements  Serializable
{


        static  private  ANALYZEPROBLEM  oInstance  =
null;
        static final long serialVersionUID = 0001L;
        public Boolean FAILED = Boolean.FALSE;
        private  ANALYZEPROBLEM (  )
        {
        }
        static public ANALYZEPROBLEM getInstance (  )
        {
                if ( null == oInstance )
                {
                        oInstance          =          new
ANALYZEPROBLEM();
                }
                return oInstance;
        }
```

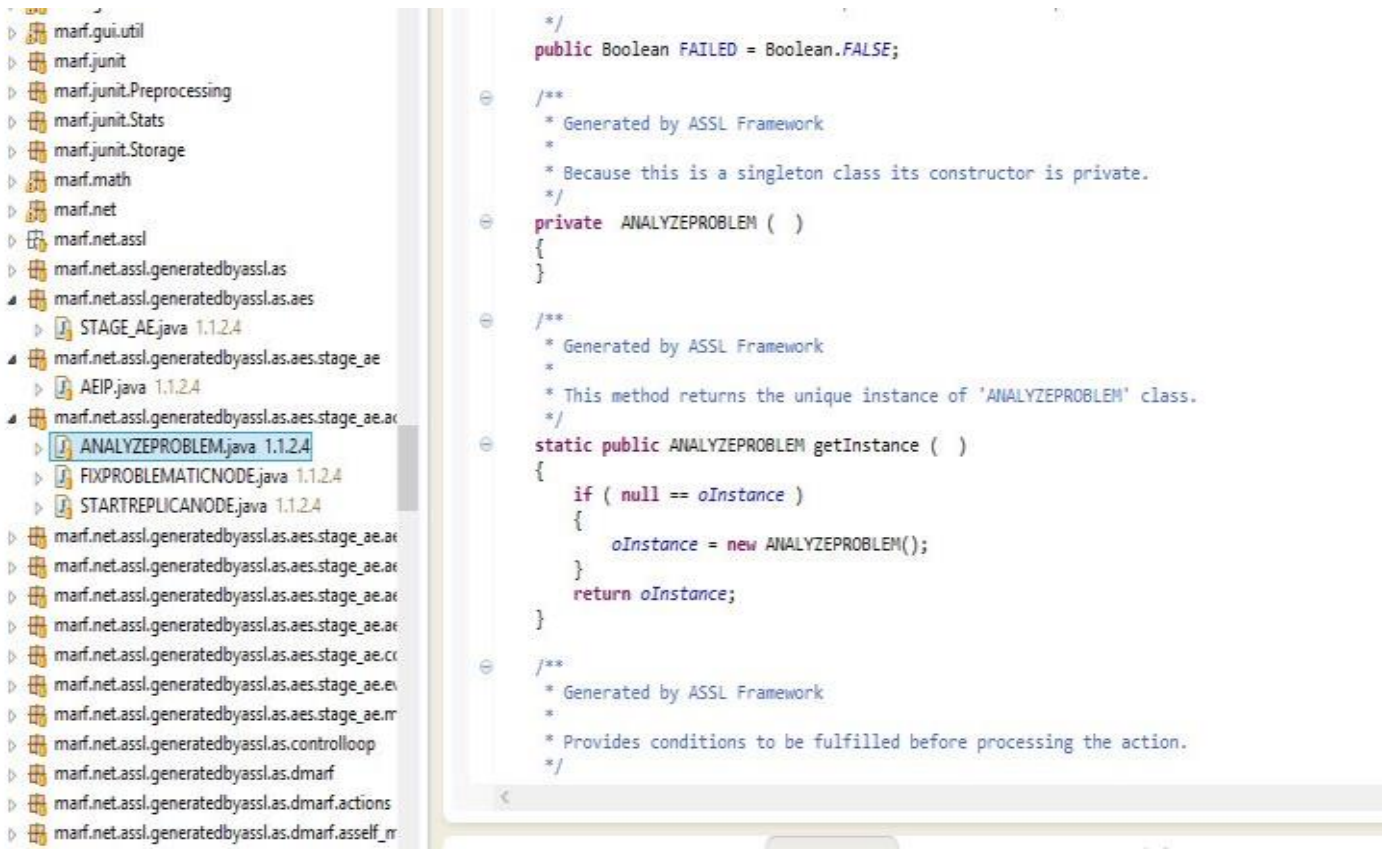

**Figure 45. Code Snippet for Singleton**

Singleton
marf.net.assl.generatedbyassl.as.aes.stage_ae.actions.ANALYZEPROBLEM
uniqueInstance:private static
marf.net.assl.generatedbyassl.as.aes.stage_ae.actions.
ANALYZEPROBLEM

Singleton pattern is one of the simplest design patterns in Java. This type of design pattern comes under creational pattern as this pattern provides one of the best ways to create an object.
This pattern involves a single class which is responsible to creates own object while making sure that only single object get created. This class provides a way to access its only object which can be accessed directly without need to instantiate the object of the class.

ANALYZEPROBLEM class provides a static method to get its static instance to outside world, our project class will use class ANALYZEPROBLEM to get an ANALYZEPROBLEM object.

**Factory**

Introduces an interface for making an object, in case subclasses decide which class to instantiate. Factory allows a class defer instantiation to subclasses. In other words, instantiates new objects when run-time decides what kind of object to be instantiated.

An object is created without exposing the creation logic to the client and refer to newly created object using a common interface. [27].

Creator: marf.util.IMARFException
Factory Method():
marf.util.IMARFException::create(java.lang.string.java.lang.exception):
matf.util.IMARFException 6

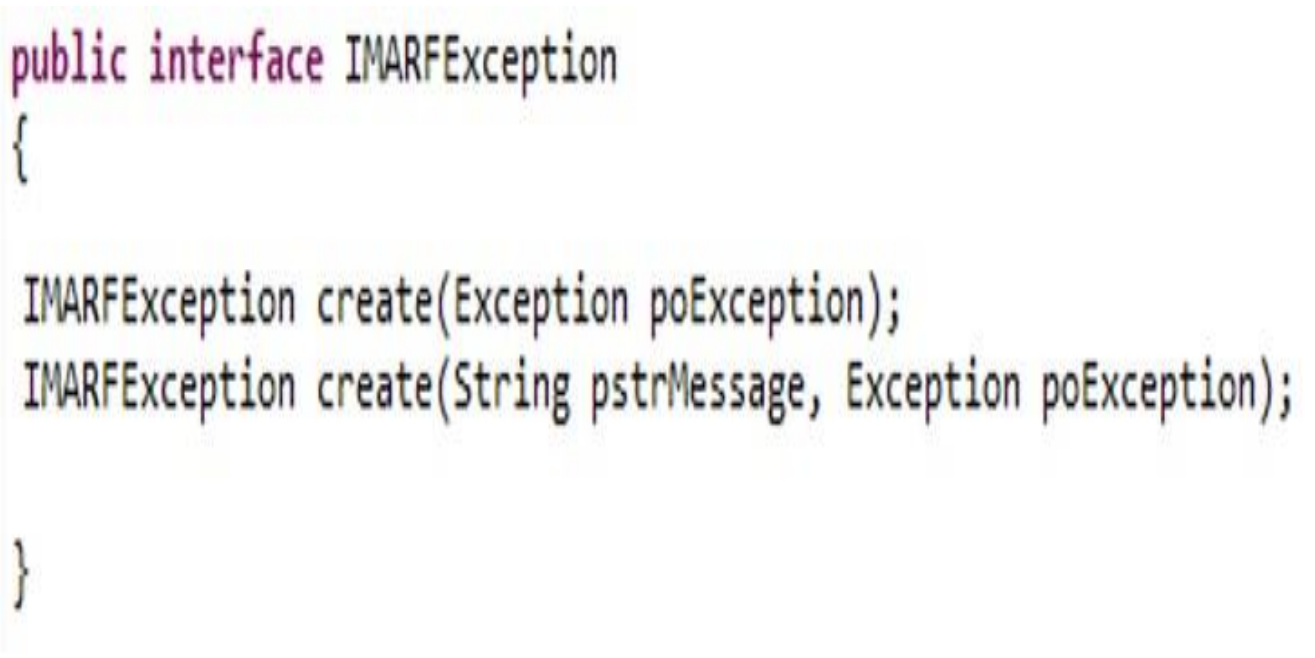

```
public interface IMARFException
{

IMARFException create(Exception poException);
IMARFException create(String pstrMessage, Exception poException);


}
```

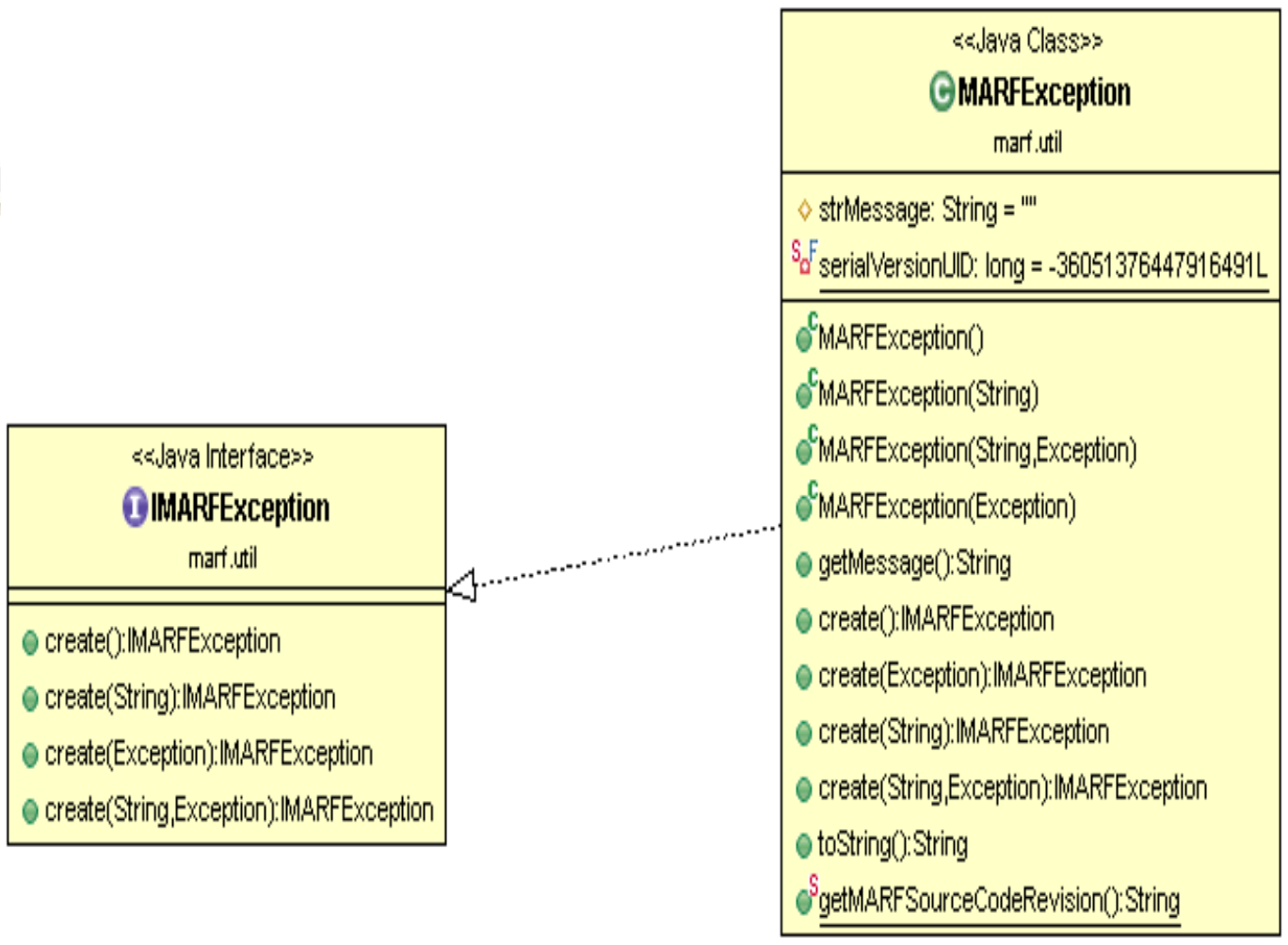

**Figure 46. Interacting classes for factory design pattern**

**Adapter**

Adapter pattern bridges between two incompatible interfaces. This pattern involves a single class which is responsible to join functionalities of independent or incompatible interfaces. This design pattern converts the original interface to another interface, through an intermediate adapter object [27] [26] [25].

Adapter consists of the following roles: Adaptee/Receiver, Adapter/ConcreteCommand, request()/execute()

This design pattern has been found in DMARF

Adaptee/Receiver: marf.Storage.Sample
Adapter/ConcreteCommand: marf.Preprocessing.Preprocessing
Request()/Execute():marf.Preprocessing.Preprocessing::removeNoise():boolean
Request()/Execute():marf.Preprocessing.Preprocessing::removeSilence():boolean
Request()/Execute():marf.Preprocessing.Preprocessing::normalize(int):boolean
Request()/Execute():marf.Preprocessing.Preprocessing::removeNoise(int,int):boolean
Request()/Execute():marf.Preprocessing.Preprocessing::removeClone():java.lang.Object

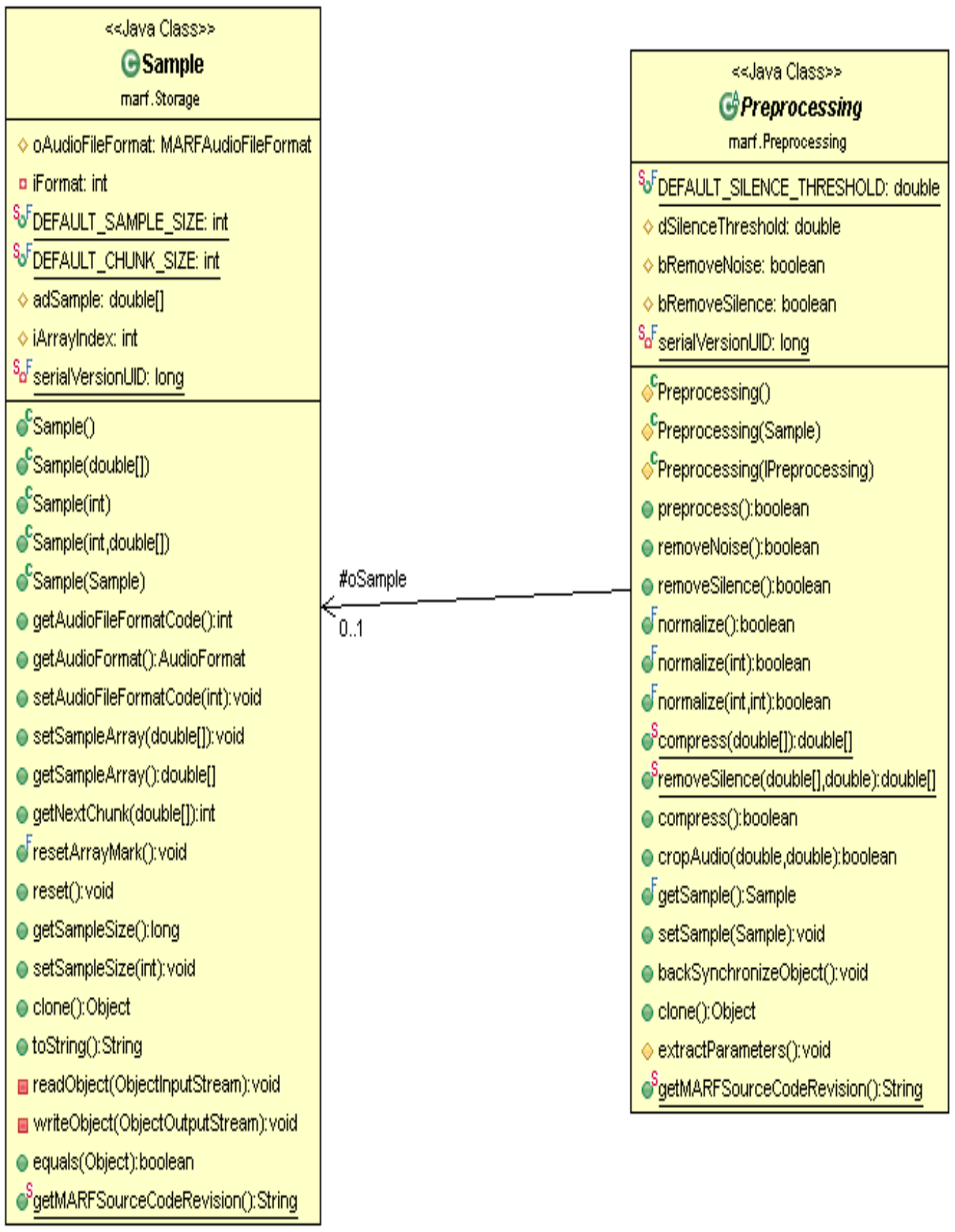

**Figure 47. UML Class Diagram for Adapter Pattern**

**Class Sample**

```
public Sample()
	{
		try
		{
	setAudioFileFormatCode(MARFAudioFileFormat.WAV);
		}
		catch(InvalidSampleFormatException e)
		{
			throw new RuntimeException(e);
		}
	}
	public synchronized int getAudioFileFormatCode()
	{
		return this.iFormat;
	}


}
```

Class Preprocessing:

```
public boolean removeNoise()
	throws PreprocessingException
	{
		LowPassFilter oNoiseRemover = new LowPassFilter(this.oSample);
		oNoiseRemover.bRemoveNoise = false;
		oNoiseRemover.bRemoveSilence = false;

		boolean bChanges = oNoiseRemover.preprocess();


	this.oSample.setSampleArray(oNoiseRemover.getSample().getSampleArray());
		oNoiseRemover = null;

		return bChanges;
	}
public boolean removeSilence()
	throws PreprocessingException
	{

	this.oSample.setSampleArray(removeSilence(this.oSample.getSampleArray(), this.dSilenceThreshold));
		return true;
	}

public final boolean normalize(int piIndexFrom)
	throws PreprocessingException
	{
```

```
if(this.oSample == null)
{
throw new PreprocessingException
(
"Preprocessing.normalize(from) - sample is not available (null)"
);
}

return normalize(piIndexFrom, this.oSample.getSampleArray().length - 1);
}
public final boolean normalize(int piIndexFrom, int piIndexTo)


public Object clone()
{
Preprocessing oCopy = (Preprocessing)super.clone();

oCopy.oSample =
this.oSample == null ?
null :
(Sample)this.oSample.clone();

return oCopy;
}
```

*b) GIPSY*

**Observer**

Defines a one-to-many dependency between objects so that when one object changes state, all its dependents are notified and updated automatically [28].

The main concept of observer design pattern is to differentiate between the independent functionality and dependent functionality. It is always preferable to model the dependent functionality with an observer hierarchy and a subject abstraction for independent functionality. All the observers should first register themselves with an observable object before they actually start their processing and then the observer will be responsible for extracting information they need from subject [29].

The respective participant's for the design are:

- **Observable** - interface or abstract class defining the operations for attaching and eliminating observers to the client. In can also be considered as **Subject**.
- **Observer** - interface or abstract class defining the operations to be used to notify this object.

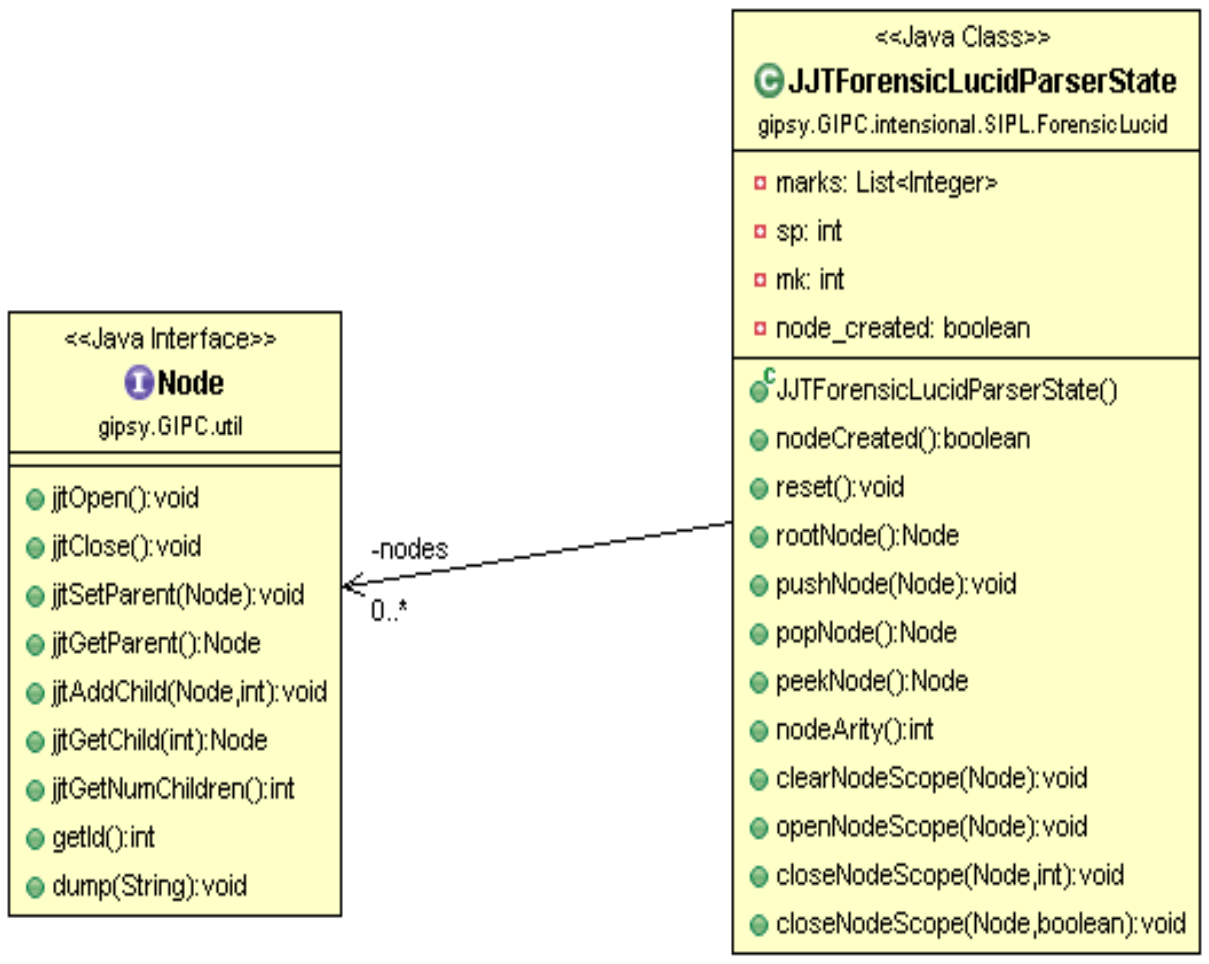

**Figure 48. Interacting Classes for Observer Pattern**

The following figure depicts the observer and as well as the subject, and necessary notifications are made automatically.

**Observer:**gipsy.GIPC.util.Node
**Subject:**gipsy.GIPC.intensional.SIPL.ForensicLucid.JJTForensicLucidParserState
**Notify():**gipsy.GIPC.intensional.SIPL.ForensicLucid.JJTForensicLucidParserState::closeNodeScope(gipsy,GIPC.util.Node, int):void
**Notify():**gipsy.GIPC.intensional.SIPL.ForensicLucid.JJTForensicLucidParserState::closeNodeScope(gipsy,GIPC.util.Node, boolean):void
This pattern mainly initiates it process by registering all the services that are capable making progress into the observable which is presented in JJTForensicParserState.java and notifies its corresponding observers when they are need for modifications which can be done in Node.java

This pattern is used in the GIPSY system to reduce the amount of complexity in providing services to its fellow components. Here all the observers are placed in **Node.Java** and all the changes are done automatically upon the notifications from **JJTForensicLucidParsesState.java**

**Corresponding observer code**

**Node.java**

```
packagegipsy.GIPC.util;
importjava.io.Serializable;
publicinterface Node
extendsSerializable
{
publicvoidjjtOpen();
publicvoidjjtClose();
publicvoidjjtSetParent(Node n);
```

```
public Node jjtGetParent();
publicvoidjjtAddChild(Node n, inti);
public Node jjtGetChild(inti);
publicintjjtGetNumChildren();
publicintgetId();
publicvoid dump(String pstrPrefix);
}
```

**Subject Code**

Normally Subject contains a method (setObserver) which allows the observer passed to it through method parameters

**JJTFOrensicLucidParserState.java**

Here the subject class contains a method jjtSetParent that gets the jjtSetParent observer in (Node.java) passed to it method parameters

```
void closeNodeScope(Node n, boolean condition) {
   if (condition) {
     int a = nodeArity();
     mk = ((Integer)marks.pop()).intValue();
     while (a-- > 0) {
             Node c = popNode();
             c.jjtSetParent(n);
             n.jjtAddChild(c, a);
     }
     n.jjtClose();
     pushNode(n);
     node_created = true;
   } else {
     mk = ((Integer)marks.pop()).intValue();
     node_created = false;
   }
 }
```

**Notifications**

The notifications are sent to the observer for corresponding updates in the class through the following method implementations

**publicvoid**closeNodeScope(Node n, **int**num)

and **publicvoid**closeNodeScope(Node n, **boolean** condition)

**Decorator**

The Decorator is known as a structural pattern, as it's used to form large object structures across many disparate objects [31]. Attach additional responsibilities to an object dynamically. Decorators provide a flexible alternative to sub classing for extending functionality. Client-specified embellishment of a core object by recursively wrapping it.Wrapping a gift, putting it in a box, and wrapping the box.[30]

Decorator pattern consists of five roles they are:Component, concreateComponent, Decorator and concreateDecorator.

Decorator: Maintains a reference to a Component object and defines an interface that conforms to Component's interface [33].

**Component** - Interface for objects that can have responsibilities added to them dynamically [33].

**Concrete Decorators** - Concrete Decorators extend the functionality of the component by adding state or adding behavior [33].

**ConcreteComponent** - Defines an object to which additional responsibilities can be added [33].

In the following diagram MARFCATDWTapp acts as a decorator class , IDemandWorker acts as a component class, DWTapp class inherits IdemandWorker class and also invokes some of its methods. MARFPCATDWT acts as a concreateDecorator class which can be used to extend the functionality of the decorator class dynamically.

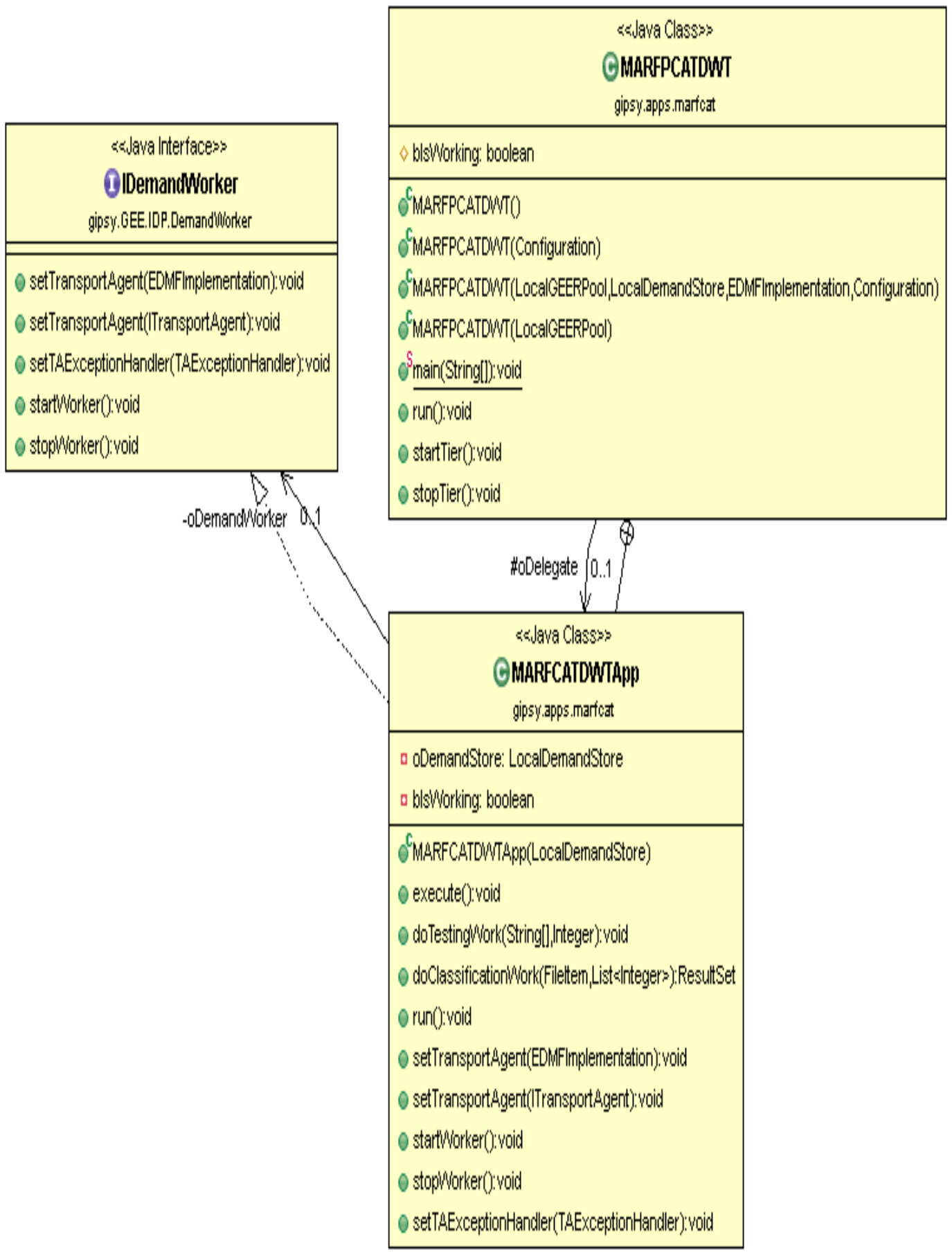

**Figure 49. UML Class Diagram for Decorator Pattern**

```
public void setTransportAgent(EDMFImplementation poDMFImp)
{
this.oDemandWorker.setTransportAgent(poDMFImp);
}

@Override
public void setTransportAgent(ITransportAgent poTA)
```

```
{
this.oDemandWorker.setTransportAgent(poTA);
}

@Override
public void startWorker()
{
this.oDemandWorker.startWorker();
this.bIsWorking = true;
}

@Override
public void stopWorker()
{
this.oDemandWorker.stopWorker();
this.bIsWorking = false;
}


public void setTAExceptionHandler(TAExceptionHandler poTAExceptionHandler)
{

this.oDemandWorker.setTAExceptionHandler(poTAExceptionHandler);
}
```

**Prototype**

Prototype design pattern refers to creating duplicate object while keeping performance in mind. This type of design pattern comes under creational pattern as this pattern provides one of the best way to create an object [36]. Specify the kinds of objects to create using a prototypical instance, and create new objects by copying this prototype .Co-opt one instance of a class for use as a breeder of all future instances. The new operator considered harmful [34].

The participants of the prototype design pattern are:
**Client** - Used to create a new object by asking a prototype to clone itself [37].
**Prototype** - declares an interface for cloning itself [37].
**Concrete Prototype** - Used to implement the operation of cloning itself [37].

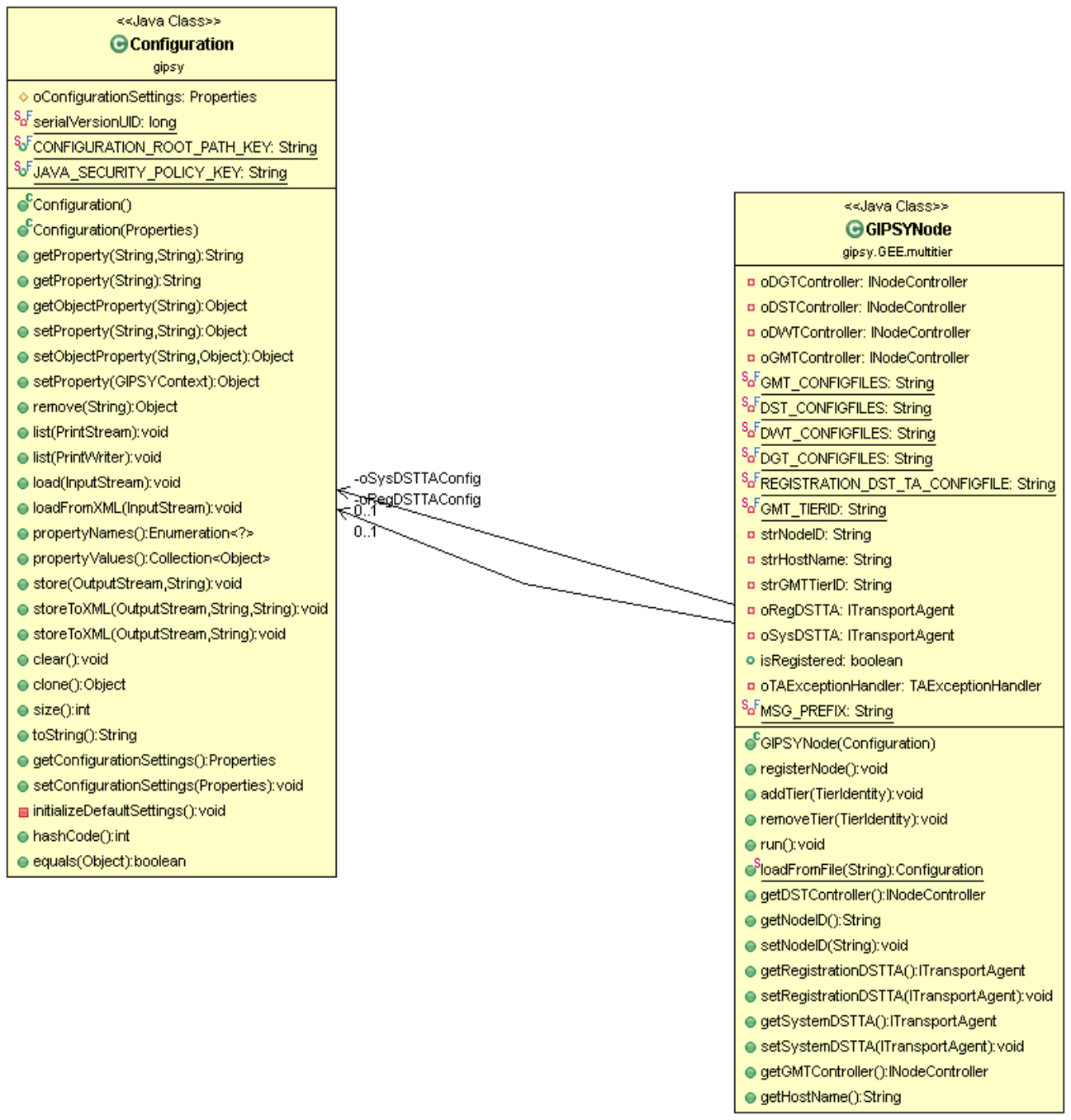

**Figure 50. Classes Implementing Prototype Pattern**

The following above figure explains the client, prototype and necessary operations of communication in the design pattern From the class diagram below are the classes which are involved in the pattern

**Client :** GIYPSYNode
**Prototype :** Configuration
**Operation():** GIYPSY method call - run()

**Corresponding Code Snippet :**
**GIYPSYNode**

```
public void run()
{
   Configuration oTierConfig = oRequest.getTierConfig();
   oTierConfig = (Configuration) oRequest.getTierConfig().clone();
}
```

**Configuration**

```
public synchronized Object clone()
        {
                Configuration oNewConfig = new Configuration();

        oNewConfig.setConfigurationSettings((Properties) this.oConfigurationSettings.clone());
                return oNewConfig;
        }
```

**Proxy**

Provide a surrogate or placeholder for another object to control access to it. Use an extra level of indirection to support distributed, controlled, or intelligent access. Add a wrapper

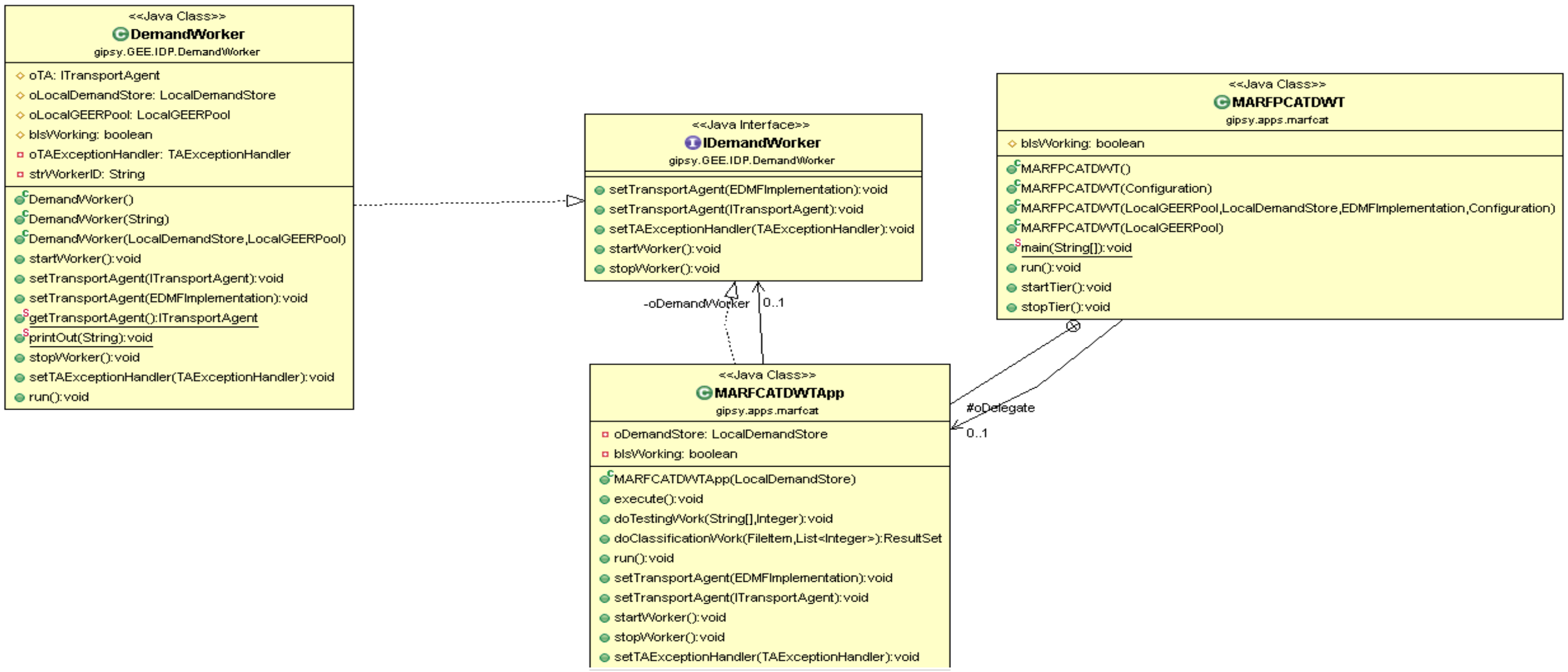

**Figure 53. UML Class Diagram for Proxy Design Pattern**

and delegation to protect the real component from undue complexity [38] (figure 53).

Observations from the class Diagram
Real Subject: Demand Worker
Subject: IDemandWorker
Proxy: MARFCATDWTApp

- Real Subject (Demand Worker) is concrete
- It can be observed that real subject (Demand Worker) inherits from the subject (IDemandWorker).
- Proxy (MARFCATDWTApp) inherits from the subject (IDemandWorker).
- Proxy (MARFCATDWTApp) implements subject (IDemandWorker) methods.

This pattern is needed to create a virtual environment for communication and also to support distributed control.

Observed Code

```
package gipsy.GEE.IDP.DemandWorker;
import gipsy.GEE.IDP.ITransportAgent;
import gipsy.GEE.multitier.EDMFImplementation;
import gipsy.GEE.multitier.TAExceptionHandler;
public interface IDemandWorker
extends Runnable
{
void setTransportAgent(EDMFImplementation poDMFImp);
void setTransportAgent(ITransportAgent poTA);
void setTAExceptionHandler(TAExceptionHandler
poTAExceptionHandler);
void startWorker();
void stopWorker();
}
```

*Tool support*

For all the patterns we have used ObjectAid UML Explorer for generating the classes and relationships between them. "ObjectAid UML is an agile and light weight code visualization tool for Eclipse IDE"[1]. This tool is helpful in generating the java source code files into a class diagram which also updates automatically to the code changes. It basically acquires the UML notations to visualize classes. The tool is simple and effective to use, we should just drag our classes of interest into the explorer to view all the properties of a typical class like methods, attributes, and method parameters and its association with other classes.
One more advantage of this tool is, whenever we refactor a particular class or modules the updates are automatically reflected on UML class diagram too. For example if we extract a method, the diagram simply reflects without going out of sync [].

## IV. Implementation

*Refactoring Changesets and Diffs*

*a) DMARF*

*Change 1/6: Extract Method in LPC.java*

**Name:** Extract Method

**Description**: If we have a code fragment that can be grouped together turn the fragment into method whose name explains the purpose of the method [41]

**Motivation:** The prime purpose of this refactoring is to eradicate long method code smell. The longer a procedure is, the more difficult it is to understand, best refactoring strategy is to shorten a method is Extract Method. If a method has lots of parameters and temporary variables, these elements get in the way of extracting methods
**Code Smell:** Long method.

**Changes Made:** Methods created

Diff Files

iWindowsNum()
oSpectrogram()
adWindowed()

```
Index:
src/marf/FeatureExtraction/LPC/LPC.java
==========================================
=========================
RCS file:
/cvsroot/marf/marf/src/marf/FeatureExtrac
tion/LPC/LPC.java,v
retrieving revision 1.41
diff -u -r1.41 LPC.java
---
src/marf/FeatureExtraction/LPC/LPC.java
+++
src/marf/FeatureExtraction/LPC/LPC.java
@@ -7,20 +7,21 @@
 import
marf.FeatureExtraction.FeatureExtractionE
xception;
 import
marf.Preprocessing.IPreprocessing;
 import marf.Storage.ModuleParams;
+import marf.Storage.StorageException;
 import marf.gui.Spectrogram;
 import marf.math.Algorithms;
 import marf.util.Debug;


 /**
- * <p>Class LPC implements Linear
Predictive Coding.</p>
+ * <p>Class LPC impleaments Linear
Predictive Coding.</p>
  *
- * $Id: LPC.java,v 1.41 2006/08/04
03:31:05 mokhov Exp $
+ * $Id: LPC.java,v 1.7 2014/08/25
00:11:20 als_ah Exp $
  *
  * @author Ian Clement
  * @author Serguei Mokhov
  *
- * @version $Revision: 1.41 $
+ * @version $Revision: 1.7 $
  * @since 0.0.1
  */
 public class LPC
@@ -106,9 +107,25 @@
        public final boolean
extractFeatures(double[] padSampleData)
        throws FeatureExtractionException
        {
+
+              /*
+               * Creating new Method to
separate Window length and sample data
+               * value initialization.
+               */
+              try
+              {
+
        oSpectrogram(padSampleData);
+
+              } catch (StorageException
s)
+              {
+
        s.printStackTrace();
+              }
+
 4 Aug 2006 03:31:05 -0000      1.41
              try
              {
 25 Aug 2014 03:35:51 -0000
        Debug.debug("LPC.extractFeatures()
has begun...");
+
+                     int iWindowsNum =
iWindowsNum(padSampleData);

                      double[] adSample =
padSampleData;

@@ -116,51 +133,29 @@
                      Debug.debug("poles:
" + this.iPoles);
                      Debug.debug("window
length: " + this.iWindowLen);

-                     Spectrogram
oSpectrogram = null;
-
-                     // For the case
when we want intermediate spectrogram
-
        if(MARF.getDumpSpectrogram() ==
true)
-                     {
-
        oSpectrogram = new
Spectrogram("lpc");
-                     }
-
                      this.adFeatures =
new double[this.iPoles];
```

```
                        double[] adWindowed
= new double[this.iWindowLen];
                        double[]
adLPCCoeffs = new double[this.iPoles];
                        double[] adLPCError
= new double[this.iPoles];

-                       // Number of
windows
-                       int iWindowsNum =
1;

                        int iHalfWindow =
this.iWindowLen / 2;

                        for(int iCount =
iHalfWindow; (iCount + iHalfWindow) <=
adSample.length; iCount += iHalfWindow)
                        {
-                               // Window
the input.
-                               for(int j =
0; j < this.iWindowLen; j++)
-                               {
-
        adWindowed[j] = adSample[iCount -
iHalfWindow + j];
-
        //windowed[j] = adSample[count -
iHalfWindow + j] * hamming(j,
this.windowLen);
-
        //Debug.debug("window: " +
windowed[j]);
-                               }
-
+                               adWindowed
= adWindowed(adSample, adWindowed,
iHalfWindow,iCount);
+

        Algorithms.Hamming.hamming(adWindo
wed);

        Algorithms.LPC.doLPC(adWindowed,
adLPCCoeffs, adLPCError, this.iPoles);

-
        if(MARF.getDumpSpectrogram() ==
true)
-                               {
-
        oSpectrogram.addLPC(adLPCCoeffs,
this.iPoles, iHalfWindow);
-                               }
-
                                // Collect
features
                                for(int j =
0; j < this.iPoles; j++)
                                {

        this.adFeatures[j] +=
adLPCCoeffs[j];

        //Debug.debug("lpc_coeffs[" + j +
"]"  + lpc_coeffs[j]);
                                }
-
-

        iWindowsNum++;
+
                        }

                        // Smoothing
@@ -174,12 +169,6 @@


        Debug.debug("LPC.extractFeatures()
- number of windows = " + iWindowsNum);

-                       // For the case
when we want intermediate spectrogram
-
        if(MARF.getDumpSpectrogram() ==
true)
-                       {
-
        oSpectrogram.dump();
-                       }
-

        Debug.debug("LPC.extractFeatures()
has finished.");

                        return
(this.adFeatures.length > 0);
@@ -190,6 +179,114 @@
                        throw new
FeatureExtractionException(e);
                }
        }
+
+       /*
+        * Method created for window
length and sample data initialization
+        */
+       private void oSpectrogram(double[]
padSampleData)throws StorageException
+       {
+               try
+               {
+                       double[] adSample =
padSampleData;
+
```

```
+                       double[]
adLPCCoeffs = new double[this.iPoles];
+
+                       int iHalfWindow =
this.iWindowLen / 2;
+
+                       Spectrogram
oSpectrogram = oSpectrogram();
+
+
        oSpectrogram(adSample,
oSpectrogram, adLPCCoeffs, iHalfWindow);
+
+               } catch (Exception e)
+               {
+
        e.printStackTrace(System.err);
+
+                       try {
+
+                               throw new
FeatureExtractionException(e);
+
+                       } catch
(FeatureExtractionException f)
+                       {
+
        f.printStackTrace();
+                       }
+               }
+       }
+
+       /*
+        * Adding separate method for
Dumping Spectrogram values
+        */
+       private void oSpectrogram(double[]
adSample, Spectrogram
oSpectrogram,double[] adLPCCoeffs, int
iHalfWindow)
+                       throws
StorageException
+       {
+               for (int iCount =
iHalfWindow; (iCount + iHalfWindow) <=
adSample.length; iCount += iHalfWindow)
+               {
+                       if
(MARF.getDumpSpectrogram() == true)
+                       {
+
        oSpectrogram.addLPC(adLPCCoeffs,
this.iPoles, iHalfWindow);
+                       }
+               }
+               if
(MARF.getDumpSpectrogram() == true) {
+
        oSpectrogram.dump();
+               }
+       }
+
+       /*
+        * Adding separate method for
Intermediate Spectrogram value creation
+        */
+       private Spectrogram oSpectrogram()
+       {
+               Spectrogram oSpectrogram =
null;
+
+               if
(MARF.getDumpSpectrogram() == true)
+               {
+                       oSpectrogram = new
Spectrogram("lpc");
+               }
+               return oSpectrogram;
+       }
+
+/*
+ * Creating Separate Method call for
retrieving Windows value from the
+ * sample data.
+ */
+       private int iWindowsNum(double[]
padSampleData)
+       {
+               double[] adSample =
padSampleData;
+
+               int iWindowsNum = 1;
+
+               int iHalfWindow =
this.iWindowLen / 2;
+
+               iWindowsNum =
iWindowsNum(adSample, iWindowsNum,
iHalfWindow);
+
+               return iWindowsNum;
+       }
+
+       /*
+        * Creating Separate Inner Method
call for retrieving Windows value from
the
+        * sample data.
+        * This method is called from the
iWindowsNum to get the value.
+        */
+
+       private int iWindowsNum(double[]
adSample, int iWindowsNum, int
iHalfWindow)
```

```
+       {
+               for (int iCount =
iHalfWindow; (iCount + iHalfWindow) <=
adSample.length; iCount += iHalfWindow)
+               {
+                       iWindowsNum++;
+               }
+               return iWindowsNum;
+       }
+
+       /*
+        * Adding new method for fetching
the windows values from the sample data
+        */
+       private double[]
adWindowed(double[] adSample, double[]
adWindowed,int iHalfWindow, int iCount)
+       {
+               for (int j = 0; j <
this.iWindowLen; j++)
+               {
+                       adWindowed[j] =
adSample[iCount - iHalfWindow + j];
+               }
+               return adWindowed;
+       }
```

*Change 2/6: Extract Class in OptionFileLoader.java*

**Name:** Extract Class
**Description**: In order to minimize the additional responsibilities to the classes we used extract class as the strategy to create new classes and reduce the complexity of Extracted Class
**Motivation:** When there is class that knows too much and does too much, then this class is tightly coupled to many other classes. If there is a big problem it cannot be solved into separate solutions which doesn't compile with a basic idea of object oriented programming. Normally such kind of classes are more difficulty to be maintained, rather than having evenly divided programming design,
**Code Smell:** God Class

**Changes Made:**
Extracted and Class created:
OptionFileLoaderExt

Diff files:

```
RCS file:
/cvsroot/marf/marf/src/marf/util/OptionFi
leLoader.java,v
retrieving revision 1.1.4.2
diff -u -r1.1.4.2 OptionFileLoader.java
--- src/marf/util/OptionFileLoader.java
17 Nov 2009 05:09:58 -0000    1.1.4.2
+++ src/marf/util/OptionFileLoader.java
25 Aug 2014 03:35:51 -0000
@@ -7,6 +7,8 @@
 import java.util.Vector;
+
+
 /**
  * <p>Loads a configuration file.</p>
  *
@@ -23,17 +25,22 @@
  * WARNING: this is not a great option
to keep passwords in memory, because
  * the data is kept as strings.</p>
  *
- * $Id: OptionFileLoader.java,v 1.1.4.2
2009/11/17 05:09:58 mokhov Exp $
+ * $Id: OptionFileLoader.java,v 1.3
2014/08/25 00:38:49 h_lao Exp $
  *
  * @author Marc-Andre Laverdiere
  * @author Serguei Mokhov
  *
  * @since 0.3.0.6
- * @version $Revision: 1.1.4.2 $
+ * @version $Revision: 1.3 $
  */
 public class OptionFileLoader
 implements IOptionProvider
 {
+
+       /**
+        * Creating Instance of the new
class for extraction -
OptionFileLoaderExt Instance
+        */
+       private OptionFileLoaderExt
optionFileLoaderExtender = new
OptionFileLoaderExt();
        /**
         * Singleton Instance.
         */
@@ -45,11 +52,6 @@
        protected Hashtable
oHashOptionValuePairTracker;

        /**
-        * Indicates that the config was
initialized.
-        */
-       protected boolean bIsInitialized;
-
-       /**
         * Default config file name.
         */
        protected static final String
DEFAULT_CONFIG_FILE_NAME = ".config";
@@ -70,7 +72,7 @@
        protected OptionFileLoader()
        {
```

```
        this.oHashOptionValuePairTracker =
new Hashtable();
-               this.bIsInitialized =
false;
+
        optionFileLoaderExtender.setBIsIni
tialized(false);
        }

        /**
@@ -94,36 +96,7 @@
        public void loadConfiguration()
        throws IOException
        {
-
        loadConfiguration(DEFAULT_CONFIG_F
ILE_NAME);
-       }
-
-       /**
-        * Loads specified configuration
file.
-        * @param pstrFileName name of the
configuration file
-        * @throws IOException on error
reading file
-        */
-       public void
loadConfiguration(final String
pstrFileName)
-       throws IOException
-       {
-               BufferedReader oReader =
new BufferedReader(new
FileReader(pstrFileName));
-
-               while(oReader.ready()){
-                       String strLine =
oReader.readLine();
-                       // For each line,
skipping end of file, empty lines
-                       if (strLine != null
&& !strLine.equals("")){
-                               strLine =
strLine.trim(); //remove whitespace
-                               String
strUncommented =
this.removeComment(strLine); //remove
comments
-
-                               // if not a
comment, extract the key and the
associated value
-                               if
(strUncommented != null){
-
        String strKey =
this.getKey(strUncommented);
-
        String strValue =
this.getValue(strUncommented);
-//
        System.out.println(strKey + "->" +
strValue);
-
        this.oHashOptionValuePairTracker.p
ut(strKey,strValue);
-                               }
-                       }
-               }
-               this.bIsInitialized =
true;
+
        optionFileLoaderExtender.loadConfi
guration(DEFAULT_CONFIG_FILE_NAME,
oHashOptionValuePairTracker, this);
        }

        /**
@@ -198,7 +171,7 @@
         */
        public boolean isInitialized()
        {
-               return
this.bIsInitialized;
+               return
this.optionFileLoaderExtender.getBIsIniti
alized();
        }


@@ -231,7 +204,7 @@
        }

        /**
-        * @see
marf.util.IOptionProvider#size()
+        * @see
tools.IOptionProvider#size()
         */
        public int size()
        {
Index:
src/marf/util/OptionFileLoaderExt.java
==========================================
=========================
RCS file:
src/marf/util/OptionFileLoaderExt.java
diff -N
src/marf/util/OptionFileLoaderExt.java
--- /dev/null   1 Jan 1970 00:00:00 -0000
```

```
+++
src/marf/util/OptionFileLoaderExt.java1
Jan 1970 00:00:00 -0000
@@ -0,0 +1,51 @@
+package marf.util;
+
+import java.io.BufferedReader;
+import java.io.FileReader;
+import java.io.IOException;
+import java.util.Hashtable;
+
+public class OptionFileLoaderExt {
+
+       private boolean bIsInitialized;
+
+       public boolean getBIsInitialized()
+       {
+               return bIsInitialized;
+       }
+
+       public void
setBIsInitialized(boolean bIsInitialized)
+       {
+               this.bIsInitialized =
bIsInitialized;
+       }
+
+       /**
+        * Moved the method from
OptionFileLoader to prevent Duplicate
initialization code and
+        * Separation of loading
configuration Logic.
+        * Loads specified configuration
file.
+        */
+
+       public void
loadConfiguration(final String
pstrFileName,Hashtable
oHashOptionValuePairTracker,
        OptionFileLoader optionFileLoader)
throws IOException
+       {
+               BufferedReader oReader =
new BufferedReader(new
FileReader(pstrFileName));
+
+               while (oReader.ready())
+               {
+                       String strLine =
oReader.readLine();
+                       if (strLine != null
&& !strLine.equals(""))
+                       {
+                               strLine =
strLine.trim();
+                               String
strUncommented =
optionFileLoader.removeComment(strLine);
+
+                               if
(strUncommented != null)
+                               {
+
        String strKey =
optionFileLoader.getKey(strUncommented);
+
        String strValue =
optionFileLoader.getValue(strUncommented)
;
+
        oHashOptionValuePairTracker.put(st
rKey, strValue);
+                               }
+                       }
+               }
+
+               this.bIsInitialized =
true;
+       }
+}
```

*Change 3/6: Move methods into Layer.java*

**Name:** Move method
**Description:** Methods that make extensive use of another class may belong in another class. Consider moving this method to the class it is so envious of [42].
**Motivation:** The primary moto is to move a method, or part of a method that clearly wants to be elsewhere. In different words, when a method references or calls too many methods or data existing in other class, we use "move method" to move it to the desired class.
**Code Smell:** Feature Envy

**Changes Made:**
Methods moved to classes
Layer.java

Diff Files

```
Index:
src/marf/Classification/NeuralNetwork/Lay
er.java
=========================================
=========================
RCS file:
/cvsroot/marf/marf/src/marf/Classificatio
n/NeuralNetwork/Layer.java,v
retrieving revision 1.9.4.2
diff -u -r1.9.4.2 Layer.java
---
src/marf/Classification/NeuralNetwork/Lay
```

```
er.java 17 Nov 2009 05:09:57 -0000
        1.9.4.2
+++
src/marf/Classification/NeuralNetwork/Lay
er.java 25 Aug 2014 03:35:51 -0000
@@ -3,7 +3,9 @@
 import java.io.Serializable;
 import java.util.ArrayList;

+import
marf.Classification.ClassificationExcepti
on;
 import marf.util.BaseThread;
+import marf.util.Debug;


 /**
@@ -15,11 +17,11 @@
  * class itself is properly
synchronized.
  * </p>
  *
- * $Id: Layer.java,v 1.9.4.2 2009/11/17
05:09:57 mokhov Exp $
+ * $Id: Layer.java,v 1.3 2014/08/25
01:03:41 s_challa Exp $
  *
  * @author Serguei Mokhov
  * @since 0.3.0.2
- * @version $Revision: 1.9.4.2 $
+ * @version $Revision: 1.3 $
  */
 public class Layer
 extends BaseThread
@@ -168,7 +170,57 @@
         */
        public static String
getMARFSourceCodeRevision()
        {
-               return "$Revision: 1.9.4.2
$";
+               return "$Revision: 1.3 $";
+       }
+
+       /**
+        * Adding Method getOutputResults
for Grasp Indirection
+        * Method is retrieved from
NeuralNetwork.java
+        */
+       public double[] getOutputResults()
+       {
+               double[] adRet = new
double[size()];
+
+               for (int i = 0; i <
size(); i++)
+               {
+                       adRet[i] =
get(i).dResult;
+               }
+               return adRet;
+       }
+
+       /**
+        * Adding Method setInputs for
Grasp Indirection
+        * Method is retrieved from
NeuralNetwork.java
+        */
+       public final void setInputs(final
double[] padInputs)throws
ClassificationException {
+
+               if (padInputs.length !=
size())
+               {
+                       throw new
ClassificationException("Input array size
not consistent with input layer.");
+               }
+
+               for (int i = 0; i <
padInputs.length; i++) {
+                       get(i).dResult =
padInputs[i];
+               }
+       }
+
+       /**
+        * Moving Method InterpretAsBinary
from Neural network
+        */
+       public final int
interpretAsBinary()
+       {
+               int iID = 0;
+
+               for (int i = 0; i <
size(); i++)
+               {
+                       iID *= 2;
+                       if (get(i).dResult
> 0.5) {
+                               iID += 1;
+                       }
+
        Debug.debug(get(i).dResult + ",");
+               }
+               Debug.debug("Interpreted
binary result (ID) = " + iID);
+               return iID;
        }
 }
```

```
Index:
src/marf/Classification/NeuralNetwork/Neu
ralNetwork.java
===========================================
===========================
RCS file:
/cvsroot/marf/marf/src/marf/Classificatio
n/NeuralNetwork/NeuralNetwork.java,v
retrieving revision 1.58.4.2
diff -u -r1.58.4.2 NeuralNetwork.java
---
src/marf/Classification/NeuralNetwork/Neu
ralNetwork.java 17 Nov 2009 05:09:57 -0000
+++
src/marf/Classification/NeuralNetwork/Neu
ralNetwork.java 25 Aug 2014 03:35:51 -0000
@@ -35,12 +35,12 @@
 /**
  * <p>Artificial Neural Network-based
Classifier.</p>
  *
- * $Id: NeuralNetwork.java,v 1.58.4.2
2009/11/17 05:09:57 mokhov Exp $
+ * $Id: NeuralNetwork.java,v 1.3
2014/08/25 01:06:13 s_challa Exp $
  *
  * @author Ian Clement
  * @author Serguei Mokhov
  *
- * @version $Revision: 1.58.4.2 $
+ * @version $Revision: 1.3 $
  * @since 0.0.1
  */
 public class NeuralNetwork
@@ -296,10 +296,10 @@

        ITrainingSample oTrainingSample =
(ITrainingSample)oTrainingSamples.get(iCo
unt);

                                        //
XXX: can be median and feature vectors
-
        setInputs(oTrainingSample.getMeanV
ector());
+
        oInputs.setInputs(oTrainingSample.
getMeanVector());

        runNNet();

-                                       int
iID = interpretAsBinary();
+                                       int
iID = oOutputs.interpretAsBinary();

 //
        dError +=
Math.abs(oCluster.getSubjectID() - iID);

        dError += dMinErr *
Math.abs(oTrainingSample.getSubjectID() -
iID);
@@ -390,8 +390,8 @@

                        // Make result...
                        // TODO: fix second
best kludge of adding the same thing
twice.
 1.58.4.2
-
        this.oResultSet.addResult(new
Result(interpretAsBinary()));
-
        this.oResultSet.addResult(new
Result(interpretAsBinary() + 1));
+
        this.oResultSet.addResult(new
Result(oOutputs.interpretAsBinary()));
+
        this.oResultSet.addResult(new
Result(oOutputs.interpretAsBinary() +
1));

                        return true;
                }
@@ -848,45 +848,6 @@

    //----------- Methods for Running
the NNet -----------------

-       /**
-        * Sets inputs.
-        * @param padInputs double array
of input features
-        * @throws ClassificationException
if the input array's length isn't
-        * equal to the size of the input
layer
-        */
-       public final void setInputs(final
double[] padInputs)
-       throws ClassificationException
-       {
-               if(padInputs.length !=
this.oInputs.size())
-               {
-                       throw new
ClassificationException
-                       (
-                               "Input
array size not consistent with input
layer."
-                       );
-               }
```

```
-
-               for(int i = 0; i <
padInputs.length; i++)
-               {
-
        this.oInputs.get(i).dResult =
padInputs[i];
-               }
-       }
-
-       /**
-        * Gets outputs of a neural
network run.
-        * @return array of doubles read
off the output layer's neurons
-        */
-       public double[] getOutputResults()
-       {
-               double[] adRet = new
double[this.oOutputs.size()];
-
-               for(int i = 0; i <
this.oOutputs.size(); i++)
-               {
-                       adRet[i] =
this.oOutputs.get(i).dResult;
-               }
-
-               return adRet;
-       }
-
     //----------- Methods for Outputting
the NNet -----------------

        /**
@@ -1090,7 +1051,7 @@
                */

               // Must setup the input
data...
-               setInputs(padInput);
+
        oInputs.setInputs(padInput);


        //if(piExpectedLength/*.length*/
!= this.oOutputs.size())
               //    throw new
ClassificationException("Expected array
size not consistent with output layer.");
@@ -1125,34 +1086,6 @@
               }
        }

-       /**
-        * Interprets net's binary output
as an ID for the final classification
result.
-        * @return ID, integer
-        */
-       private final int
interpretAsBinary()
-       {
-               int iID = 0;
-
-               for(int i = 0; i <
this.oOutputs.size(); i++)
-               {
-                       // Binary
displacement happens to not have any
-                       // effect in the
first iteration :-P
-                       iID *= 2;
-
-                       // Add 1 if the
resulting weight is more than 0.5
-
        if(this.oOutputs.get(i).dResult >
0.5)
-                       {
-                               iID += 1;
-                       }
-
-
        Debug.debug(this.oOutputs.get(i).d
Result + ",");
-               }
-
-               Debug.debug("Interpreted
binary result (ID) = " + iID);
-
-               return iID;
-       }
-
     /* From Storage Manager */

        /**
@@ -1387,7 +1320,7 @@
         */
        public static String
getMARFSourceCodeRevision()
        {
-               return "$Revision:
1.58.4.2 $";
+               return "$Revision: 1.3 $";
        }
 }
```

*b) GIPSY*

*Change 4/6: Adding new classes in GMTInfoKeeper*

**Name:** Type Checking

**Description:** Creating a number of concrete strategy classes equal to the number of conditional branches inside the type checking code.

**Motivation**: The code will be more difficult to understand and maintain when a complicated conditional statements present. In this regard, our goal should be eliminating type checking condition statements by applying strategy refactoring when applying object oriented paradigm. We should take advantage of polymorphism instead of using conditional statements to simulate dynamic dispatch and later binding.

Type Checking- Replace Strategy Pattern Introduced -----> GMTInfoKeeper

**Added classes:**

IdentityType

RemoveDst

RemoveDgtOrDwt

*Change 5/6: Extract Method in SemanticAnalyzer*

**Name:** Extract Method

**Description**: If we have a code fragment that can be grouped together turn the fragment into method whose name explains the purpose of the method [41]

**Motivation:** The prime purpose of this refactoring is to eradicate long method code smell. The longer a procedure is, the more difficult it is to understand, best refactoring strategy is to shorten a method is Extract Method. If a method has lots of parameters and temporary variables, these elements get in the way of extracting methods

**Code Smell:** Long method.
Long Method - Multiple method call introduced ----> SemanticAnalyzer

Methods:

getFPnum

getFDnum

getFtemp2

getP2value

getP1value

getFp1value

getFpvalue

*Change 6/6: Move methods to classes*

McCabe / logiscope - GIPSYGMTOperator ----> Information Expert

**Name:** Move method to classes
**Code Smell:** Information Expert

**Changes Made:**

**Methods moved**

Graph panel

gipsygmtcontroller

## CONCLUSION

We have presented the specifications and capabilities of DMARF and GIPSY and their uses. We also demonstrated their architectural view models, architectural styles and frameworks. We analyzed the code, and then applied some pattern recognition support tools. Many code smells have been identified and their corresponding refactoring methods for each case study. ObjectAid UML Explorer has been used as a reverse engineering tool to derive the actual architecture of the two case studies in order to be compared with conceptual architecture that the team members have shaped.
Since both software were designed to be able to merge and fuse each other, we demonstrated the conceptual fused architecture of both systems in order to show the ability of DMARF to use GIPSY's runtime for distributed computing instead of its own.
JDeodorant, SonarQube have been used by the team members to analyze the quality of the case studies with reference to its source code. Finally, implemented four refactoring for each case study with supporting test cases and corresponding results are interpreted**.**

# Appendix A

## SonarQube Results

DMARF

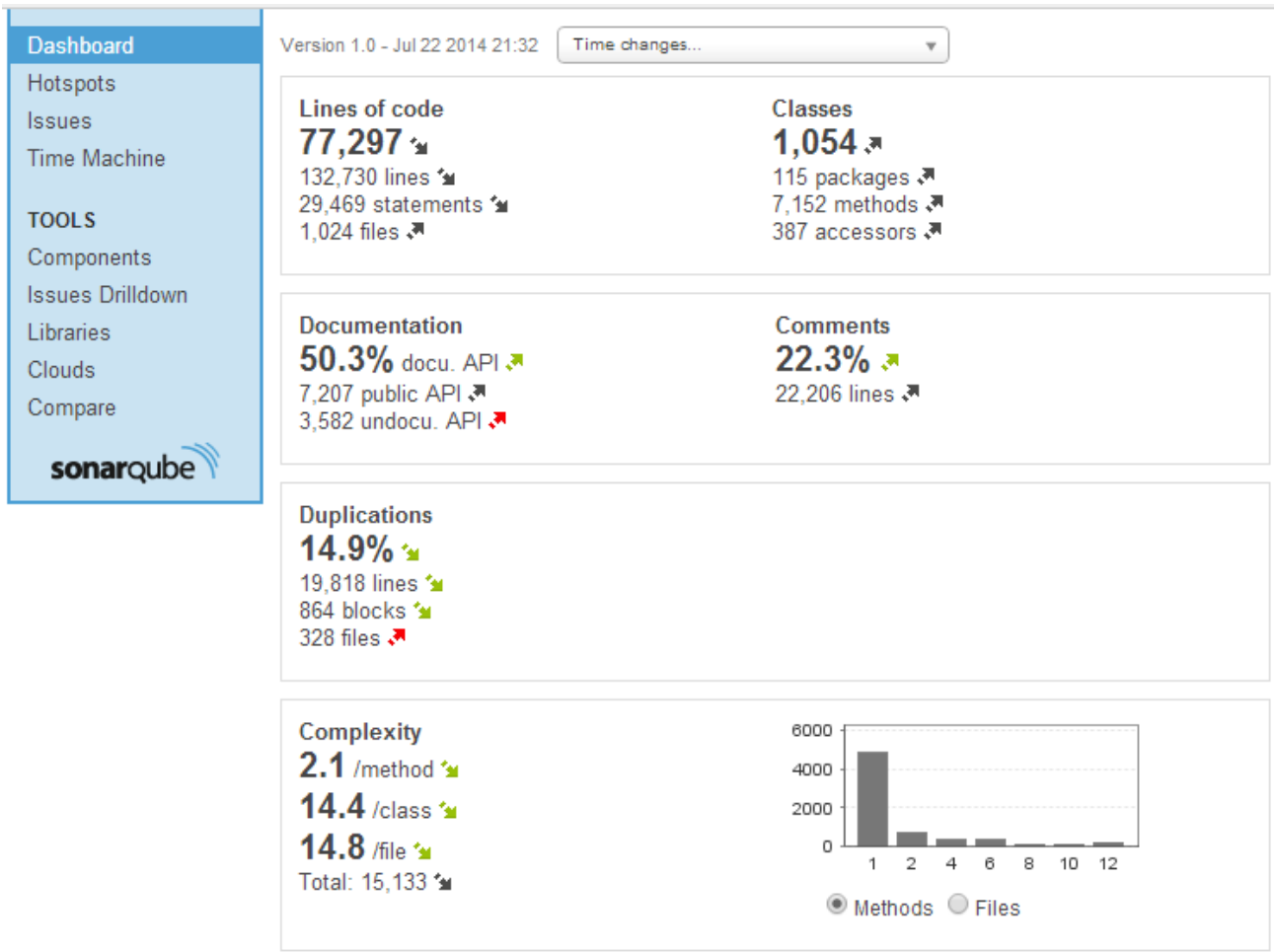

GIPSY

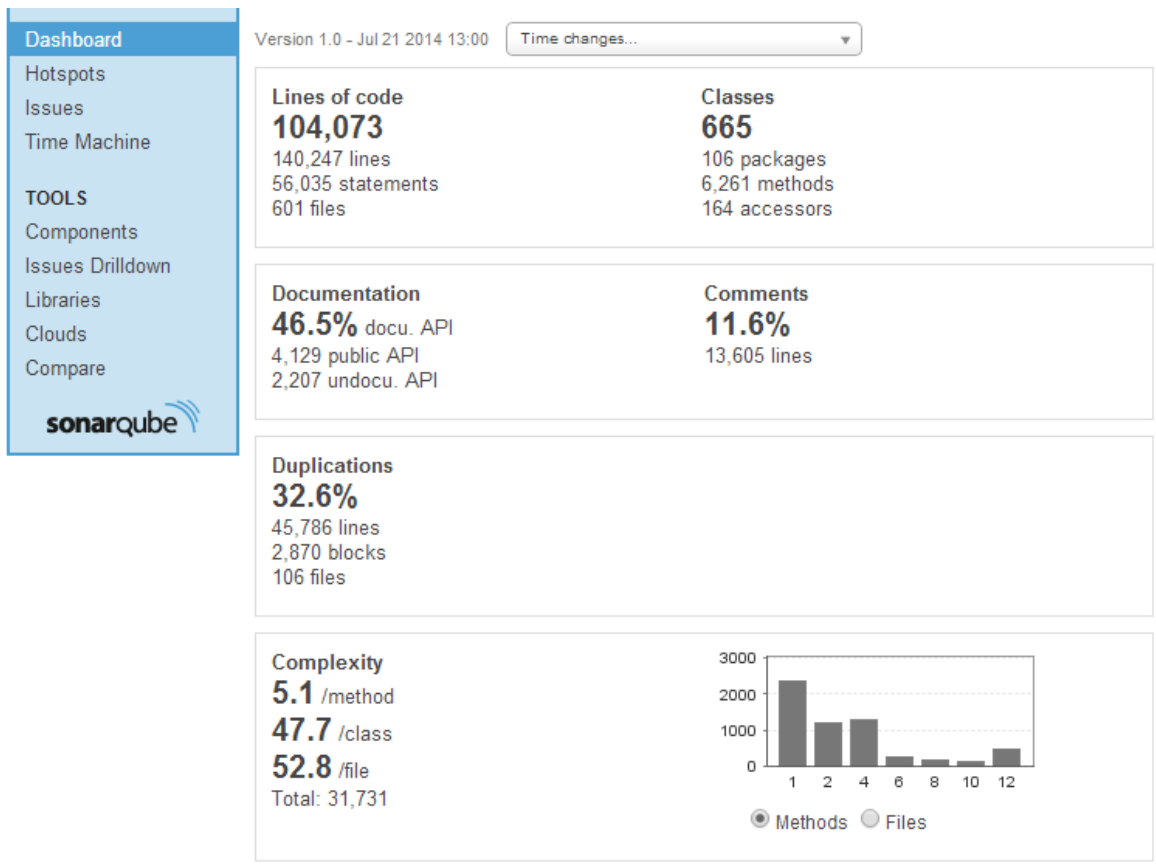

Appendix B

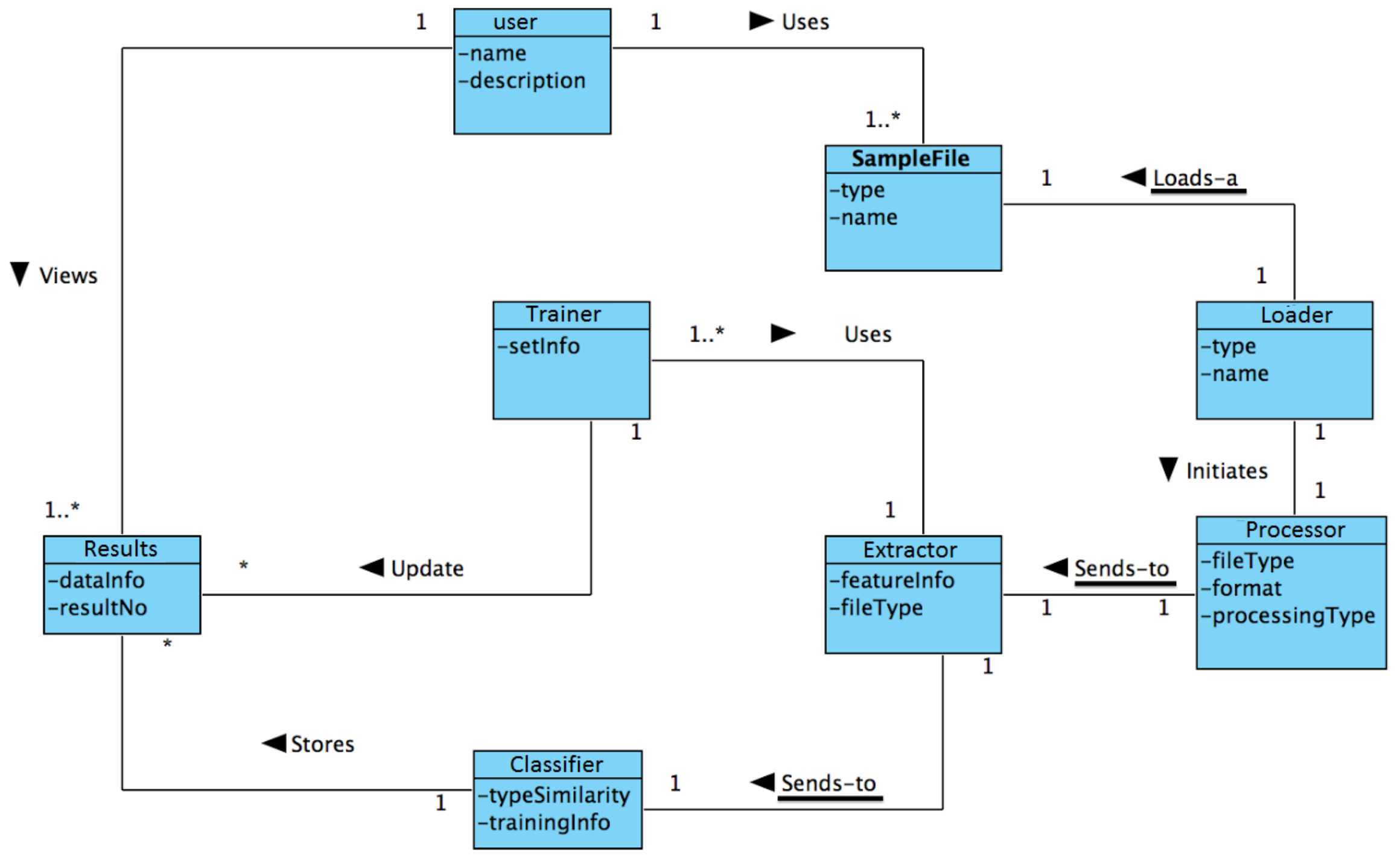

**DMARF Domain Diagram**

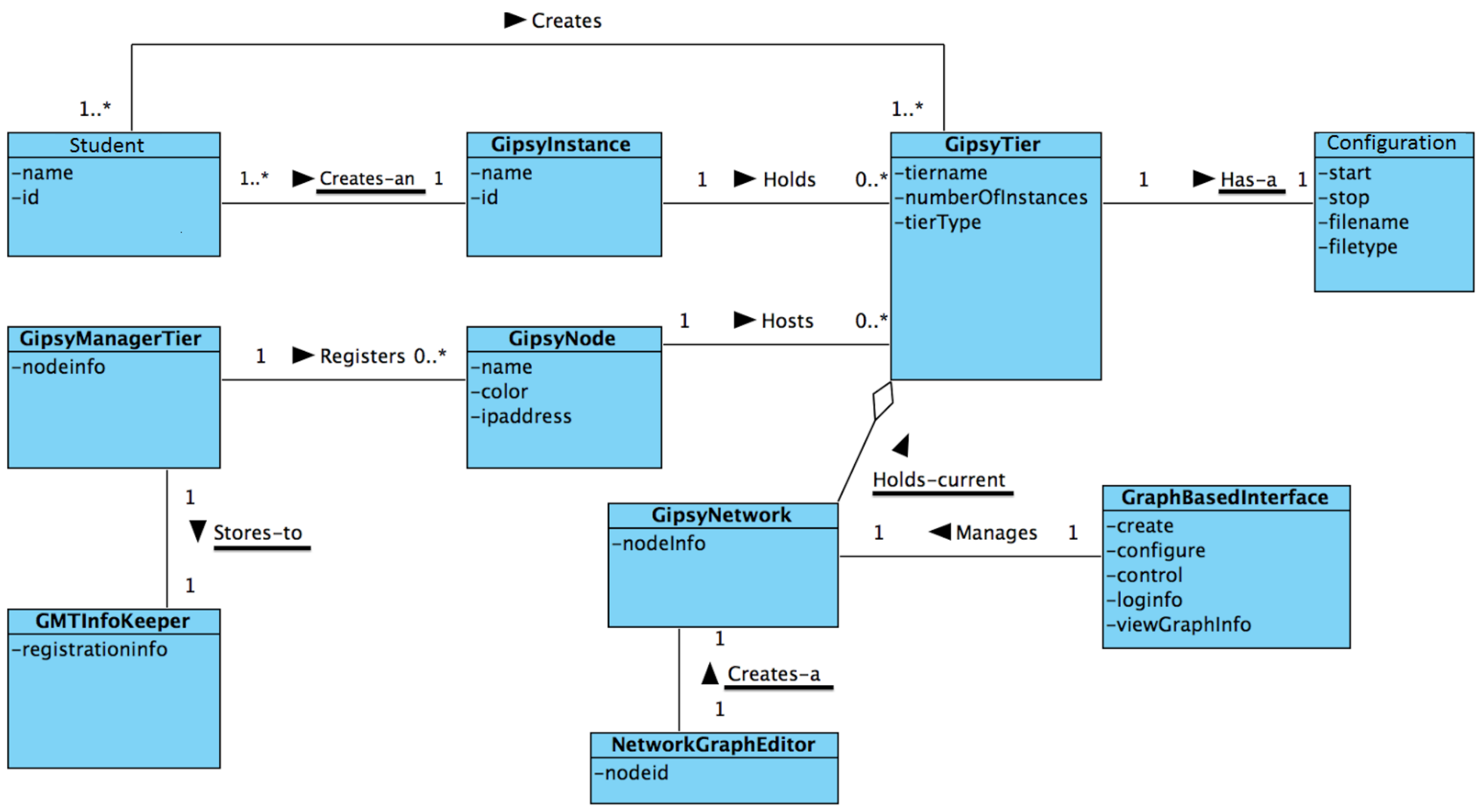

**GIPSY Domain Diagram**

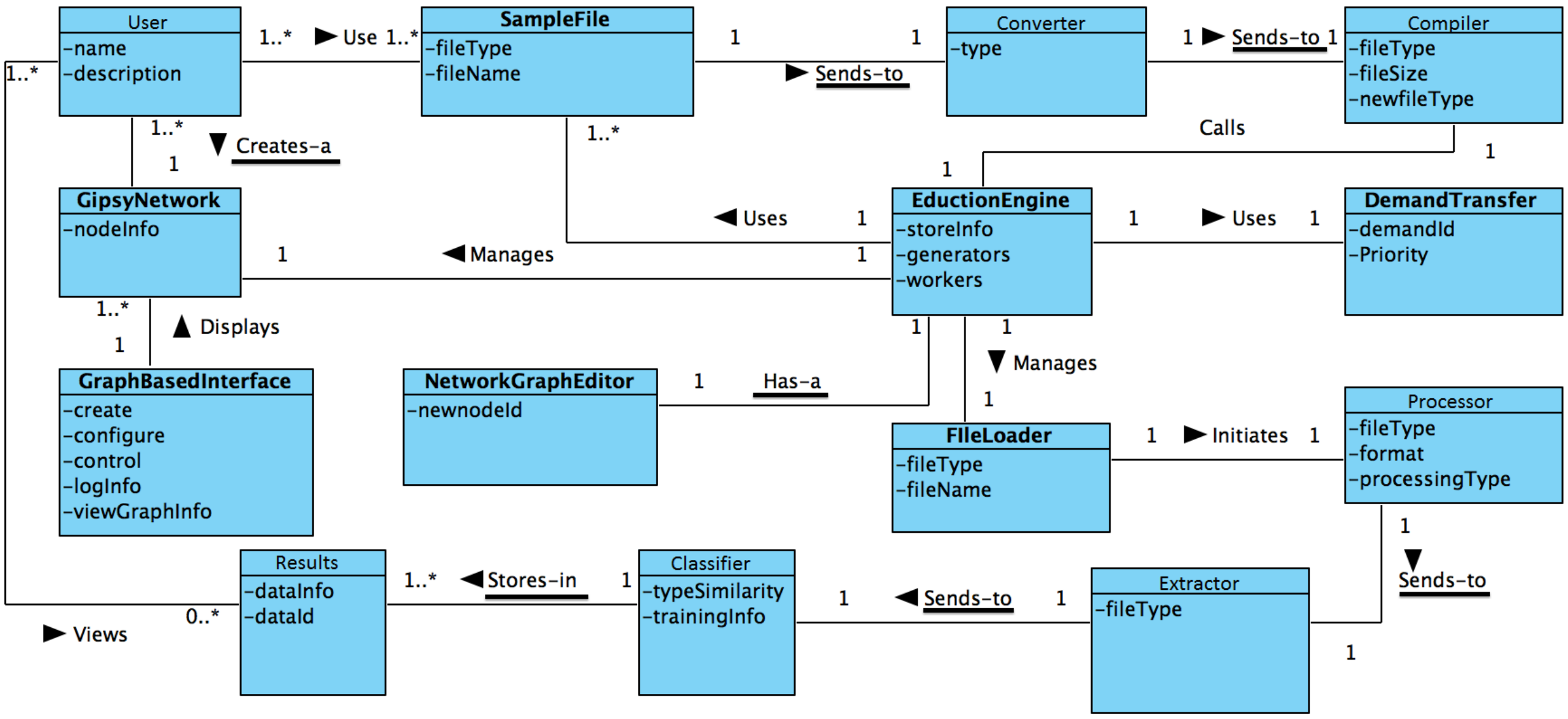

**FUSED DMARF- over- GIPSY Domain Diagram**

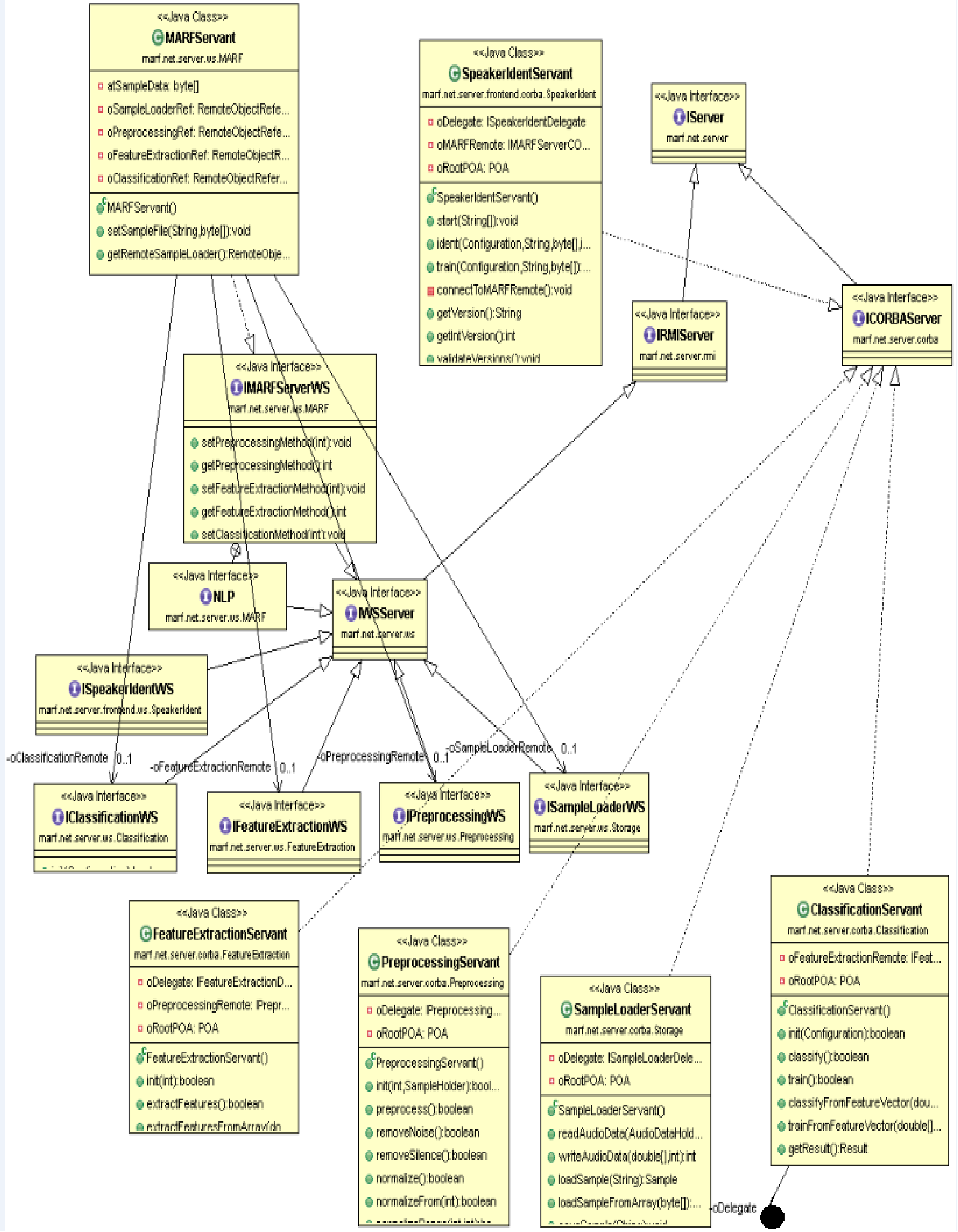

**Class diagram for DMARF**

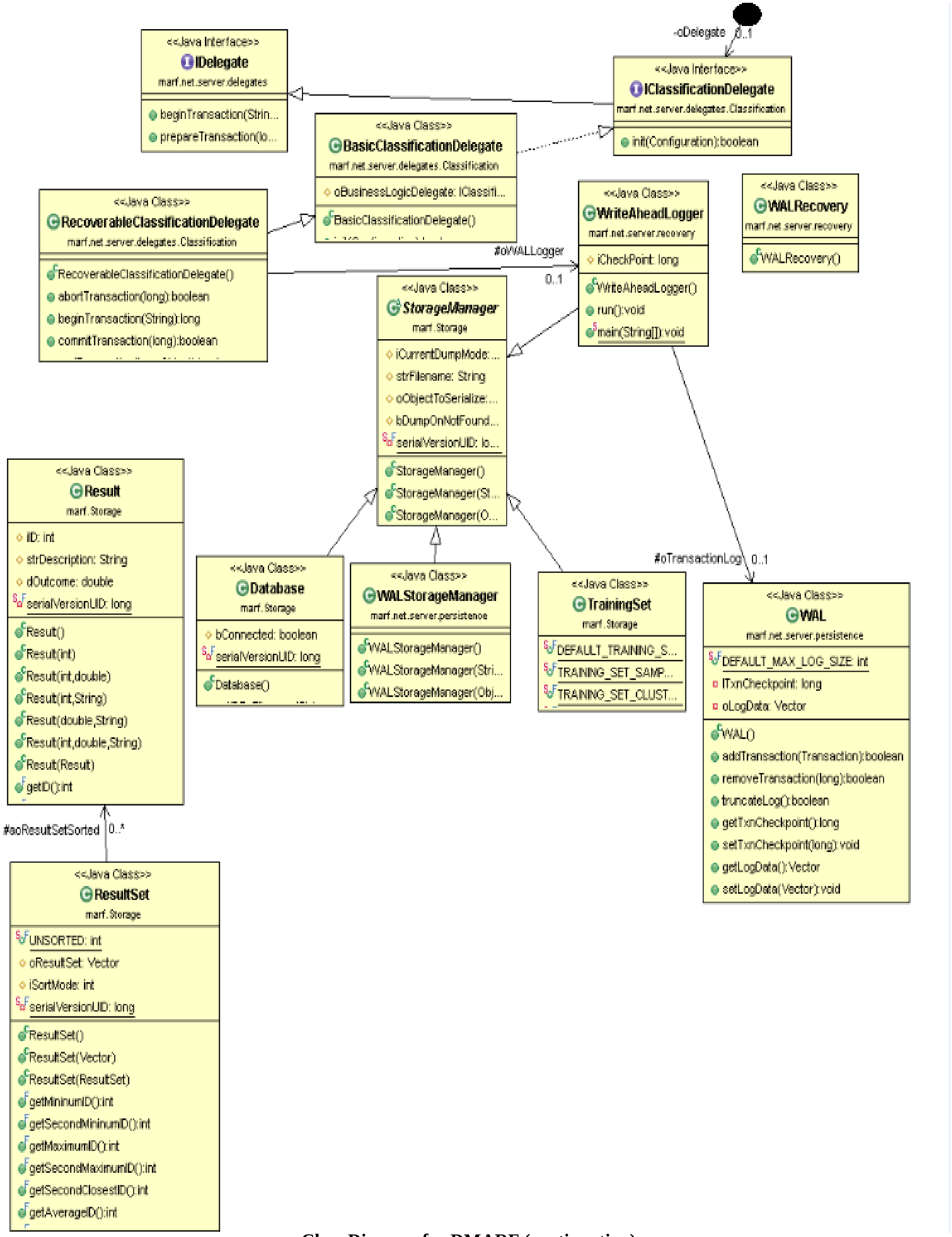

**Class Diagram for DMARF (continuation)**

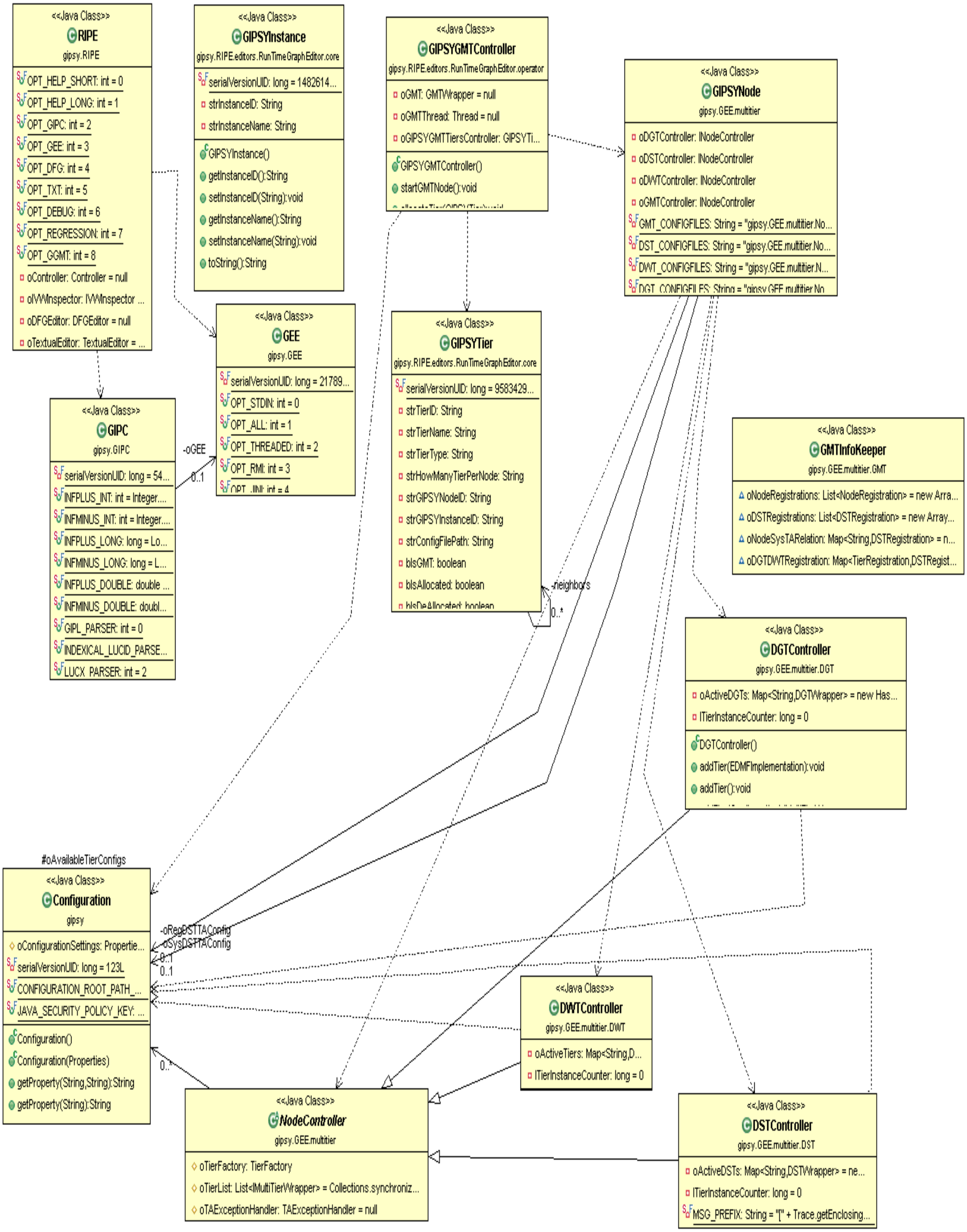

**Class Diagram for GIPSY**

**CONTRIBUTION LIST PM1: TEAM 4**

| | Name | Student ID | Subject |
|---|---|---|---|
| 1 | Afshin Somani | 6765793 | **DMARF**: Developing Autonomic Properties for Distributed Pattern-Recognition Systems with ASSL: A Distributed MARF Case Study<br>**GIPSY**: An Interactive Graph-Based Automation Assistant: A Case Study to Manage the GIPSY's Distributed Multi-tier Run-Time System |
| 2 | Ahmad Al-Sheikh Hassan | 6735029 | **DMARF:** DMARF and its Applications over Web Services<br>**GIPSY:** Scalability Evaluation Of The GIPSY Runtime System |
| 3 | Anurag Reddy Pedditi | 6862322 | **DMARF:** Towards a Self-Forensics Property in the ASSL Toolset<br>**GIPSY:** Distributed Eductive Execution of Hybrid Intensional Programs |
| 4 | Challa Sai Sukesh Reddy | 6847250 | **DMARF:** Towards Autonomic Specification of Distributed MARF with ASSL: Self-healing<br>**GIPSY**: A General Architecture for Demand Migration in a Demand-Driven Execution Engine in a heterogeneous and Distributed Environment |
| 5 | Vijay Nag Ranga | 6745814 | **DMARF:** Self-Optimization Property In Autonomic Specification Of Distributed MARF With ASSL<br>**GIPSY:** Towards Autonomic GIPSY |
| 6 | Saravanan Iyyaswamy Srinivasan | 7090838 | **DMARF:** Autonomic specification of Self-Protection for DMARF with ASSL<br>**GIPSY:** GIPSY Architecture |
| 7 | Hongyo Lao | 6871240 | **DMARF**: On design and implementation of distributed modular audio recognition frame-work: Requirements and specification design document<br>**GIPSY:** Using the General Intensional Programming System (GIPSY) for evaluation of higher-order intensional logic (HOIL) expressions |
| 8 | Zhu Zhili | 6954618 | **DMARF**: Managing distributed MARF with SNMP<br>**GIPSY**: Advances in the design and implementation of a Multi-tier architecture in the GIPSY environment with Java. |

**CONTRIBUTION LIST PM3 (Design Patterns): TEAM 4**

| | Name | Student ID | Design Patterns |
|---|---|---|---|
| 1 | Afshin Somani | 6765793 | Decorator Pattern |
| 2 | Ahmad Al-Sheikh Hassan | 6735029 | Adapter Pattern |
| 3 | Anurag Reddy Pedditi | 6862322 | Proxy Pattern |
| 4 | Challa Sai Sukesh Reddy | 6847250 | Factory Pattern |
| 5 | Vijay Nag Ranga | 6745814 | Observer Pattern |
| 6 | Saravanan Iyyaswamy Srinivasan | 7090838 | Prototype Pattern |
| 7 | Hongyo Lao | 6871240 | Singleton Pattern |
| 8 | Zhu Zhili | 6954618 | State Pattern |